\documentclass[twoside,12pt]{elsarticle}
\pdfoutput=1
\usepackage{hyperref}
\usepackage{epsfig,amsfonts,amsthm}
\usepackage{amsmath,amssymb}
\usepackage{graphicx,color}
\usepackage[utf8]{inputenc}

\newcommand{\be}{\begin{equation}}
\newcommand{\ee}{\end{equation}}
\newcommand{\bea}{\begin{eqnarray}}
\newcommand{\eea}{\end{eqnarray}}

\renewcommand{\Im}{\mathrm{Im }}

\newcommand{\fd}{\phi^\dagger}
\newcommand{\f}{\phi}
\newcommand{\doublet}[2]{ \left(\! \begin{array}{c}#1 \\ #2 \end{array}\!\right) }

\newcommand{\lr}[1]{ \langle #1 \rangle}
\newcommand{\Tr}{\mathrm{Tr}}
\newcommand{\Z}{\mathbb{Z}}
\newcommand{\R}{\mathbb{R}}
\renewcommand{\S}{\mathbb{S}}
\newcommand{\mmmatrix}[9]{ \left(\! \begin{array}{ccc}#1 & #2 & #3\\ #4 & #5 & #6\\ #7 & #8 & #9\\ \end{array}\!\right) }
\newcommand{\mmatrix}[4]{ \left(\! \begin{array}{ccc}#1 & #2 \\ #3 & #4 \\ \end{array}\!\right) }
\newcommand{\toCP}{\xrightarrow{CP}}

\journal{Progress in Particle and Nuclear Physics}

\def\lsim{\mathrel{\rlap{\lower4pt\hbox{\hskip0pt$\sim$}}
    \raise1pt\hbox{$<$}}}         
\def\gsim{\mathrel{\rlap{\lower4pt\hbox{\hskip0pt$\sim$}}
    \raise1pt\hbox{$>$}}}         

\begin{document}

\begin{frontmatter}

\title{ \vspace{1cm} Building and testing models with extended Higgs sectors}
\author{Igor P. Ivanov\footnote{Electronic address: igor.ivanov@tecnico.ulisboa.pt}}
\address{CFTP, Instituto Superior T\'ecnico, Universidade de Lisboa\\
Avenida Rovisco Pais 1, 1049-001 Lisboa, Portugal}

\begin{abstract} 
Models with non-minimal Higgs sectors represent a mainstream direction
in theoretical exploration of physics opportunities beyond the Standard Model.
Extended scalar sectors help alleviate difficulties of the Standard Model
and lead to a rich spectrum of characteristic collider signatures and astroparticle consequences.
In this review, we introduce the reader to the world of extended Higgs sectors.
Not pretending to exhaustively cover the entire body of literature,
we walk through a selection of the most popular examples: the two- and multi-Higgs-doublet models,
as well as singlet and triplet extensions.
We will show how one typically builds models with extended Higgs sectors,
describe the main goals and the challenges which arise on the way,
and mention some methods to overcome them.
We will also describe how such models can be tested, what are the key observables one focuses on,
and illustrate the general strategy with a subjective selection of results.
\end{abstract}

\end{frontmatter}

\tableofcontents



\section{Introduction}

\subsection{The quest for New Physics}

Search for the physics beyond the Standard Model (SM) is today the top-priority task 
for the high-energy physics (HEP) community. The way it feels 
can be paralleled to the early decades of the Age of Discoveries, when Europeans were sailing to open ocean in search 
for new overseas lands. They knew that these lands must exist but had little idea how far they were
and what it would take to reach them.
Today, the HEP community, too, is navigating in open seas. We know that New Physics must exist 
but we have no definitive sign of this new continent on the horizon. 
The recent discovery of the 125 GeV boson at the LHC \cite{Chatrchyan:2012xdj,Aad:2012tfa}
seems to make part of the ``old world'' rather than a passage to New Physics,
and the 750 GeV mirage dissipated upon closer inspection.

On the one hand the Standard Model, the gauge theory of electromagnetic, weak, and strong interactions 
of quarks and leptons, is still extremely successful in describing all laboratory measurements
of how fundamental particles behave. The celebrated Brout-Englert-Higgs mechanism (the Higgs mechanism for short) 
\cite{Englert:1964et,Higgs:1964pj,Higgs:1964ia,Guralnik:1964eu} is a cornerstone of the
electroweak part of the SM \cite{Glashow:1961tr,Weinberg:1967tq,Salam:1968rm}.
The gauge interactions do not require the scalar sector to be minimal,
and this inspires hope to see New Physics through the Higgs window.

On the other hand, we know that the SM is an incomplete theory. 
The strongest indications come from its failure to accommodate dark matter 
and to provide a viable explanation of the baryon asymmetry of the Universe.
It also has several unattractive features in its own construction.
The fundamental scalar field suffers from quadratically divergent loop corrections 
to its mass, which, even if absorbed in the renormalization counterterms, 
lead to quadratic sensitivity to the scale of any possible New Physics. 
The Yukawa sector, which is often called the ugly part of the SM, is just postulated as it is. 
The quark and lepton masses as well as the quark mixing matrix display several hierarchies,
and the SM offers no explanation to any of them. 
Neutrino masses are non-zero but shockingly small, 
and their mixing pattern is quite different from quark mixing. 
Although the minimal Standard Model assumes neutrinos to be strictly massless,
their non-zero neutrino masses can be accommodated by adding right-handed neutrinos
and assuming they get their masses via the same Yukawa-type interactions as charged fermions.
However this would be extremely unnatural, and many see the non-zero neutrino masses
as the first direct indication of New Physics.

In short, many intriguing features cry for explanations, which the SM does not offer.
This motivates the physics community to theoretically explore various constructions 
beyond the SM (bSM). 
One big part of this activity is model building based on extended Higgs sectors.
In this bottom-up approach to New Physics, one does not try to guess the theory 
at the highest energy scale but focuses on its consequences around the electroweak scale.
The scalar sector, which we just start to explore,
is still very weakly constrained, at the ten percent level at best, and 
it can hide new physics treasures waiting to be discovered.
It is this hope of existence of a close ``Higgs continent'' that drives
most of the works on extended scalar sector model building.

This review is an attempt to give a concise and digestible introduction
to several variants of extended Higgs sectors.
As this topic is huge and the literature is immense, this review certainly cannot list all models 
with non-minimal Higgs sectors constructed up to now.
Nor does it pretend to give an exhaustive coverage of all works within the selected models.
The goal of the review is to convey to the reader 
the feeling of how one typically builds and tests extended Higgs sectors;
to describe the ideas, to set up the basic notation,
and to overview the methods.
With several popular models chosen as illustrative examples,
we will dig deeper into the challenges which arise 
and into the ways to overcome them.

The non-minimal scalar sectors selected here are two- and multi-Higgs-doublet models
and singlet and triplets extensions, with a few other examples mentioned in passing. 
A justification of this choice is that we wish to focus on those
bSM models in which the non-minimal Higgs sector plays the defining role.
We will not dwell on supersymmetric models as well as gauge unification 
models, as their construction and phenomenology is strongly shaped by
other ingredients.
Nor will we consider composite Higgs models, extra dimensions, or technicolor, 
since they represent very distinct branches of the non-minimal
Higgs physics and definitely require introduction of other techniques.
An extensive discussion of such models can be found in the pre-LHC era collective report
\cite{Accomando:2006ga}.

The rest of this Section will provide a short literature guide, a brief historical account
of the Higgs sector developments, and a quick summary of the popular computer tools
used to explore bSM models.
In Section~\ref{section-SM}, we start with a brief pedagogical recap 
of the Brout-Englert-Higgs mechanism in the SM.
Section~\ref{section-2HDM} is devoted 
to the two-Higgs-doublet model (2HDM), a very popular non-minimal Higgs sector.
This is followed in Section~\ref{section-NHDM} by a discussion of more than two Higgs doublets,
with the emphasis on novel features absent in 2HDM. 
Section~\ref{section-beyond-doublets} deals with scalar sectors
in which the Higgs doublet is accompanied by extra singlets and triplets.
Finally, in Section~\ref{section-astro} we collect several astroparticle 
and cosmological consequences of extended Higgs models.

\subsection{Literature overview}

The Higgs mechanism is an essential part of all
textbooks, reviews, and lectures on the Standard Model, on various bSM models, 
and on their phenomenological manifestations.
Apart from the general-purpose textbooks on HEP,
we highlight the very pedagogical accounts by Okun \cite{Okun:1982ap} and Rubakov \cite{Rubakov:2002fi},
and the famous Higgs Hunter's Guide by Gunion, Haber, Kane, and Dawson \cite{Gunion:1989we}.
The monograph of Branco, Lavoura, and Silva on $CP$-violation \cite{CPV-book} provides a detailed
account of $CP$-violating and flavor physics effects which arise in multi-Higgs-doublet models.

Back in 1998, Spira published the classical review \cite{Spira:1997dg} on the Higgs boson production and decay channels 
in the SM and in the much anticipated MSSM, the minimal supersymmetric extension of the SM.
Very recently, he completed a substantial update of that review \cite{Spira:2016ztx}.
In 2002, as the LHC project hit its construction stage, Carena and Haber presented a detailed description 
of the Higgs sector in the SM and MSSM \cite{Carena:2002es}.
Appearing a few years later, the monumental two-volume review by Djouadi describes in meticulous details
the expected Higgs boson properties in the SM \cite{Djouadi:2005gi} and in MSSM \cite{Djouadi:2005gj}.
The workshop report \cite{Accomando:2006ga} offers a very broad and detailed update 
on the status of many extended scalar sectors, both mainstream and exotic ones, 
ranging from 2HDM and supersymmetric models to little Higgs models, extended gauge groups, extra dimensions, 
composite, and Higgsless models.
Those volumes essentially summarize the state of the art in the pre-LHC era.
As for the recent literature, we highlight the very detailed review on the two-Higgs-doublet models 
by Branco et al \cite{2HDM-review}, the recent workshop report \cite{Contino:2016spe}
on the prospects of the Higgs and EW symmetry breaking studies at a 100 TeV pp collider,
and the fourth report of the LHC Higgs cross section Working Group \cite{deFlorian:2016spz},
which includes calculations for several benchmark models with extended scalar sectors.
In fact, the website of the LHC Higgs Cross Section Working Group for BSM Higgs \cite{LHCHXSWG3}
is a useful source of information on mainstream models with extended Higgs sectors.

Another valuable pedagogical resource is the constant flux of lecture notes on various aspects
of the Higgs mechanism and its experimental signatures, both within and beyond the SM.
A recent and very detailed account of the SM structure and its experimental tests at colliders 
was written by Altarelli \cite{Altarelli:2013tya}.
The summary of the LHC Run 1 experimental program which led to the 125 GeV boson discovery 
was given by Bernardi and Herndon \cite{Bernardi:2012gb}.
The lectures by Vysotsky \cite{Vysotsky:2011zz} deal with the details of the Higgs mechanism, with an emphasis on
practical calculations of loop corrections.
Other introductory lectures of the SM Higgs mechanism include \cite{Pich:2005mk,Langacker:2009my,Iliopoulos:2013rna}.

Many lectures accompany the SM Higgs mechanism with a selection of bSM models 
with extended scalar sectors \cite{Reina:2005ae,Wells:2009kq,Vysotsky:2011zz,Logan:2014jla,Pich:2015lkh}.
Lectures from the pre-LHC era are often dominated by the MSSM Higgs physics and the prospects of its experimental discovery;
a nice and detailed account is given in 2004 TASI lectures by Reina \cite{Reina:2005ae}.
Wells \cite{Wells:2009kq} gives a thorough description of the Higgs bosons in the 2HDM and MSSM, 
and offers an introduction to models in which the Higgs mixes with exotic scalars such as the ``hidden world'' and extra dimensions models.
A very gentle introduction to the Brout-Englert-Higgs mechanism 
and Higgs boson(s) properties in the SM, 2HDM, and Higgs triplet models
is presented in the 2013 TASI lectures by Logan \cite{Logan:2014jla}.
The report of lectures given in 2015 at the Cracow School of Theoretical Physics 
by Pich \cite{Pich:2015lkh} covers multi-doublet models and singlets.
A very recent update on $CP$ and flavor violation in bSM models, with the new Higgses playing essential role, 
was given by Gori \cite{Gori:2016lga}.
An interesting attempt to explain the essence of the Higgs mechanism to the mathematical audience
can be found in \cite{Hamilton:2015sit}.
Finally, we mention the educational paper by Maldacena \cite{Maldacena:2014uaa} 
where he draws an enlightening socioeconomic analogy of the Higgs mechanism,
which is worth reading even for experts.

\subsection{Historical excursion}\label{section-historical}

Many entry-level materials on the Brout-Englert-Higgs mechanism 
put it that in 1964, independently from each other,
Brout and Englert \cite{Englert:1964et},
Higgs \cite{Higgs:1964pj,Higgs:1964ia}, and Guralnik, Hagen and Kibble \cite{Guralnik:1964eu}
came up with a mechanism which gives masses to $W$ and $Z$ bosons and to fundamental fermions,
and that this theoretical insight was finally confirmed in 2012 with the discovery of the Higgs boson
at the LHC \cite{Chatrchyan:2012xdj,Aad:2012tfa}.
The real situation was of course much more intricate than this oversimplified scenario.
The groundbreaking papers did not actually address the electroweak theory because the theory itself did not exist yet.
In those years, the physics community was trying to figure out the underlying physical picture 
behind the proliferation of hadrons in experiments. 
It was exactly in this context that Nambu and Jona-Lasinio brought the idea of spontaneous symmetry
breaking from condensed matter physics to particle physics \cite{Nambu:1961tp,Nambu:1961fr}.
In their own words, ``{\em The basic principle underlying
the model is the idea that field theory may admit ... extraordinary
(nontrivial) solutions that have less symmetries than are built into the Lagrangian.}''
This description took its roots in the Bardeen-Cooper-Schrieffer microscopic 
theory of superconductivity and specifically in the fact that the photon does acquire an effective mass 
inside superconductor in spite of the underlying theory being gauge-invariant \cite{Nambu:1960tm}.
Goldstone \cite{Goldstone:1961eq} considered field theories with this ``superconductor'' solutions
and noticed that if a model possesses a continuous symmetry 
and if the scalar field acquires a non-zero vacuum expectation value (vev)
which breaks this symmetry, then a massless scalar appears.
He conjectured it to be a general result for Lorentz-invariant theories, 
which was proved in three different ways in the subsequent paper 
by Goldstone, Salam, and Weinberg \cite{Goldstone:1962es}. 
Since no massless Goldstone boson is observed in experiment,
this mechanism of mass generation seemed to be at odd with reality. 
The work of Nambu also falls in this category,
though it is not a fundamental scalar field but rather a scalar bilinear of the fermion field that acquires vev.

In the next two years, amidst the flurry of short papers 
\cite{Schwinger:1962tn,Anderson:1963pc,Klein:1964ix,Gilbert:1964iy,Englert:1964et,Higgs:1964pj,Higgs:1964ia,Guralnik:1964eu}
and intensive discussions at the Seminar on Unified theories of elementary particles in Rochester in 1963, the solution emerged.
Already Anderson puts in \cite{Anderson:1963pc} this vision explicitly:
''{\em It is likely ... that the way is now open for a degenerate-vacuum
theory of the Nambu type without any
difficulties involving either zero-mass Yang-Mills gauge
bosons or zero-mass Goldstone bosons. These two
types of bosons seem capable of "canceling each other
out" and leaving finite mass bosons only.}''
In parallel, in the Soviet Union, --- in the ``island'' poorly connected to the rest of the scientific world \cite{Polyakov:1992yi}, ---
Migdal and Polyakov, still undergraduate students at that time, were trying too to find the solution.
They figured it out in 1965 and for more than a year were having hard time breaking through 
the system until finally publishing it \cite{Migdal:1966tq}.

So, the idea was in the air and just waiting to be cast in an appropriate mathematical form.
The sequence of events that led to the proposal of the Brout-Englert-Higgs-Hagen-Guralnik-Kibble mechanism
of electroweak symmetry breaking, --- which is almost universally shortened to the Higgs mechanism, ---
together with valuable personal recollections was covered in a number of historic talks and reviews
by the protagonists themselves \cite{Polyakov:1992yi,Brout:1998qb,Higgs:2002ht,Guralnik:2009jd,Kibble:2009zz,Englert:2013bra,Kibble:2015mwa},
including the Nobel lectures by Englert \cite{Englert:2014zpa} and Higgs \cite{Higgs:2014aqa}, 
and by others \cite{Ellis:2012sy,Quigg:2015cfa,Wells:2016exe}. 
Perhaps, the most in-depth account of what was happening around 1964 was given, from a personal perspective,
by Guralnik in \cite{Guralnik:2009jd}.
A valuable historic account was also presented by Bernstein in his review \cite{Bernstein:1974rd} back in 1974.

In the years that followed, Higgs \cite{Higgs:1966ev} looked more closely into the phenomenology of the Abelian Higgs model,
including the properties of the new scalar boson,
and explicitly mentioned that a certain variable change can render the original gauge symmetry hidden rather than broken.
In 1967, Kibble \cite{Kibble:1967sv} investigated the phenomenon for non-Abelian gauge symmetries,
paying special attention to the equivalence of the physical picture in different gauges.
In the same year, Weinberg \cite{Weinberg:1967tq} and, a bit later, Salam \cite{Salam:1968rm} 
finally put together the electroweak theory, the Brout-Englert-Higgs mechanism,
and the fermions to create what would come to be known as the Standard Model.
Developments in the early 1970s culminating in the proof of the renormalizability of 
massive Yang-Mills theories \cite{'tHooft:1971rn} and the development of the new regularization
and renormalization techniques \cite{'tHooft:1972fi} opened the way to precise calculations.
The model based on the $SU(2)_L \times U(1)_Y$ gauge symmetry was intensively explored,
but for several years it had to coexist with rival theories. 
After the weak neutral current discovery in 1973 by the Gargamelle experiment \cite{Hasert:1973ff}
the SM became widely accepted by the physics community.

Concerning the scalar sector, the situation was somewhat ambiguous in those early days.
Today the students are taught that the Higgs boson, being the only ``scalar witness'' of the Higgs mechanism, 
is a perfect experimental probe of the mechanism itself.
However in the early 1970's many viewed the presence of a fundamental scalar field as an awkward feature of the model
rather than its strength. The spirit of arguments against the fundamental scalars was nicely summarized 
by Wells \cite{Wells:2016exe}.
It was only in 1975 that the first detailed investigation into the phenomenological properties
of the Higgs boson and its search strategies was presented \cite{Ellis:1975ap}.
The hunt had begun, and a very useful guide by Gunion, Haber, Kane, and Dawson \cite{Gunion:1989we} 
helped hunters to stay focused. 

Mid-70s were also revolutionary for flavor physics.
The fourth quark anticipated on the basis of the Glashow–Iliopoulos–Maiani mechanism \cite{Glashow:1970gm}
was discovered in 1974 \cite{Aubert:1974js,Augustin:1974xw}.
The fifth, bottom quark as well as the $\tau$-lepton followed in 1977.
This did not come as a complete surprise: in fact, back in 1973 Kobayashi and Maskawa \cite{Kobayashi:1973fv}
had already realized, on mathematical grounds, that three fermion generations 
can naturally induce $CP$-violation via the charged weak currents.
With the discovery of the open-flavor mesons and the quickly accumulating systematics
of their decays, the Cabibbo-Kobayashi-Maskawa (CKM) matrix of quark mixing started to take its shape.
By the end of 1970s, the hierarchical structure of the quark masses and mixing patterns
became evident and, without any explanation from the SM,
it grew into the so-called flavor puzzle \cite{Feruglio:2015jfa}. 
Attempts to discern a rationale in these patterns and to come up with a (more) natural
source of $CP$-violation called upon various models beyond the Standard Model,
including models with extended Higgs sectors.

It was then that came the first big wave of model-building activity with non-minimal
Higgs sectors. The two most popular frameworks were the standard $SU(3)_c \times SU(2)_L\times U(1)_Y$ gauge theory 
with several Higgs doublets,
and unification models with larger gauge groups such as the Pati-Salam model \cite{Pati:1973uk}
with the gauge group $SU(4) \times SU(2)_L \times SU(2)_R$, as well as its low-energy limit, 
the left-right symmetric model \cite{Mohapatra:1974gc,Senjanovic:1975rk},
and the Georgi-Glashow $SU(5)$ model \cite{Georgi:1974sy}.
The new fundamental fields including new Higgs fields proliferated in these models, 
but if they stay heavy, one recovers the SM in the low energy limit.
These models often equipped with discrete symmetries provided insights into
fermion masses and mixing.

Since in this review we focus on models with extended scalar sector but with the SM gauge group,
we will not dwell on the developments in gauge-unified theories, 
and proceed with the brief history of multi-Higgs-doublet models.
As said above, the two driving forces behind these early efforts were 
the attempts to find an appealing explanation for the $CP$-violation in weak interactions
and to relate the quark mixing pattern with their masses.
In 1973, T.~D.~Lee \cite{Lee:1973iz,Lee:1974jb} suggested
that there could exist not one but two Higgs doublets
and that the $CP$-violation can originate 
from the relative phase between their vevs.
The lagrangian of the model is perfectly $CP$-invariant, 
no complex parameter is present anywhere,
yet the vacuum configuration breaks this invariance spontaneously, as a result of minimization,
leading to $CP$-violating Higgs boson exchanges.
Notice that at that time the Kobayashi-Maskawa three-generation idea had just appeared 
and by no means was an accepted paradigm.
Thus, Lee's model should be regarded as an alternative to CKM rather than an additional source of $CP$-violation.

It was immediately understood, though, that multi-doublet models generically lead to 
dangerously large flavor-changing neutral currents (FCNC), which would easily violate the experimental bounds on kaon mixing.
To remedy this problem, the idea of natural flavor conservation (NFC) 
was put forth in 1976 by Glashow and Weinberg \cite{Glashow:1976nt} and Paschos \cite{Paschos:1976ay}. 
One requires that all right-handed fermions of the same charge couple only to one Higgs doublet. 
Technically, this can be guaranteed 
by an extra global $\Z_2$ symmetry which flips the signs of the Higgs doublet
together with the corresponding right-handed fermions. However, imposing this symmetry on Lee's 2HDM
immediately blocks the possibility of $CP$-violation, either spontaneous or explicit.
With three Higgs doublets, this clash between NFC and $CP$-violation can be resolved,
as shown first by Weinberg in 1976 \cite{Weinberg:1976hu}
and later, in the form of spontaneous $CP$-violation,
by Branco \cite{Branco:1979pv,Branco:1980sz}. 
Weinberg's model stirred up much interest,
but around 1981 various authors reported an unacceptably large $CP$ parameter $\epsilon'/\epsilon \approx -0.05$
which it seemed to predict \cite{Deshpande:1981na,Sanda:1981ca,Donoghue:1981sd}.
More accurate analysis of flavor observables performed by Branco, Buras, and Gerard in 1985 \cite{Branco:1985ch,Branco:1985pf} 
resurrected the model, granting it an extra decade of a reasonably healthy life,
until it was finally declared dead --- at least as the only source of $CP$-violation --- at the turn of the century \cite{Chang:2000ck}.

In parallel, an equally intensive model-building activity focused on relating the quark mixing angles
with the mass ratios across the generations in the lepton, up-quark and down-quark sectors. 
All these patterns are hierarchical, --- the neutrino mixing pattern being completely unknown at that time, --- 
and none is explained within the SM.
It was tempting to link them either by finding a natural reason why the mass ratios should
shape the mixing, or by finding the origin of the two hierarchies in the same symmetry-based construction.

Apart from highly influential but rather {\em ad hoc} ideas such as the texture-based approach \cite{Fritzsch:1977za} 
and the Froggatt-Nielsen framework \cite{Froggatt:1978nt},
this activity began within multi-Higgs-doublet models in late 1970s, 
first with the two-generation quark mixing and later with the full CKM.
The main theme was imposing an extra global symmetry group acting on scalars and fermions. 
One starts with the lagrangian which is invariant under this flavor symmetry group, 
but the vev alignment of the scalar fields breaks it. Symmetry-constrained Yukawa matrices
describing coupling between fermions and Higgs fields collapse after the Higgs potential minimization
into the quark mass matrices $M_d$ and $M_u$, in which the symmetry is not manifest anymore.
The hope is to find such a symmetry setting, --- the group itself and the representation assignment, ---
which would simultaneously reduce the number of free parameters, 
break the flavor symmetry in order to reproduce the observed masses and mixing,
and lead to extra relations among measurable quantities beyond the SM or predict new phenomena.
In short, the hunt was for models which are neither too restrictive nor too loose.

The first attempt was, naturally, to stick to the attractive NFC hypothesis
and devise a model with the Cabibbo mixing angle calculable in terms of mass ratios.
Gatto and collaborators showed that to be impossible 
within the $SU(2)_L \times U(1)_Y$ gauge group \cite{Barbieri:1978qh}, even if one applies NFC 
after electroweak symmetry breaking \cite{Gatto:1979mr}.
The models were too rigid: whenever they predicted the Cabibbo angle, its value was fixed by the symmetry itself,
independent of the quark mass ratios, and was off by far from the experimental result.

A way out was to relax the strict NFC principle and allow for tree-level FCNCs.
Plenty of possibilities arose along this way. The initial attempts stuck to permutation
symmetry groups with various representation assignments, in particular $S_3$
\cite{Derman:1977wq,Derman:1978nz,Derman:1978rx,Pakvasa:1977in} and $S_4$
\cite{Pakvasa:1978tx,Yamanaka:1981pa,Brown:1984mq} symmetry groups.
Larger discrete groups were also contemplated \cite{Segre:1978ji}.
Quite remarkably, such models led to interesting relations between mixing and masses,
which in the beginning on 1980s were compatible with experiment. 
We will cover these early examples in section~\ref{section-NHDM-examples}.

As the measurements of the flavor physics parameters were getting more precise,
it became increasingly difficult to accommodate them in a minimalistic, symmetry based construction.
Another source of worries was that the top-quark mass was predicted by most of the early
symmetry-based models to lie in the 20-30 GeV range, which was soon ruled out experimentally.
More sophisticated models such as Ma's four-Higgs-doublet model with $S_3 \times \Z_3$ symmetry 
\cite{Ma:1990qh,Ma:1991eg} were introduced.
Also, the experience gained with the symmetry-based multi-Higgs models made it clear
that the global symmetries must eventually break completely if one wants to produce
the physical quark masses, mixing, and $CP$ violation \cite{Leurer:1992wg}.
It all suggested that the relation between symmetries and observables was less immediate
and probably less beautiful than initially anticipated.
The search carried on but with reduced intensity, also taking into account that 
the gauge unification models and supersymmetry attracted much attention in the 80-90's.
It is from the latter, however, that an extra support arrived and motivated many to look deeper into 
the phenomenology of 2HDM, which reproduces the scalar sector 
of MSSM, the minimal supersymmetric extension of the SM \cite{Gunion:1984yn,Gunion:1986nh,Gunion:1989we}.
Interest in the phenomenology of models with more than two doublets
was also revived, see \cite{Cheng:1987rs,Lavoura:1994fv,Botella:1994cs,Grossman:1994jb} 
to mention a few key works.

In the meantime, models with Higgs fields in other $SU(2)_L$ representations such as singlets \cite{Silveira:1985rk,McDonald:1993ex}
and triplets 
\cite{Konetschny:1977bn,Magg:1980ut,Schechter:1980gr,Cheng:1980qt,Georgi:1985nv,Chanowitz:1985ug,Gunion:1989ci} 
started to appear in 1980's.
Triplets were summoned as one of the sources of neutrino masses 
\cite{Konetschny:1977bn,Magg:1980ut,Schechter:1980gr,Cheng:1980qt,Mohapatra:1980yp,Gelmini:1980re}
and also in the search of novel phenomenological signals, especially
when the large vevs are compatible with the $\rho$-parameter equal to one, 
as in the Georgi-Machacek model \cite{Georgi:1985nv,Chanowitz:1985ug}.
In contrast, gauge singlets lead to less remarkable phenomenology but have interesting 
astroparticle and cosmological consequences such as natural dark matter (DM) candidates and
a strong electroweak phase transition in the early Universe.

One of motivations to build models with extended scalar sectors was the search for predictive
models of neutrino masses and mixing. 
The evidences for neutrino oscillations were accumulating through the 80-90's \cite{Fisher:1999fb},
which necessitated explanations of the tiny neutrino masses \cite{Weinberg:1979sa}. 
The seesaw paradigm emerged \cite{Minkowski:1977sc,Mohapatra:1979ia,GellMann:1980vs} 
and the Higgs triplets appeared to be a convenient means
of giving small masses to Majorana neutrinos through the seesaw Type II mechanism 
\cite{Magg:1980ut,Schechter:1980gr,Cheng:1980qt,Mohapatra:1980yp,Lazarides:1980nt}.
As for the neutrino mixing pattern, two possibilities for the solar oscillation parameters coexisted for some time: 
the small-angle and large-angle solutions.
Many believed that neutrino mixing pattern was hierarchical similarly to the CKM matrix, 
and the models of that time reflected this expectation, see e.g. the $S_3\times \Z_3$ model \cite{Ma:1991eg}.
Early 2000's brought a strong shift of the perspective, as the large-angle solution was confirmed, 
and the neutrino oscillation parameters started to display the tri-bimaximal pattern, hinting at a possible
symmetry-based origin \cite{Harrison:2002er,Harrison:2002kp}.
A detailed description of the plethora of neutrino mass models, many of which were based
on several Higgs doublets or triplets as well as on multiple gauge-singlet scalars called flavons, can be found
in the reviews of that epoch \cite{King:2003jb,Altarelli:2004za} and in the more recent ones 
\cite{Altarelli:2010gt,King:2013eh,King:2015aea,King:2017guk}.
Although such models usually involve other fields apart from extra Higgses,
the Higgs sector often plays an important role.

At present, models with extended scalar sectors often aim to simultaneously explain 
several weak points of the SM. Typically, when proposing a specific model,
one looks at its collider predictions, checks its compatibility with the LEP and LHC searches,
with the electroweak precision tests, with the flavor physics data including
rare $B$-meson decays, and with the Higgs boson measurements.
It is of course desirable that the model alleviates one or several a-few-sigma deviations, 
which are still standing.
If the model contains DM candidates, one calculates their relic abundance, scattering on visible matter,
decay and annihilation patterns, and compares the results with cosmological data,
with the direct DM detection constraints, and checks for possible astrophysical signals.

Two decades ago, when one was comparing a model with experimental data, one would typically
show functional dependences of the observables on the free parameters of the model.
With the present day computer power, it is customary to perform extensive numerical scans
of the multi-dimensional parameter space and present the results in the form of {\em scatter plots}.
When doing so, one should keep in mind that numerical scans of multi-dimensional spaces 
with non-trivial boundaries may miss special ``pockets'' of parameter space with non-typical predictions for observables.
Also, in the last decade, if one suggests that experiments check a specific New Physics model,
one often produces ``benchmark points'': a selection of parameter sets
displaying the most characteristic features of the model. 

Building concrete models and comparing them with the data is not the only direction of research.
Many works, staying on a more formal side, aim to understand fundamental and mathematical 
symmetry-related issues at the interplay of family symmetries and $CP$-invariance, 
explicit and spontaneous breaking of various symmetries,
fermion masses and mixing, structures in the scalar potential, and others.
This complementary activity looks to derive rigorous results 
which can be applied to broad classes of models. Bypassing the trial-and-error method, 
they should guide the phenomenology-oriented bSM model building activity
through the meander of subtle mathematical properties of their models.
Examples of these works will be mentioned in sections~\ref{section-2HDM} and \ref{section-NHDM}.

\subsection{Computer tools}

Computer tools play a vital role in focused exploration
of collider phenomenology and astroparticle signals of bSM models.
These tools can be grouped into several categories. ``Theory front-end'' calculators
implement specific bSM models, generate vertices, 
and transfer them to event generators.
Versatile general-purpose matrix element generators such as 
\texttt{MadGraph} \cite{Alwall:2014hca}, \texttt{CalcHEP} \cite{Belyaev:2012qa}, and \texttt{Sherpa} \cite{Gleisberg:2008ta} 
construct and evaluate Feynman diagrams corresponding to a desired process, 
and then feed the resulting hard process to hadronization codes such as \texttt{Pythia} \cite{Sjostrand:2014zea}.
On the dark matter front, \texttt{MicrOMEGAs} \cite{Belanger:2013oya} computes DM observables in generic 
extensions of the SM. 
Practical introduction into some of these tools can be found in lectures by 
Kong \cite{Kong:2012vg} and Vicente \cite{Vicente:2015zba}.

Specialized tools check consistency of predictions of a given model with the available experimental results.
Models with extended scalar sectors typically contain a variety of novel Higgs bosons
which can be produced at colliders and predict modifications of the 125 GeV Higgs properties 
with respect to the SM. Checking the compatibility of these predictions
with the body of experimental results from LEP, Tevatron, and the LHC is conveniently done via 
public codes \texttt{HiggsBounds}, the latest version 4 described in \cite{Bechtle:2013wla}, 
and \texttt{HiggsSignals} \cite{Bechtle:2013xfa}. 

The software package \texttt{SARAH} \cite{Staub:2013tta,Staub:2015kfa} based on the computer algebra 
system Mathematica allows the user to implement and conveniently explore their favorite SUSY and non-SUSY models. 
It analytically calculates mass matrices, tadpole equations, vertices, one-loop corrections
to tadpoles and self-energies, as well as two-loop renormalization group equations (RGE). 
Its output can be interfaced to numerical codes such as \texttt{SPheno} \cite{Porod:2011nf} 
which solves RGE and calculates the mass spectra and various key phenomenological quantities,
or \texttt{Vevacious} \cite{Camargo-Molina:2013qva},
a dedicated tool which efficiently minimizes complicated scalar potentials. It first finds 
all extrema of the tree-level potential via homotopy continuation method \cite{Maniatis:2012ex} and
then uses them as seeds for gradient-based minimization of the one-loop effective potential.
For metastable minima, it evaluates the tunneling time into the deeper one.
The analysis of numerous lepton and quark flavor observables is automatized 
with the interface tool \texttt{FlavorKit} \cite{Porod:2014xia} built upon \texttt{SARAH} and \texttt{SPheno}.

Two other packages, \texttt{HEPfit} \cite{HEPfit} and \texttt{GAMBIT} \cite{GAMBIT},
which are being developed now, will provide public versions of fitting codes to 
facilitate testing user's bSM models against various experimental constraints. 

Although many extended scalar sectors are already included in general-purpose event generators,
there exist several ``theory front-end'' calculators designed for specific Higgs sector.
$CP$-conserving two-Higgs-doublet model was among the first to be implemented in a specialized tool \texttt{2HDMC}
\cite{Eriksson:2009ws}. \texttt{ScannerS} \cite{Coimbra:2013qq} is another code aiming for detailed analysis of parameter spaces 
in generic multi-scalar potentials. It minimizes the potential, checks for tree-level unitarity constraints,
and when doing so, it takes into account symmetries of the potential.
As for now, the code includes complex singlet extension, with and without DM candidates, and 2HDM.

The Georgi-Machacek model, a custodial-symmetry preserving triplet model, 
was implemented in the specialized calculator \texttt{GMCALC} \cite{Hartling:2014xma}. 
Together with \texttt{FeynRules} and \texttt{MadGraph5}, 
a fully automated tool chain was built in \cite{Degrande:2015xnm}
and used to generate the NLO distributions at the LHC and to 
explore in detail the LHC signals of the Georgi-Machacek model.


\section{The Brout-Englert-Higgs mechanism in the Standard Model}\label{section-SM}

\subsection{Abelian Higgs model}

To keep the exposition pedagogical and to set up notation, 
we begin with a recap of the Brout-Englert-Higgs mechanism
in the simplest setting known as the Abelian Higgs model. This is the quantum electrodynamics (QED) 
with the spontaneous symmetry breaking phenomenon to render the gauge boson massive.
For completeness, we mention another path to the massive $U(1)$ gauge theory based on the Stueckelberg mechanism, 
see \cite{Ruegg:2003ps} for a review, but it cannot be easily incorporated in the standard electroweak picture.

Consider the QED lagrangian:
\begin{equation}
{\cal L}_{QED}=-\frac{1}{4}F_{\mu\nu}F^{\mu\nu}+\bar \psi (iD_\mu\gamma^\mu -m)\psi\,, \label{QED}
\end{equation}
where $D_\mu=\partial_\mu -ig q A_\mu$ is the covariant derivative, with $A_\mu(x)$ being the gauge field,
and $F_{\mu\nu} = \partial_\mu A_\nu - \partial_\nu A_\mu$.
The field $\psi(x)$ describes the Dirac electron with charge $q$ in units of the coupling constant $g$.
This theory has the $U(1)$ gauge symmetry: the lagrangian is invariant under the local phase rotations
of the electron field and simultaneous shifts of the gauge field:
\be
\psi(x) \to e^{iq\alpha(x)}\psi(x)\,, \quad
A_\mu(x) \to A_\mu(x)+\frac{1}{g}\partial_\mu \alpha(x)\,.\label{gauge-QED}
\ee
Quantization of the electromagnetic field $A_\mu$ requires introduction of a gauge-fixing term and leads to the massless photon.

We now want a similar theory based on gauge symmetry but with a massive photon. Adding the mass term by hand 
${\cal L}_{mass}=-m_A^2 A_\mu A^\mu/2$ does not work because this term is not gauge invariant.
The Abelian Higgs mechanism is a way out in this situation.
We introduce a complex scalar field $\Phi(x)$ with charge $q_\phi$ which participates in the gauge
interactions and interacts with itself via the scalar potential $V$:
\be
{\cal L}={\cal L}_{QED}+(D_\mu\Phi)^*(D^\mu \Phi)-V(\Phi)\,.\label{L0}
\ee
Here, $D_\mu=\partial_\mu -iq_\phi g A_\mu$ when it acts on the scalar field. 
We build $V(\Phi)$ as polynomial of $\Phi^\dagger \Phi$, and to keep the theory renormalizable, 
we limit ourselves by quadratic and quartic terms:
\be
V(\Phi)=- \mu^2|\Phi|^2+\lambda |\Phi|^4\,.\label{singlePhi}
\ee
Positive $\lambda$ is a must if we want the potential to be bounded from below and
have a stable vacuum. The coefficient $\mu^2$ can have either sign;
negative $\mu^2$ leads to the usual theory of a massive scalar field and a massless gauge boson,
while for positive $\mu^2$, the vacuum state of the theory changes and the Higgs phenomenon takes place.
To obtain the ground state of the theory, we minimize the potential (\ref{singlePhi}) and find
$\langle\Phi(x)\rangle = ve^{i\alpha}/\sqrt{2}$, where $v \equiv \sqrt{\mu^2/\lambda}$.
The phase $\alpha$ is unconstrained but its exact value is unphysical:
whatever value we choose, we can set $\alpha = 0$ by the global symmetry transformation.
At this moment, the global symmetry present in the initial scalar potential is broken spontaneously.
The model passes from the ``symmetric'' to the ``Higgs'' phase.
When the vacuum expectation value (vev) is fixed, we can expand $\Phi(x)$ around it as
\be
\Phi(x) = {1\over\sqrt{2}}[v + h(x) + i \eta(x)]\,,\label{Phi1}
\ee
where the two real neutral fields $h$ and $\eta$ describe deviations in the radial and angular directions 
on the complex $\Phi(x)$ plane.
To analyze the physical degrees of freedom and their masses, we expand the lagrangian 
up to second order in fields:
\be
{\cal L} = {\cal L}_{QED} + {1 \over 2}(\partial_\mu h)^2 + \mu^2 h^2 + 
{1\over 2} (\partial_\mu \eta - g q_\Phi v A_\mu)^2 + \mbox{interactions},\label{Lexp2}
\ee
where the omitted interaction terms are cubic and quartic combinations of fields $h$, $\eta$, and $A_\mu$.
The real field $h(x)$ describes excitations with mass $m_h^2 = 2 \mu^2$,
but $\eta$ is intrinsically coupled with $A_\mu$,
and the two fields can directly transform one into another.

A brief but important remark: if the theory were not gauged, then the last term in (\ref{Lexp2})
would be just the kinetic term for $\eta$. Since the potential does not produce the $\eta^2$
term, this field is massless. This is the simplest case of the Goldstone theorem \cite{Goldstone:1961eq,Goldstone:1962es}:
spontaneous breaking of a continuous global symmetry leads to massless scalar fields.

In the gauged theory, both $\eta$ and $A_\mu$ are dynamical fields which mix with each other.
As a result, there emerges a way to recast the scalar field $\eta$ in the form of an additional polarization 
state of the vector field $A_\mu$.
This is most efficiently done if, instead of (\ref{Phi1}), we express $\Phi(x)$ in the Higgs phase as
\be
\Phi(x) = {1\over\sqrt{2}}[v + h(x)] e^{i\eta(x)/v}\,.\label{Phi2}
\ee
Although we use here the same letters $h$ and $\eta$ and the linear terms of the expansion
around the vacuum coincides with (\ref{Phi1}), these fields differ from those in (\ref{Phi1})
in their interactions.
The lagrangian now resembles (\ref{Lexp2})
\be
{\cal L} = {\cal L}_{QED} + {1 \over 2}(\partial_\mu h)^2 + 
{1\over 2} (\partial_\mu \eta - g q_\Phi v A_\mu)^2 \left(1 + {h \over v}\right)^2 - V\,,\label{Lexp3}
\ee
where $V$ depends only on $v+h$ but not on $\eta$. 
This is now an exact expression, not a second order approximation as (\ref{Lexp2}).
The new fields $h$ and $\eta$ have simple gauge transformation properties:
the law (\ref{gauge-QED}) is accompanied by the change $\eta(x) \to \eta(x) + q_\Phi v \alpha(x)$
[which is to be read as $\eta(x) = \eta_{\mathrm new}(x) + q_\Phi v \alpha(x)$],
while $h(x)$ remains unchanged. Whatever the scalar field configuration we consider,
we can always remove $\eta$ [that is, set $\eta_{\mathrm new}(x)=0$]
by selecting the gauge transformation parameter $\alpha(x) = \eta(x)/q_\Phi v$.
In this way, the field $\eta$ {\em completely disappears from the lagrangian}.
The disturbing term becomes simply the mass term for the vector field $A_\mu$:
$m_A = g q_\Phi v$. The bracket $(1+h/v)^2$ leads to $hA_\mu A^\mu$ and $hhA_\mu A^\mu$ vertices
being directly proportional to $m_A^2$.

In short, the physical degrees of freedom reorganize themselves: we started with 2 real d.o.f.
in $\Phi$ and 2 transverse d.o.f. in $A_\mu$ and ended up with one real $h$
and 3 d.o.f. in $A_\mu$.
The gauge boson becomes massive, while the massless scalar dictated by the Goldstone theorem disappears
(``eaten'' by the vector boson).
This is the hallmark feature of the Brout-Englert-Higgs mechanism.
Note that when removing $\eta(x)$ everywhere, we fixed the gauge via condition $\eta_{\mathrm new}(x)=0$: 
although $\eta$ disappeared from the lagrangian,
it echoes in the gauge-fixing condition. The good feature of this gauge choice
is that all interactions the lagrangian encodes involve only physical, propagating, detectable degrees
of freedom. The evolution of any initial state will automatically be unitary, 
giving the name ``unitary'' to this gauge \cite{Weinberg:1971fb,Weinberg:1973ew}.

Although the unitary gauge nicely reveals the physical degrees of freedom, 
it leads in practical calculations to the same problems as the ordinary massive vector field 
whose mass is introduced by hand. Green's function for vector boson propagation
\be
G_{\mu\nu} = {1 \over k^2 - M^2}\left(g_{\mu\nu} - {k_\mu k_\nu \over M^2} \right)\label{green1}
\ee
does not fall as $1/k^2$ as $k^2 \to \infty$, leading to dangerous accumulation of extra powers of momentum
in loops. If the massive vector boson interacts with a conserved current,
then the contribution of the dangerous term in (\ref{green1}) vanishes and we are safe.
But if the bosons interact with non-conserved currents, as it happen with weak interactions, 
the theory becomes nonrenormalizable. 
Since the numerator of (\ref{green1}) is essentially the polarization density matrix, 
one can identify the source of the troubles:
the longitudinal polarization. For highly relativistic boson,
it becomes nearly aligned with $k_\mu/M$, its components linearly grow with energy
and contribute to extra divergences in loops.

Nevertheless, the Higgsed gauge theory is renormalizable,
as proved by `t Hooft and Veltman \cite{'tHooft:1971rn,'tHooft:1972fi};
this fact is just obscured by an unfortunate gauge choice.
A better gauge-fixing procedure, known as the renormalizable $R_\xi$ gauge,
is to complement the lagrangian (\ref{L0}) with the following gauge-fixing term
\be
{\cal L}_{gf} = - {1 \over 2\xi} (\partial_\mu A_\mu + g q_\Phi v \xi \eta)^2\,.
\ee
Combined with (\ref{Lexp2}), it puts the mixing terms
in the form of full derivative which, assuming the appropriate boundary conditions, 
can be dropped out.
The quadratic terms become diagonal and describe two fields, the massive $A_\mu$
and the massive unphysical Goldstone field $\eta$,
whose Green's functions are
\be
G_{\mu\nu} = {1 \over k^2 - M^2}\left(g_{\mu\nu} - (1-\xi){k_\mu k_\nu \over k^2 -\xi M^2} \right)\,,
\quad
G_\eta = {1 \over k^2 - \xi M^2}\,.
\label{green2}
\ee
For any finite $\xi$, all propagators behave as $1/k^2$ for $k^2 \to \infty$.
The presence of gauge parameter $\xi$ is tolerated in the intermediate calculations;
one just needs to make sure that it drops out in physical observables.
In the limit $\xi \to \infty$
we approach the unitary gauge: we formally recover the ordinary heavy vector boson propagator (\ref{green1})
while the scalar field becomes infinitely massive and drops out.
Thus, the apparently pathological properties of loop diagram in the unitary gauge
arise from the fact that we took $\xi \to \infty$ limit before evaluating the loop integral.
If we do it after loop integration, then the result, being $\xi$-independent, allows for this limit.

\subsection{The Brout-Englert-Higgs mechanism of the SM}

The Glashow-Weinberg-Salam theory of electroweak interactions 
\cite{Glashow:1961tr,Weinberg:1967tq,Salam:1968rm,Weinberg:1971fb} 
is based on the gauge group $SU(2)_L\times U(1)_Y$.
The covariant derivative is written generically as
\be
D_\mu = \left(\partial_\mu - i g' {Y \over 2}B_\mu\right)\mathbb{I} - i g T_i W^i_\mu\,,\label{gauge-SM}
\ee
where $g$ and $g'$ are the gauge coupling constants, $T_i$ are the $SU(2)$ generators in the corresponding
representation (absent for singlets and $\sigma_i/2$ for doublets), and $\mathbb{I}$ is the identity matrix.
By construction, the $SU(2)_L$ subgroup of transformations affects the left-handed fermions, which form a doublet
$L = (\nu_L,\, e_L)^T$ for leptons, $Q_L = (u_L, d_L)^T$ for quarks (we assume one fermion generation 
for the moment), but not the right-handed fermions, which are singlets. 
The triplet of gauge fields $W^{i}_{\mu}$, $i=1,2,3$, is in the adjoint representation of $SU(2)_L$.
The gauge field $B_\mu$ for the $U(1)_Y$ subgroup 
couples to the left- and right-handed fermions via their respective hypercharges $Y$, 
which are summarized in Table~\ref{table:Y}.
\begin{table}[h]
\begin{center}
\begin{tabular}{ccrrrrrrcr}
\hline
&& $u_L$ & $d_L$ & $u_R$ & $d_R$ & $\nu_L$ & $e_L$ & [$\nu_R$] & $e_R$\\[1mm] \hline \\[-3mm]
$Y$ && 	${1 \over 3}$ & ${1 \over 3}$ & ${4 \over 3}$ & $-{2 \over 3}$ & $-1$ & $-1$ & 0 & $-2$\\[2mm]
$T_3$ &&	${1 \over 2}$ & $-{1 \over 2}$ & 0 & 0 & ${1 \over 2}$ & $-{1 \over 2}$ & 0 & 0 
\\[1mm] 
\hline
\end{tabular}
\end{center}
\caption{Hypercharge and $T_3$ assignments for the SM fermions.}\label{table:Y}
\end{table}
The pattern of fermion couplings to the $SU(2)_L$ fields is fixed by the group structure,
but the Abelian group $U(1)_Y$ does not restrict the values of fermion hypercharges at the classical level. 
When constructing the theory, we set them as free parameters, and only later, after the electric charge
is derived in terms of $Y$ and $T_3$, see Eq.~(\ref{Nishijima}), we recover the values of $Y$. 
In that sense, Table~\ref{table:Y} uses {\em a posteriori} information of fermion properties. 
In this Table $[\nu_R]$ indicates the hypothetical right-handed neutrino, which is absent in the minimal SM,
but which, even if added, is electroweak-blind. 
The Higgs sector of the SM is built upon one doublet of complex scalar fields of hypercharge $Y = 1$,
which is usually written as $\Phi = (\phi^+,\, \phi^0)^T$. The superscripts ``$+$'' and ``$0$'' 
allude to the electric charges of the corresponding fields.
Again, this is an {\em a posteriori} piece of information, which emerges after choosing the vacuum;
for the moment, one should just view them as two independent components of $\Phi$.

With all these conventions, the electroweak lagrangian of the minimal SM for a single lepton generation takes the following form:
\bea
{\cal L} &=& -{1 \over 4} W^i_{\mu\nu} W^{\mu\nu, i} -{1 \over 4} F_{\mu\nu} F^{\mu\nu}
+ |D_\mu \Phi|^2 - V(\Phi) \nonumber\\
&+& i \bar L D_\mu \gamma^\mu L + i \bar e_R D_\mu \gamma^\mu e_R 
-f_e(\bar L e_R \Phi + \Phi^\dagger  \bar e_R L)\,.\label{SMlagrangian}
\eea
The tensors $W_{\mu\nu}^i = \partial_\mu W_\nu^i - \partial_\nu W_\mu^i + g \epsilon_{ijk} W_\mu^j W_\nu^k$ and 
$F_{\mu\nu} = \partial_\mu B_\nu - \partial_\nu B_\mu$ 
are the field strengths for the $SU(2)_L$ and $U(1)_Y$ groups,
respectively. Unlike in the $U(1)$ case, no mass term for the fermions is now allowed, 
as the left- and right-handed fields transform differently.

Just as in the Abelian Higgs model,
the Higgs potential of the SM is assumed to have the same shape as in (\ref{singlePhi}):
the famous ``Mexican hat'' in the 4-real-dimensional space of scalar fields.
Its minimum is attained on the 4-sphere of radius $v$ parametrized as
\be
\langle \Phi\rangle =e^{i \alpha_0 + i \alpha_i \sigma_i} \, {1 \over \sqrt{2}} 
\left(\begin{array}{c} 0 \\ v \end{array}\right)\,,\label{vev}
\ee
which spontaneously breaks the global symmetry.
Parameters $\alpha_0$, $\alpha_i$ are unconstrained by minimization,
but whatever values they take, they can be removed by a global $SU(2)\times U(1)$ rotation. 
Setting $\alpha_0 = \alpha_i = 0$, we fix the Higgs vacuum and expand the scalar field
around it.
It is only after this choice that the superscripts ``$+$'' and ``$0$'' in the Higgs field components 
and the labels $\nu_L$ and $e_L$ inside the doublet $L$ become justified.

Since the field $\lr{\phi^0}$ with $Y = 1$ and $T_3 = -1/2$
acquires a non-zero vev, these two quantum numbers are not conserved separately.
What is conserved is their linear combination 
\be
Q = T_3+Y/2\,,\label{Nishijima} 
\ee
which is zero for the vacuum state and which can now be calculated for all other fields.
Coupled to this charge is the residual gauge interaction, a specific subgroup
of $SU(2)_L\times U(1)_Y$ denoted as $U(1)_{EM}$, which leaves the vacuum invariant.
It is identified with the electromagnetic interactions described
with the field $A_\mu$, which couples other fields proportionally to their electric charges $Q$.

To explore the bosonic sector, we express the Higgs doublet as
\be
\Phi(x) =e^{i \eta_0(x)/v + i \eta_i(x) \sigma_i/v} \, {1 \over \sqrt{2}} 
\left(\begin{array}{c} 0 \\ v + h(x) \end{array}\right)\,,\label{Phi-expanded}
\ee
insert it into the Higgs kinetic term $|D_\mu\Phi|^2$, and keep track of terms quadratic in fields. 
We find, in addition to $(\partial_\mu h)^2/2$, the following expression 
\be
{1\over 2}\left({gv \over 2}W^1_{ \mu} - \partial_\mu \eta_1\right)^2
+ {1\over 2}\left({gv \over 2}W^2_{\mu} - \partial_\mu \eta_2\right)^2 
+ {1\over 2}\left({gv \over 2}W^3_{\mu} - {g'v \over 2} B_\mu + \partial_\mu \eta_0 - \partial_\mu \eta_3\right)^2\,.
\label{DPhi2}
\ee
The field combination in the last bracket is usually written as $\bar g Z_\mu /2$, where
$Z_\mu = c_W W^3_{\mu} - s_W B_\mu$ and the weak angle $\theta_W$ is defined via
\be
c_W \equiv \cos\theta_W = g/\bar g\,, \quad 
s_W \equiv \sin\theta_W = g'/\bar g\,, \quad \bar g^2 \equiv g^2 + g^{\prime 2}\,.
\ee
In the unitary gauge, we use gauge transformation of $W^1_{\mu}$, $W^2_{\mu}$, and $Z_\mu$
to eliminate $\eta_1$, $\eta_2$, and $\eta_0-\eta_3$.
This fixes the gauge up to the unconstrained $\eta_0 + \eta_3$,
which {\em de facto} was absent in (\ref{Phi-expanded}).
Then Eq.~(\ref{DPhi2}) produces the mass terms for the vector bosons
$Z_\mu$ and $W^\pm_\mu \equiv (W^1_{\mu} \mp i W^2_{\mu})/\sqrt{2}$:
\be
{g^2 v^2 \over 4}W_\mu^- W^{+\mu} + {\bar g^2 v^2 \over 8} Z_\mu Z_\mu
\equiv
m_W^2 W_\mu^- W^{+\mu} + {1 \over 2} m_Z^2 Z_\mu Z_\mu\,,\quad
m_Z = {\bar g v \over 2}\,, \quad m_W = {g v \over 2}\,. \label{SM-WZmasses}
\ee
The combination $A_\mu = s_W W^3_\mu + c_W B_\mu$
orthogonal to $Z_\mu$ remains massless and is identified with the photon.
For the inventory of the bosonic degrees of freedom, we write the Higgs doublet as
\begin{equation}
\Phi = \langle \Phi\rangle + {1 \over \sqrt{2}}
\left(\begin{array}{c} \phi_1 + i \phi_2 \\ h + i \phi_3 \end{array}\right)\,.\label{Phiexp}
\end{equation}
In the starting lagrangian we had 4 (Higgs doublet), $3\times 2$ (massless $W^i_\mu$), and $1\times 2$ (massless $B_\mu$) 
= 12 degrees of freedom. In the Higgs phase, they regroup as
1 (the Higgs boson), $3\times 3$ (massive $W^\pm$ and $Z$), and $1\times 2$ (massless photon).

Next, we check how the reshuffling of the bosonic degrees of freedom affects the fermionic kinetic sector.
To identify the residual gauge symmetry with the electromagnetic interactions, 
we need to demonstrate the vector-like coupling of the fermions to $A_\mu$.
We take the fermion kinetic terms in the lagrangian (\ref{SMlagrangian}), express all gauge fields in terms
of $A_\mu$, $Z_\mu$, and $W^\pm_\mu$, and keeping track of $A_\mu$ interactions only, obtain
\be
{\cal L}_{EM} = 
i \bar e_L  \left(\partial_\mu - i \bar g c_W s_W {Y_L-1 \over 2} A_\mu\right) \gamma^\mu  e_L +
i \bar e_R  \left(\partial_\mu - i \bar g c_W s_W {Y_{e_R} \over 2} A_\mu\right) \gamma^\mu  e_R\,.
\ee
With $Y_L = -1$ and $Y_{e_R} = -2$, the two terms have equal coefficients
and can be grouped as $i \bar e (\partial_\mu - i e Q A_\mu) \gamma^\mu e$,
with the new coupling constant $e =  \bar g c_W s_W = gg'/\bar g$ and with the electron charge $Q=-1$.
In the same way, one derives that $Z_\mu$ couples to fermions proportionally 
to the combination $T_3 - s_W^2Q$, which is different for left- and right-handed fields.
As the result, the weak neutral current is
\be
{\cal L}_{NC} = {\bar g \over 2} \bar \psi \gamma^\mu \left[(T_3 - 2s_W^2 Q) - T_3\gamma_5\right]\psi\, Z_\mu
= {\bar g \over 2}J_{NC}^\mu Z_\mu \,,
\ee
where $\psi$ is either $\nu$ or $e$.
The charged currents mediated by the $W^\pm$ bosons affect only the left-handed fermions and arise 
from the non-diagonal elements of the covariant derivative matrix:
\be
{\cal L}_{CC} = {g \over 2\sqrt{2}} \bar \nu \gamma^\mu (1 - \gamma_5) e W^+_\mu + h.c. =
{g \over 2\sqrt{2}} (J_{CC}^\mu W^+_\mu + J_{CC}^{\mu\, \dagger} W^-_\mu)\,. 
\ee
In the low-energy limit, one recovers the phenomenological $V-A$ theory of weak interaction.
The $W$-exchange between two charged currents reduces to the pointlike current-current interaction
with the coupling $g^2/8m_W^2 = G_F/\sqrt{2}$, which allows to determine the Higgs vev $v \approx 246$ GeV.
Knowing $e = g s_W = g' c_W$ and measuring $\theta_W$ via neutrino scattering,
one can obtain $g$ and $g'$ and predict the masses of the $W$ and $Z$-bosons.
These predictions were nicely confirmed in 1983 by the UA1 and UA2 experiments 
at CERN in \cite{Arnison:1983rp,Bagnaia:1983zx}.

One can also use the neutral currents to establish the same link. 
The $Z$-boson exchange leads in the low-energy limit
to the neutral-current interaction
\be
{\cal L}_{NC}^{eff} = {\bar g^2 \over 8 m_Z^2} J_{NC}^\mu J_{NC \mu}^\dagger \equiv 
{G_F \over \sqrt{2}} \,\rho\, J_{NC}^\mu J_{NC \mu}^\dagger\,.
\ee
The $\rho$-parameter defined by this relation quantifies the possible difference between
the strength of the neutral and charged weak currents at low energies.
The SM predicts \cite{Ross:1975fq} that 
\be
\rho = {m_W^2 \over m_Z^2}\cdot {\bar g^2 \over g^2} = 1\label{rho}
\ee 
up to loop corrections, which was experimentally verified to high accuracy: $\rho_{\mathrm{exp}} = 1.00037(23)$ \cite{PDG}.
The tree-level result $\rho=1$ is the manifestation of a hidden approximate symmetry
in the SM Higgs sector called custodial symmetry. 
Recall that the anti-fundamental representation of the group $SU(2)$ can be mapped to the fundamental one: 
$\tilde \Phi = i\sigma_2 \Phi^* = (\phi^{0*}, -\phi^-)^T$. 
It has a different hypercharge than $\Phi$, but in the limit $g'\to 0$ this distinction becomes irrelevant.
One can then construct the bidoublet
\be
\hat\Phi = (\tilde \Phi, \Phi) = \left(\begin{array}{cc}
\phi^{0*} & \phi^+ \\
-\phi^{+*} & \phi_0
\end{array}\right)\,,
\label{bidoublet}
\ee
and rewrite the Higgs lagrangian in terms of $\hat \Phi$, for example $\Phi^\dagger \Phi = \Tr[\hat\Phi^\dagger \hat\Phi]/2$.
Now the lagrangian has global symmetry group $SU(2)_L \times SU(2)_R$: 
$\hat\Phi \to \exp(i \alpha_L^i \sigma_i/2) \hat\Phi \exp(-i \alpha_R^i \sigma_i/2)$. 
The $SU(2)_L$ group is promoted to the gauge group, while in $SU(2)_R$, 
only one subgroup $U(1)_Y$ is gauged. 
In the Higgs phase, the bidoublet gets vev: $\lr{\hat\Phi} = \mathrm{diag}(v,\, v)/\sqrt{2}$,
and since it is proportional to the unit matrix, it is invariant under the diagonal subgroup
$SU(2)_L \times SU(2)_R \to SU(2)_V$ with $\alpha^i_L = \alpha^i_R$.
This residual symmetry group is the custodial symmetry. In the limit $g' \to 0$
it leads to the $SO(3)$-invariance in the space of massive $W^i_\mu$ fields, 
and eventually sets $\rho = 1$, which survives even in the real case when custodial symmetry is approximate.
For other scalar sectors, this construction can fail which we will see in section~\ref{section-triplets}.

Turning to the fermion masses, we notice that 
the toy model (\ref{QED}) already contained the $U(1)$-invariant mass term $m\bar\psi \psi$
for the Dirac fermion $\psi(x)$. 
In terms of chiral fields, $\psi_L$ and $\psi_R$, it is written as
\begin{equation}
m\bar\psi \psi = m(\bar\psi_L \psi_R + \bar \psi_R \psi_L)\,,\label{mass-fermion-QED}
\end{equation}
and it mixes $\psi_L$ and $\psi_R$. 
This mixing does not cause any problem for vector-like interactions.
But now $\psi_L$ and $\psi_R$ live in different representations of $SU(2)_L$
and have different hypercharges, so that the mass term (\ref{mass-fermion-QED}) 
is not gauge invariant anymore.
However the same scalar field $\Phi$ can also take care of the fermion mass.
If we introduce Yukawa interaction via
\begin{equation}
{\cal L}_f= y_\psi \bar \psi_L \psi_R \Phi + y_\psi \bar \psi_R \psi_L \Phi^*\,,
\end{equation}
where $y_\psi$ is a real Yukawa coupling constant,
then the Higgs mechanism changes it to 
\begin{equation}
{\cal L}_f= m_\psi \bar \psi_L \psi_R + {m_\psi \over v} \bar \psi_L \psi_R h + h.c.
\end{equation}
The fermion gains mass $m_\psi = y_\psi v/\sqrt{2}$ and, in addition,
couples to the Higgs boson proportionally to its mass.

The full flavor sector of the SM contains three copies of leptons and quarks:
$L_i$, $\ell_{Ri}$, $Q_{Li}$, $d_{Ri}$, $u_{Ri}$, $i = 1,2,3$.
The fermion kinetic and gauge part is flavor-diagonal: 
\be
{\cal L}_f = i \sum_i \left(\bar L_i D_\mu \gamma^\mu L_i + \bar \ell_{Ri} D_\mu \gamma^\mu \ell_{Ri} 
+ \bar Q_{Li} D_\mu \gamma^\mu Q_{Li} + \bar d_{Ri} D_\mu \gamma^\mu d_{Ri}
+ \bar u_{Ri} D_\mu \gamma^\mu u_{Ri}\right)\,.\label{SM-fermion-kin}
\ee
It enjoys the large $[U(3)]^5$ global symmetry group and does not involve any new free parameters.
The Yukawa sector, in complete contrast, 
comes with many free parameters and does not possess non-trivial global symmetries.
It is usually parametrized as
\be
- {\cal L}_Y = \bar L_i \Pi_{ij} \ell_j \Phi + \bar Q_{Li} \Gamma_{ij} d_{Rj} \Phi + \bar Q_{Li} \Delta_{ij} u_{Rj} \tilde\Phi
+ h.c.
\label{SM-Yukawa}
\ee
with three $3\times 3$ complex matrices $\Pi$, $\Gamma$, $\Delta$.
Notice that up-quark Yukawa interactions involve the same $\tilde\Phi$ as in (\ref{bidoublet})
with $Y = -1$.
In the Higgs phase, we replace $\Phi \to \lr{\Phi}$ and obtain the mass matrices:
\be
M_\ell = {v \over\sqrt{2}} \Pi\,, \quad
M_d = {v \over\sqrt{2}} \Gamma\,, \quad
M_u = {v^* \over\sqrt{2}} \Delta\,,\label{SM-mass-matrices}
\ee
whose mass eigenstates are the physical fermions.
They are diagonalized with biunitary transformations:
\be
V^\dagger_{dL} M_d V_{dR} = D_d = \mathrm{diag}(m_d, m_s, m_b)\,, \quad
V^\dagger_{uL} M_u V_{uR} = D_u = \mathrm{diag}(m_u, m_c, m_t)\,, \label{SM-fermion-diag}
\ee
In the minimal SM, neutrinos are massless, and the diagonalization of the charged lepton matrix
completes the lepton sector.
For quarks, we have four independent unitary transformations, $V_{dL}$, $V_{uL}$, $V_{dR}$, $V_{uR}$.
The transformations of the right-handed fields do not lead to any physical effects as they
cancel out, by virtue of the singlet nature of the right-handed fermions, in the corresponding terms of Eq.~(\ref{SM-fermion-kin}).
The mismatch between $V_{dL}$ and $V_{uL}$ does have physical significance. The non-diagonal coupling between components 
of $Q_L$ in Eq.~(\ref{SM-fermion-kin}), if written in terms of physical fermions
$d_{L}^{\mathrm{ph.}} = V^\dagger_{dL} d_L$ and  
$u_{L}^{\mathrm{ph.}} = V^\dagger_{uL} U_L$, becomes
\be
{g \over \sqrt{2}}\left(\bar u_{L} \gamma^\mu d_{L} W^+_\mu + h.c. \right) =
{g \over \sqrt{2}}\left(\bar u^{\mathrm{ph.}}_{L} \gamma^\mu V d^{\mathrm{ph.}}_{L} W^+_\mu + h.c. \right)\,,
\ee
where the Cabibbo-Kobayashi-Maskawa matrix
\be
V \equiv V^\dagger_{uL} V_{dL}
\ee
describes the quark-family mixing in charged currents.
The experimentally known $V$ \cite{Porter:2016mps} has three mixing angles and one complex phase,
which leads to $CP$-violation \cite{CPV-book}. 
In contrast, all tree-level neutral currents remain flavor-diagonal.
Indeed, rotation matrices cancel in $\bar u_L \gamma^\mu u_L$ and in $\bar d_L \gamma^\mu d_L$,
therefore both $Z_\mu$ and $A_\mu$ do not mix families.
The Higgs boson $h$ appears in the Yukawa interactions in the combination $v+h$,
and once fermion mass matrices are diagonalized, the Higgs-fermion couplings are automatically diagonal.
In short, the flavor-changing neutral currents (FCNC) are absent in the SM at the tree-level and appear only in loops.

\subsection{What is broken in the Higgs phenomenon?}

At the basic level, one views the Brout-Englert-Higgs mechanism as a phase transition.
One passes from the electroweak-symmetric phase to the Higgs phase, in which the vacuum breaks 
the full electroweak symmetry to the electromagnetic subgroup: $SU(2)_L \times U(1)_Y \to U(1)_{EM}$.
In the case of Higgs phenomenon, the vev of the Higgs field plays the role of the order parameter,
a quantity which usually accompanies phase transitions. 
The symmetric phase 
corresponds to $\lr{\Phi} = 0$, while the broken phase has $\lr{\Phi} \not = 0$.
This view is reflected in the often used terminology electroweak symmetry breaking (EWSB)
or spontaneous symmetry breaking (SSB), 
which alludes to spontaneous partial breaking of the gauge symmetry.

But is the gauge symmetry actually broken spontaneously?
In the above exposition of the Higgs mechanism, there were two instances when a symmetry was broken.
First, when we selected one minimum out of infinite amount of equivalent minima,
a spontaneous breaking indeed took place, but only of a global symmetry.
This minimum represents a vacuum, and in order to perturbatively describe the quantum field theory,
we need to quantize the fields. Quantization of gauge field theories requires introduction of a gauge-fixing procedure,
and during this procedure we break the gauge symmetry by hand, explicitly, not spontaneously.
Thus, the two notions, EWSB and SSB, are in certain sense correct,
but they do not refer to the same symmetry.

If electroweak symmetry is not broken spontaneously, will it be possible to describe the phenomenon
without breaking it altogether?
Already in 1966, Higgs showed \cite{Higgs:1966ev} that the lagrangian of the Abelian Higgs model,
at least in the classical theory, can be written in terms of gauge-invariant fields.
Indeed, let us slightly modify Eq.~(\ref{Phi2}) by writing it as $\Phi(x) = \rho(x)e^{i\beta(x)}/\sqrt{2}$.
Then, focusing on the bosonic part of the theory, we write lagrangian (\ref{Lexp3}) as
\be
{\cal L} = - {1 \over 4} F_{\mu\nu}F^{\mu\nu} + {1 \over 2}(\partial_\mu \rho)^2 - V(\rho) 
+ {1\over 2} \rho^2 g^2 q_\Phi^2 \left(A_\mu - {1 \over g q_\Phi}\partial_\mu \beta\right)^2\,,\label{Lexp10}
\ee
The gauge invariance in the form of correlated shifts of $A_\mu$ and $\beta$ is still intact. 
Let us now introduce the {\em gauge invariant field} 
$B_\mu \equiv A_\mu - \partial_\mu \beta/g q_\Phi$. If $\Phi(x) \not = 0$ everywhere,
then $\beta(x)$ is uniquely defined, and this change of variables is regular.
The gauge field kinetic terms is simply $F_{\mu\nu} = \partial_\mu B_\nu - \partial_\nu B_\mu \equiv F^{(B)}_{\mu\nu}$
(in general, this transition will contain isolated singularities $[\partial_\mu, \partial_\nu] \beta(x)$ located
at points of vanishing $\Phi(x)$ \cite{Chernodub:2008rz}). Thus, the classical lagrangian becomes
\be
{\cal L} = - {1 \over 4} F^{(B)}_{\mu\nu}F^{(B)\, \mu\nu} + {1 \over 2}(\partial_\mu \rho)^2 - V(\rho) 
+ {1\over 2} \rho^2 g^2 q_\Phi^2 B_\mu B^\mu\,.\label{Lexp11}
\ee
It contains only manifestly gauge invariant fields $\rho$ and $B_\mu$.
The original $U(1)$ gauge symmetry --- which is not a physical symmetry but
just a redundancy of description, --- is completely banished from the theory.
Nothing is transformed, no symmetry is involved, no symmetry breaking happens.
The model still has two phases, with $\lr{\rho} = 0$ or $\lr{\rho} \not = 0$,
but this has nothing to do with gauge symmetry breaking. 
As Englert says in his Nobel lecture \cite{Englert:2014zpa}: 
``{\em ... The vacuum is no more
degenerate and strictly speaking there is no spontaneous
symmetry breaking of a local symmetry. The reason why
the phase with nonvanishing scalar expectation value is often
labeled SSB is that one uses perturbation theory to select at
zero coupling with the gauge fields a scalar field configuration
from global SSB; but this preferred choice is only a
convenient one.}''.

Note however that this nice change of variable works only for the bosonic part of the theory.
How to extend this gauge-invariant formalism to the fermionic part of the lagrangian is not clear.
For the scalar field, its angular variable can be traded for the would-be Goldstone boson,
while for the fermion there is nothing to trade it for.

One might suspect that if one passes to the quantum theory, one will need to re-introduce 
gauge transformations in order to fix the vacuum and the gauge-fixing procedure
to describe the propagation of the gauge bosons. 
In 1972-1973, B.~Lee and Zinn-Justin published a series of four papers whose titles 
explicitly mentioned ``spontaneously broken gauge theories'', and 
the perturbative expansion was the only tool available at that time \cite{Lee:1974zg}.
However, Wilson showed in 1974 how
to work with the quantum gauge theory without introducing gauge-fixing procedure \cite{Wilson:1974sk}.
Using it, Elitzur proved in 1975 the general theorem that truly spontaneous breaking of a local gauge symmetry 
is impossible \cite{Elitzur:1975im}. The argument relies on the fact that vacuum averaging
of gauge-dependent operators such as $\lr{\Phi}$ involves, in the absence of explicit gauge-fixing
term, averaging over the full gauge orbit, and leads to $\lr{\Phi} =0$.
Only gauge-invariant operators can have non-zero vacuum averaging.
The exploration of Higgsed gauge theories on the lattice began,
and in 1980, Fröhlich, Morchio, and Strocchi \cite{Frohlich:1980gj,Frohlich:1981yi}
finally clarified this somewhat confusing situation. 
They constructed the full set of gauge-invariant composite operators
in the SM and demonstrated that their mass gaps correspond to the masses 
of elementary excitations above the vacuum in the standard, gauge-fixed formulation.
For the simplest example, the gauge-invariant composite operator
$\Phi^\dagger(x) \Phi(x)$ has in the Higgs phase the perturbative expansion $v^2/2 + \Phi^\dagger_0 h(x) + \dots$,
and therefore $\lr{\Phi^\dagger(x) \Phi(x)}$ exhibits a mass gap equal to the mass
of the Higgs boson $h$. The fermions are also incorporated in this picture,
since one works here with asymptotic states defined as gauge-invariant correlators,
not individual fermionic fields.

With modern computer power, this equivalence between masses of gauge-dependent
fields (the Higgs boson, $W$ and $Z$) and the mass gaps in the spectra of the corresponding 
gauge-invariant operators can be actually checked in numerical lattice gauge theory computations
\cite{Torek:2016ede}.
This work was recently undertaken by Maas and collaborators within the SM \cite{Maas:2014pba}; 
the first steps towards checking it in the two-Higgs-doublet model were also done \cite{Maas:2016qpu}.
In SM, the results of the gauge-invariant perturbation theory of 
Fröhlich, Morchio, and Strocchi agree with the conventional perturbation theory and are confirmed by the lattice
simulations. However, this matching is not expected to be universal,
as it uses the special properties of the $SU(2)$ group. 
There exist examples of Higgsed gauge theories with larger gauge groups where the two
approaches lead to {\em different} predictions \cite{Maas:2016ngo,Torek:2016ede},
and the lattice simulations appear to confirm the predictions of the gauge-invariant perturbation theory.
If this result gets stronger support, it will send a radical message to the bSM community: 
``{\em If this evidence could be substantially improved, it appears necessary to reevaluate candidate
theories for new physics.}'' \cite{Torek:2016ede}.

In the last decade quite a few publications appeared which again bring up the meaning and significance
of spontaneous gauge symmetry breaking.
Chernodub, Faddeev, and Niemi recast in 2008 the bosonic sector of the SM in the gauge-invariant formulation 
stressing the inapplicability of Elitzur's theorem once the gauge freedom disappears \cite{Chernodub:2008rz}.
Once again, the fermionic part was not included.
Other insightful discussions of the proper definition of gauge transformations, gauge symmetry 
and its possible spontaneous breaking, accompanied with delicate mathematics, 
include \cite{Struyve:2011nz} and especially the recent talk by Strocchi 
\cite{Strocchi:2015uaa}, where he reiterates the basic points exposed al length in his book \cite{Strocchi:2005yk}.


\section{Two-Higgs-doublet models}\label{section-2HDM}

\subsection{Several Higgs doublets: generic features }\label{section-NHDM-generic-features}

Within the SM, the single Higgs doublet is overstretched. 
It takes care simultaneously of the masses of the gauge bosons
and of the up and down-type fermions.
$N$-Higgs-doublet models (NHDM), which are among the 
simplest extensions of the SM Higgs sector, relax this requirement. 
They are based on the simple suggestion that the notion of generations can be brought to the Higgs sector. 
Since the gauge structure of the SM does not restrict the number of Higgs ``generations'',
it needs to be established experimentally, and anticipating future experimental
results, one can investigate this possibility theoretically.
As a bonus, NHDMs lead to remarkably rich phenomenology.
They allow for signals which are impossible within the SM,
such as several Higgs bosons, charged and neutral, 
modification of the SM-like Higgs couplings, FCNC at tree level,
additional forms of $CP$-violation from the scalar sector, 
and opportunities for cosmology 
such as scalar DM candidates and modification of the phase transitions in early Universe.
Also, many bSM models including supersymmetry (SUSY), gauge unification models, 
and even string theory constructions naturally lead to several Higgs doublets 
at the electroweak scale.

The NHDM lagrangian uses $N$ Higgs doublets $\phi_i$ (we now switch to the small Greek letter $\phi$ for the doublets), 
$i = 1,\dots, N$, all having the same hypercharge $Y=1$.
This echoes in the scalar potential and Yukawa interactions.
The renormalizable Higgs self-interaction potential can generically be written as
\be
V(\phi) = Y_{ij} \fd_i \f_j + Z_{ijkl} (\fd_i \f_j)(\fd_k \f_l)\,.\label{V-NHDM}
\ee
The hermiticity of $V$ implies that the coefficients satisfy 
$Y_{ij} = Y^*_{ji}$, $Z_{ijkl}=Z^*_{jilk}$, $Z_{klij}=Z_{ijkl}$.
For $N$ doublets, the general potential contains $N^2 + N^2(N^2+1)/2$ real parameters:
14 for 2HDM, 54 for 3HDM, etc. These parameters must also comply with the requirements
that the potential be bounded from below to insure existence of the global minimum
and that the minimum be neutral and not charge-breaking, which is another exotic possibility absent in the SM.
The values of the quartic coefficients $Z_{ijkl}$ cannot be too large in order not to overshoot the perturbative unitarity 
bounds \cite{Lee:1977yc,Lee:1977eg}.
The Yukawa sector is given by $N$ copies of Eq.~(\ref{SM-Yukawa}):
\be
- {\cal L}_Y = \bar L \Pi_i \ell \phi_i + \bar Q_{L} \Gamma_i d_{R} \phi_i + \bar Q_{L} \Delta_{i} u_{R} \tilde\phi_i
+ h.c.
\label{NHDM-Yukawa}
\ee
where we suppressed the fermion flavor indices.
In the most general case, all matrices $\Pi_i$, $\Gamma_i$, $\Delta_i$ are independent,
with the number of free parameters skyrocketing to $54N$.
Even though many of them can be removed by a scalar and fermion space basis change,
the dimension of the physical parameter space remains huge.

Once the scalar potential is written, the procedure for the SSB closely follows the SM. 
One minimizes the potential and performs the global symmetry transformation
to bring the vevs to the following form:
\be
\lr{\phi_1} = {1 \over \sqrt{2}}\doublet{u}{v_1}\,,\quad 
\lr{\phi_i} = {1 \over \sqrt{2}}\doublet{0}{v_i}, \quad i= 2,\dots,N\,.\label{vacuum-NHDM}
\ee
In order for the vacuum to be neutral, not charge-breaking, 
$u$ must be zero. There is no fine-tuning associated with this requirement;
neutrality of the vacuum can be guaranteed if
the coefficients in front of terms $(\fd_i \f_i)(\fd_j \f_j)-(\fd_i \f_j)(\fd_j \f_i)$ are positive.
Since all doublets couple to the gauge-bosons in the same way,
the $W$ and $Z$ masses are determined by the single value $v^2 = v_1^2 + \cdots + |v_N|^2$,
and, just as in the SM, the $\rho$-parameter remains equal to one at tree level.
Inserting the scalar vevs in the Yukawa sector (\ref{NHDM-Yukawa})
produces the fermion mass matrices: 
\be
M_\ell = {1 \over\sqrt{2}} \sum_i \Pi_i v_i\,, \quad
M_d = {1 \over\sqrt{2}} \sum_i \Gamma_i v_i\,, \quad
M_u = {1 \over\sqrt{2}} \sum_i \Delta_i v_i^*\,.\label{NHDM-fermion-masses}
\ee
Their diagonalization proceeds in the same way as in the SM leading to the fermion masses
and mixing. The big difference now is that diagonalizing mass matrices
does not necessarily diagonalize individual $\Gamma$'s and $\Delta$'s. 
As a result, even the SM-like Higgs boson can couple to fermion pairs in a way
which differs from their mass pattern and can induce tree level FCNCs.
In a generic NHDM Yukawa sector, these effects are so large that
one needs to restrict them in order to comply with experimental bounds.

Multi-doublet models offer novel opportunities for $CP$-violation.
Within the SM, it is put by hand: it arises entirely from the Yukawa matrices which 
must be complex. In multi-doublet models, a relative phase between
vevs can arise just as a result of the minimization of the potential.
This spontaneous $CP$-violation was the original motivation by T.~D.~Lee to consider a model
with more than one Higgs doublet \cite{Lee:1973iz}.
More subtle forms of $CP$-violation are also possible, see section~\ref{section-NHDM-CPV}.

Once the minimum of the Higgs potential is known, one can expand the potential around it and find the spectrum
of the physical Higgs bosons. Separation between the physical Higgses and the Goldstone modes is especially
clear in the so-called Higgs basis. Starting from the generic neutral vev alignment (\ref{vacuum-NHDM}),
one can perform a global $SU(N)$ transformation 
that brings vev to the first one $\lr{H_1^0} = v/\sqrt{2}$ and sets all other $\lr{H_i} = 0$.
In this basis, the Goldstone modes fill the first doublet, just like in the SM,
while the physical Higgses populate the other doublets.
Physical Higgses can, of course, mix across the doublets, leading to 
$N-1$ pairs of charged Higgses $H_i^\pm$
and $2N-1$ neutral scalars, one of which is the experimentally observed 125 GeV scalar $h_{125}$.
If the model is $CP$-conserving and if the vacuum respects this $CP$ symmetry,
then the neutral scalars are classified as $CP$-even and $CP$-odd ones;
for rare exceptions to this rule, see sections~\ref{section-IDM} and \ref{section-NHDM-CPV}. 

Expressing all interactions via physical Higgses, one can proceed with calculation of observables.
The phenomenology of the resulting model depends to a large extent on how strong the mixing is for the neutral scalars, 
in the Higgs basis, between the first doublet and the remaining ones. 
It often happens in NHDM that all Higgses apart from the lightest one $h_{125}$ are heavy, 
as they are governed by different free parameters.
A large mass difference typically leads to an alignment in the scalar sector:
the 125 GeV Higgs $h_{125}$ resides almost entirely in the first doublet and is very SM-like. 
The couplings of the other Higgses with the SM fields are suppressed,
and the extra scalars essentially decouple from the electroweak scale phenomenology 
\cite{Haber:1989xc,Haber:1994mt,Ginzburg:2001ss,Gunion:2002zf}.
Notice however that alignment can also take place without decoupling, which was illustrated for the 
2HDM in \cite{Carena:2013ooa,Carena:2014nza,Dev:2014yca}.
In this case, the extra scalars, being moderately light, are essentially gaugeophobic and, as a result, 
are poorly produced at colliders and can avoid New Physics searches.

\subsection{Building 2HDMs: the scalar sector}

The two-Higgs-doublet model (2HDM) is the oldest and the most studied version of NHDM.
Although suggested by T.~D.~Lee back in 1973 \cite{Lee:1973iz,Lee:1974jb}, 
it attracted in those early days less attention than more elaborate scalar sectors, see section~\ref{section-historical}.
Its exploration intensified in 1990s and especially in the modern, LHC-dominated era.
One strong motivation for this model was that the minimal supersymmetric extension of the SM,
the MSSM, contains two Higgs doublets \cite{Gunion:1984yn,Gunion:1986nh,Gunion:1989we,Djouadi:2005gj}.
Even taken out of the MSSM context, the 2HDM {\em per se} 
emerges as an interesting playground for the bSM scalar sector.
Within 2HDM, one can identify different regimes, which 
lead to distinct phenomenologies and can be studied analytically and numerically.
In the last decade it grew into a key bSM reference model
checked when one discusses New Physics searches.

2HDMs were extensively covered in the recent review \cite{2HDM-review};
the properties of the MSSM Higgs bosons were described in great detail by Djouadi in \cite{Djouadi:2005gj}.
Here, we will only repeat its main features and give a selection of recent developments. 
We stress again that we do not pretend to exhaustively cover the recent progress;
we just want to demonstrate the challenges and the methods involved.

The Higgs potential of the most general 2HDM is conventionally parametrized as
\bea
V &=& m_{11}^2 \fd_1\f_1 + m_{22}^2 \fd_2 \f_2 - m_{12}^2 \fd_1 \f_2 - (m_{12}^2)^* \fd_2 \f_1
+ {\lambda_1 \over 2}(\fd_1\f_1)^2 + {\lambda_2 \over 2}(\fd_2\f_2)^2 \nonumber\\
&& + \lambda_3 (\fd_1\f_1) (\fd_2\f_2) + \lambda_4 (\fd_1\f_2) (\fd_2\f_1) 
+ {\lambda_5 \over 2}(\fd_1\f_2)^2 + {\lambda_5^* \over 2}(\fd_2\f_1)^2\nonumber\\
&& + \left[ \lambda_6 (\fd_1\f_1)(\fd_1\f_2) + \lambda_7 (\fd_2\f_2)(\fd_1\f_2) + h.c.\right].\label{V-2HDM}
\eea
It contains 4 independent free parameters in the quadratic part 
(real $m_{11}^2$, $m_{22}^2$ and complex $m_{12}^2$) 
and 10 free parameters in the quartic part (real $\lambda_{1,2,3,4}$ and complex $\lambda_{5,6,7}$). 
Envisioning renormalization procedure, \cite{Ginzburg:2004vp,Ginzburg:2008kr} argued that the most general 2HDM 
scalars sector should also include the off-diagonal kinetic terms $\varkappa(D_\mu \phi_1)^\dagger (D^\mu \phi_2) + h.c.$.
They can of course be eliminated by a non-unitary basis change at any given order of perturbation series
so that $\varkappa$ does not enhance the space of physical possibilities.
The parameters of potential (\ref{V-2HDM}) must satisfy stability constraints (the potential must be bounded from below),
the neutrality of the vacuum,  and, in order for the perturbative expansion to make sense
and be reliable, perturbative unitarity constraints \cite{Lee:1977yc,Lee:1977eg}. 
Analyzing the most general 2HDM potential with the straightforward algebra is technically challenging,
and in order to proceed further one either simplifies it by imposing symmetries 
or resorts to more elaborate mathematical methods.

\subsubsection{Symmetries of the 2HDM scalar sector}

The number of free parameters can be reduced, and the model becomes analytically tractable, 
if one imposes additional global symmetries on the Higgs doublets.
Anticipating the NFC principle for the Yukawa sector in 2HDM, one often
considers the simplified 2HDM potential with $\lambda_6=\lambda_7 = 0$.
The quartic part of this potential is then invariant under 
\be
\Z_2: \quad \phi_1 \to \phi_1\,,\quad \phi_2 \to - \phi_2\,,\label{Z2-2HDM}
\ee
but the $m_{12}^2$ term in the quadratic part, either real or complex, softly breaks this $\Z_2$ symmetry.
With this simplified quartic potential, its stability in the strong sense \cite{Maniatis:2006fs} implies
\cite{Deshpande:1977rw,2HDM-review}:
\be
\lambda_1 > 0\,, \quad \lambda_2 > 0\,, \quad \sqrt{\lambda_1\lambda_2} + \lambda_3 > 0\,, \quad
\sqrt{\lambda_1\lambda_2} + \lambda_3 + \lambda_4 - |\lambda_5|> 0\,.\label{stability-2HDM} 
\ee
If in addition one sets $m_{12}^2=0$, the $\Z_2$ symmetry becomes exact;
the inert doublet model (IDM) described in section~\ref{section-IDM} is based on such a choice.
Furthermore, if $\lambda_5 = 0$, the model acquires the global $U(1)$ symmetry 
$\phi_1 \to \phi_1$, $\phi_2 \to e^{i\alpha}\phi_2$ known as Peccei-Quinn symmetry \cite{Peccei:1977hh}.
In total, there are six classes of family and $CP$-type symmetries which can be imposed
on the Higgs sector of 2HDM without producing accidental symmetries \cite{Ivanov:2006yq,Ferreira:2010yh}.
If one disregards $U(1)_Y$ gauge interactions, then even larger global basis changes are allowed, 
mixing $\phi_i$ with $\tilde\phi_i$. 
Within 2HDM, the full classification of such symmetries was given in \cite{Battye:2011jj,Pilaftsis:2011ed};
the phenomenology of the so called maximally symmetric 2HDM based on the softly broken $SO(5)$
symmetry group was presented in \cite{Dev:2014yca}.

\subsubsection{Basis-invariant methods} 

Coming back to the general 2HDM, we remark
that not all of the 14 parameters in Eq.~(\ref{V-2HDM}) are physical. 
If one focuses on the scalar sector only, then there exists a large freedom
to change the basis in the $(\phi_1, \phi_2)$ space, leading to 
reparametrization of the Higgs potential (the explicit expressions can be found in \cite{Ginzburg:2004vp,Davidson:2005cw}). 
It allows one to remove three among 14 free parameters without affecting the physical observables.

The freedom of basis changes brings up the issue of hidden symmetries.
The potential can be invariant under
a certain non-obvious reflection in the $(\phi_1, \phi_2)$ space, which after an appropriate
basis change becomes the $\Z_2$-symmetry (\ref{Z2-2HDM}), 
but its presence is hard to detect in the original basis.
In order to cope with this challenge, one needs to describe the model in terms of {\em basis-invariant quantities}.
The presence of a symmetry leads to certain relations among basis invariants (for example, setting some of them to zero),
which can be checked in arbitrary bases.
Technically, the basis invariants in the scalar sector are constructed as various fully contracted products of tensors $Y_{ij}$
and $Z_{ijkl}$ in Eq.~(\ref{V-NHDM}). 
This formalism was developed in \cite{Branco:2005em,Davidson:2005cw,Gunion:2005ja,Haber:2006ue,Haber:2010bw}
and applied, in particular, to establishing the basis-invariant conditions of explicit $CP$-violation.
A weak point of the tensorial combinatorics is that, when building invariants with higher and higher powers of
$Y_{ij}$ and $Z_{ijkl}$ and contracting their indices in many ways, one does not immediately know when to stop.
Basis invariants form a finitely generated algebraic system (a ring), and any basis invariant
can be expressed as a polynomial of a finite set of algebraically independent invariants.
It is this minimal set of invariants that needs to be checked in any basis if one wants to obtain
the necessary and sufficient conditions for the symmetry to exist.
However, determining this set is a difficult task.
Within 2HDM, it can be done in a relatively simple way with the geometric bilinear approach described below.
For more than two Higgs doublets, this minimal set is not yet known,
although the tensorial approach itself to constructing $CP$-odd basis invariants 
was recently extended to $N$ Higgs doublets and other non-minimal sectors \cite{Varzielas:2016zjc}.

Another set of basis-invariant methods was developed in mid-2000s by the Heidelberg group 
\cite{Nagel:2004sw,Maniatis:2006fs,Maniatis:2007vn}, Ivanov \cite{Ivanov:2005hg,Ivanov:2006yq,Ivanov:2007de,Ivanov:2008er}, 
and Nishi \cite{Nishi:2006tg,Nishi:2007dv}, although first hints at this methods appeared even earlier \cite{Sartori:2003ya}. 
This powerful technique is based on Higgs field bilinears and on geometric constructions in their space. 
It was used to answer several questions
pertinent to all 2HDMs, which were too difficult for the straightforward algebra. 
In this technique, one introduces the real-valued four-vector of bilinears
\be
r^\mu = (r_0, r_a)\,,\quad r_0 = \fd_i \f_i\,,\quad r_a = \fd_i (\sigma^a)_{ij} \f_j\,,\quad a=1, 2, 3\,,
\label{bilinears-2HDM}
\ee
and rewrites the Higgs potential (\ref{V-2HDM}) as
\be
V = M_\mu r^\mu + {1 \over 2}\Lambda_{\mu\nu}r^\mu r^\nu\,,\label{V-bilinears-2HDM}
\ee
with the usual Minkowski space convention for index contraction (e.g. $M_\mu r^\mu = M_0 r_0 - M_a r_a$).
The four free parameters $m_{ij}^2$ are now grouped into components of $M_\mu$,
\be
M_\mu = (M_0, -M_a)
=
\left({m_{11}^2 + m_{22}^2 \over 2},\,
-\textrm{Re}\left( m_{12}^2\right),\,
\textrm{Im}\left( m_{12}^2\right),\,
{m_{11}^2 - m_{22}^2 \over 2}
\right),
\label{M-2HDM}
\ee
and the ten quartic free parameters $\lambda_i$ fill the entries of the real symmetric matrix $\Lambda_{\mu\nu}$:
\be
\Lambda_{\mu\nu}
=
{1 \over 2}
\left(
\begin{array}{cccc}
	\ {\lambda_1 + \lambda_2 \over 2} + \lambda_3\  &
	\ \textrm{Re}\left( \lambda_6 + \lambda_7 \right)\  &
	\  -\textrm{Im}\left( \lambda_6 + \lambda_7 \right)\  &
	\ {\lambda_1 - \lambda_2 \over 2}\ \\*[1mm]
	\ \textrm{Re}\left( \lambda_6 + \lambda_7 \right)\  &
	\  \lambda_4 + \textrm{Re}\left( \lambda_5 \right)\  &
	\  -\textrm{Im}\left( \lambda_5 \right)\  &
	\ \textrm{Re}\left( \lambda_6 - \lambda_7 \right)\  \\*[1mm]
	\ -\textrm{Im}\left( \lambda_6 + \lambda_7 \right)\  &
	\  - \textrm{Im}\left( \lambda_5 \right)\  &
	\ \lambda_4 - \textrm{Re}\left( \lambda_5 \right)\  &
	\ -\textrm{Im}\left( \lambda_6 - \lambda_7 \right)\ \\*[1mm]
	\ {\lambda_1 - \lambda_2 \over 2}\  &
	\ \textrm{Re}\left( \lambda_6 - \lambda_7 \right)\  &
	\ -\textrm{Im}\left( \lambda_6 - \lambda_7 \right)\  &
	\ {\lambda_1 + \lambda_2 \over 2} - \lambda_3\
\end{array}
\right).
\label{Lambda-2HDM}
\ee
Each $r^\mu$ in (\ref{bilinears-2HDM}) is in one-to-one correspondence with the electroweak gauge
orbit in the space of doublets $\phi_i$. The map (\ref{bilinears-2HDM}) from doublets $\phi_i$
to $r^\mu$ does not cover the entire $1+3$-dimensional $r^\mu$ space but rather a region in it called the orbit space.
In 2HDM, the orbit space is the cone defined by $r_0 \ge 0$, $r_\mu r^\mu = r_0^2 - r_a^2 \ge 0$.
The minimum of the Higgs potential is represented by a point on or in this cone:
the apex corresponds to the electroweak-symmetric vacuum, 
a point on the surface corresponds to the normal neutral
vacuum, a point in the interior leads to a charge-breaking vacuum.

When passing from Higgs doublets to bilinears, the space we work in becomes more complicated
but the {\em object} we study gets simpler. 
The potential $V$ is now a quadratic rather than quartic function of variables.
This makes it possible to write the stability conditions for the general 2HDM 
in closed, compact, and basis-invariant form \cite{Maniatis:2006fs,Ivanov:2006yq},
derive necessary and sufficient conditions for $CP$-violation, 
both explicit and spontaneous \cite{Ivanov:2005hg,Nishi:2006tg,Maniatis:2007vn,Ferreira:2010yh}, 
compute the number of extrema and minima \cite{Nagel:2004sw,Maniatis:2006fs,Ivanov:2007de}, list all possible global symmetries 
and symmetry-breaking patterns the 2HDM potential can have \cite{Ivanov:2006yq,Ivanov:2007de,Ferreira:2010yh}
and study their behavior under renormalization \cite{Ma:2009ax,Maniatis:2011qu},
to compute the masses of the physical Higgs bosons \cite{Degee:2009vp},
and construct the full tree-level phase diagram of 2HDM \cite{Ivanov:2008er}.
The search for the minimum can be recast in geometric \cite{Ivanov:2007de} or algebraic form \cite{Nagel:2004sw,Maniatis:2006fs},
and can be conveniently implemented in computer codes using the Groebner bases technique \cite{Maniatis:2006jd}
or polynomial homotopy continuation method \cite{Maniatis:2012ex}.
In short, practically all structural issues of the tree-level Higgs potential in 2HDM can be efficiently dealt with
using the bilinear formalism.
One aspect which has not yet been treated with bilinears is the explicit expression 
for the perturbative unitarity constraints \cite{Lee:1977yc,Lee:1977eg}. 
For softly broken $\Z_2$ symmetry they take rather simple form \cite{Akeroyd:2000wc}. 
In the most general 2HDM, they were derived as constraints
on the eigenvalues of four by four scattering matrices in various 
two-boson scattering channels \cite{Ginzburg:2005dt},
but its basis-invariant expressions are not known.

\subsubsection{Minimization of the potential. Global vs. local minimum}

The global minimum of the potential gives the vevs of the Higgs fields:
\be
\lr{\phi_1^0} = {v_1\over \sqrt{2}} = {v \over \sqrt{2}}\cos\beta\,,\quad 
\lr{\phi_2^0} = {v_2\over \sqrt{2}}e^{i\xi} = {v \over \sqrt{2}}\sin\beta e^{i\xi}\,, 
\quad
\tan\beta = {v_2 \over v_1}\,,
\label{vacuum-2HDM}
\ee
with real $v_1$, $v_2$. Without loss of generality, the angle $\beta$ can be limited to the first quadrant \cite{Carena:2002es}.
In principle, given the potential (\ref{V-2HDM}), the location of the global minimum must follow from 
the minimization conditions. 
However, in the most general 2HDM it cannot be explicitly expressed in the parameters. 
Setting all first derivatives to zero leads to a system of coupled non-linear equations or
to a single algebraic equation of order six \cite{Nagel:2004sw,Maniatis:2006fs}.

If the direct minimization of a given potential is hard,
one can revert the problem by first choosing the desired vevs 
and then fixing the quadratic parameters $m_{ij}^2$ through the minimization condition,
see \cite{Pilaftsis:1999qt} and Eq.~(195) of \cite{2HDM-review}. 
In this way, one ends up with a 2HDM potential whose minimum is known by construction. 
But it leads to another problem: the same tree-level scalar potential can have another, deeper minimum.
This second minimum, even if it exists, cannot be charge-breaking:
as shown in \cite{Ferreira:2004yd,Barroso:2005sm}, the presence of a neutral minimum 
makes any possible charge-breaking extremum a saddle point.
However, the 2HDM tree-level potential can possess at least two \cite{Barroso:2007rr} 
--- and at most two \cite{Ivanov:2007de} --- neutral minima at different depths. 
Thus, it is not guaranteed that the minimum one starts with is the global one.
If we knew the location of the second minimum, it would be straightforward to check 
which one is the deepest, but finding it is no easier than the initial problem of minimization of a given potential.

Fortunately, the same bilinear formalism provides a practical method to check
whether the selected minimum of the tree-level potential is indeed global
\cite{Barroso:2012mj,Barroso:2013awa,Ivanov:2015nea}.
For example, within the $CP$-conserving 2HDM with softly broken $\Z_2$,
one simply needs to evaluate the sign of the following discriminant $D$ \cite{Barroso:2013awa}:
\be
D = m_{12}^2(m_{11}^2 - k^2 m_{22}^2)(\tan\beta - k)\,, \quad k \equiv \sqrt[4]{\lambda_1/\lambda_2}\,.\label{discriminant}
\ee
If $D > 0$, then the minimum is definitely global, being either the only one or the deepest among the two.
If $D< 0$, we have chosen the local minimum, which is metastable.
This quick decision uses only local information, as it avoids the need to search for the second minimum, 
and it is convenient for numerical scans of the parameter space.
In the general 2HDM, the expression for $D$ is more involved,
but still the presence of an extra deeper minimum can be detected via a quick and efficient algorithm \cite{Ivanov:2015nea}.

\subsubsection{Physical Higgs spectrum}

The physical scalar sector of 2HDM consists of three neutral scalars
and a pair of charged Higgs bosons $H^\pm$ \cite{Gunion:1989we}.
Rephasing the second doublet to make its vev real and positive \cite{Ginzburg:2004vp},
one expands the Higgs doublets around minimum as
\be
\phi_1 = \doublet{\varphi_1^+}{{1\over\sqrt{2}}(v_1 + \eta_1 + i \chi_1)}\,, \quad 
\phi_2 = \doublet{\varphi_2^+}{{1\over\sqrt{2}}(v_2 + \eta_2 + i \chi_2)}\,.
\ee
The Goldstone modes are $G^\pm = c_\beta \varphi_1^\pm + s_\beta \varphi_2^\pm$
and $G^0 = c_\beta \chi_1 + s_\beta \chi_2$.
The orthogonal combinations are the physical charged Higgs boson 
$H^\pm = -s_\beta \varphi_1^\pm + c_\beta \varphi_2^\pm$,
and the neutral field $\eta_3 = -s_\beta \chi_1 + c_\beta \chi_2$.
In general, the three neutral scalars $\eta_i$ mix giving rise to the three neutral Higgs bosons: 
$(H_1, H_2, H_3)^T = {\cal R}(\eta_1, \eta_2, \eta_3)^T$, with the rotation matrix ${\cal R}$
parametrized with three angles $\alpha_1$, $\alpha_2$, $\alpha_3$.
For the most general 2HDM, the neutral mass matrix can be written in different bases,
see \cite{Pilaftsis:1999qt,Ginzburg:2004vp} and two different forms in section 5.10 and Appendix C of \cite{2HDM-review},
as well as in the basis independent form \cite{Haber:2010bw} 
and with the bilinear technique \cite{Ivanov:2006yq,Degee:2009vp}.
Its explicit analytical diagonalization in terms of the parameters is cumbersome
but still manageable \cite{Pilaftsis:1999qt}.

\begin{figure}[!htb]
   \centering
\includegraphics[height=5cm]{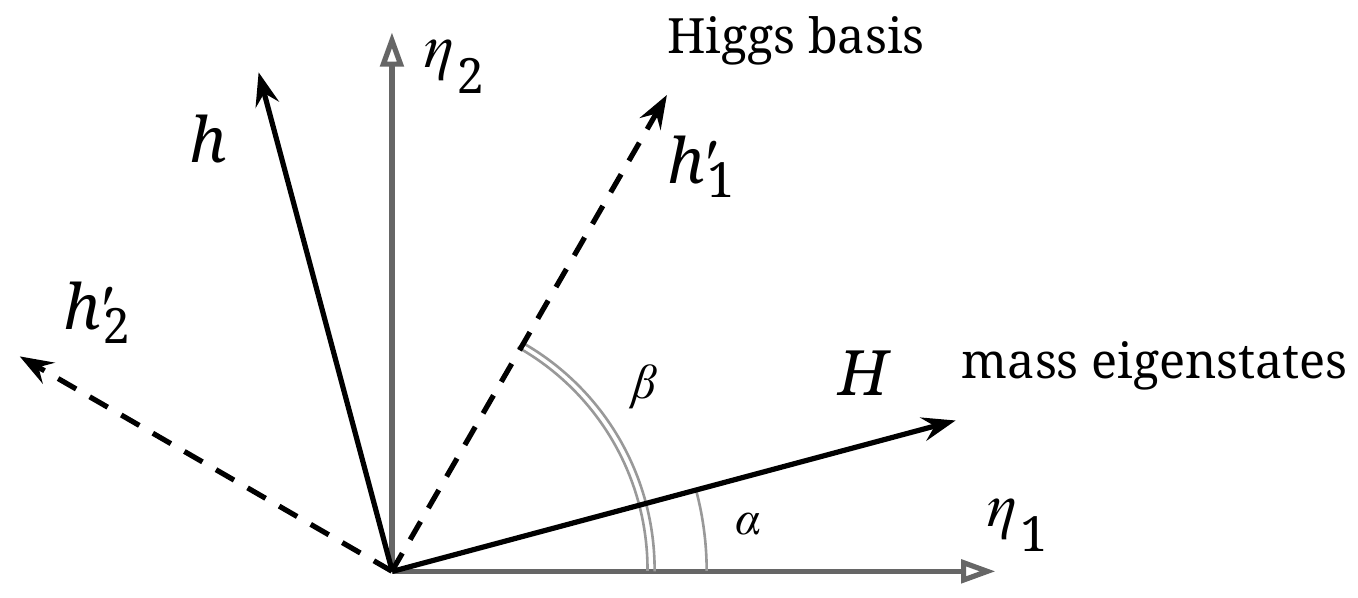}
\caption{Three bases in the space of $CP$-even neutral scalar fields:
the original basis $(\eta_1, \eta_2)$, the Higgs basis $(h_1', h_2')$, and 
the basis of mass eigenstates $(H, h)$.}
   \label{fig-2HDM-bases}
\end{figure}

If the model is $CP$-conserving, then $\eta_3 \equiv A$ decouples in the mass matrix from the first two
and becomes the $CP$-odd pseudoscalar. 
The two $CP$-even scalars mix and give rise
to the mass eigenstates $H = c_\alpha \eta_1 + s_\alpha \eta_2$ 
and $h = - s_\alpha \eta_1 + c_\alpha \eta_2$, with the convention that $h$ is the ligher of the two.
The mixing angle $\alpha$ is usually defined from $-\pi/2$ to $\pi/2$;
alternative definitions are sometimes used.
The relation among the three bases in the space of $CP$-even neutral fields 
is visualized in Fig.~\ref{fig-2HDM-bases}.
The original basis $(\eta_1, \eta_2)$, in which the fermion sector featuring natural flavor conservation takes the simplest form, 
is rotated by angle $\beta$ to obtain the Higgs basis $(h_1', h_2')$.
In this basis, the vev is present only in the first doublet, so that the gauge bosons
interact only with $h_1'$ but not $h_2'$.
The physical Higgses $(H, h)$ form the third basis, rotated by $\alpha$ with respect to the original one
or by $\beta-\alpha$ with respect to the Higgs basis.
From here, one obtains the Higgs couplings to gauge bosons relative to the SM $\xi_{VV}^{h_i}$:
\be
\xi_{VV}^{h} = \sin(\beta-\alpha)\,, \quad 
\xi_{VV}^{H} = \cos(\beta-\alpha)\,, \quad 
\xi_{VV}^{A} = 0\,. \label{2HDM-hVV}
\ee
One immediately sees that 2HDMs, and in general NHDMs, predict that all such couplings are less than one
and they satisfy the following sum rule: $\sum_{i} (\xi^{h_i}_{VV})^2 = 1$ \cite{Gunion:1990kf}.
In the $CP$-conserving case, the pseudoscalar boson $A$ decouples from the gauge boson pairs at the tree level,
which is easily seen in the Higgs basis. The same logic also leads to the absence of the tree-level
coupling $H^+W^-Z$ \cite{Grifols:1980uq,Gunion:1989we}; such vertex can appear at tree level only
in models with new scalars in higher-dimensional representations of $SU(2)_L$. 
Finally, expressions for the trilinear and quartic scalar interactions are listed
in \cite{Ginzburg:2004vp,2HDM-review}.

The alignment limit corresponds to $\beta - \alpha \to \pi/2$. In this case, the lightest neutral $h$ is SM-like Higgs
and is identified with the experimentally observed $h_{125}$. The second neutral Higgs $H$, even if it is relatively light,
decouples from the gauge-boson pairs.
In the other limit,
$\beta \approx \alpha$, the heavier state $H$ is SM-like, and a second, lighter Higgs is predicted.

\subsubsection{Reconstructing the potential via physical parameters}

As mentioned above, instead of directly minimizing a given 2HDM potential, 
it is more convenient in practice to fix the desired vevs and the masses of the physical Higgs bosons
and then to reconstruct the potential in terms of experimentally observable quantities.
For example, the $CP$-conserving model with softly broken $\Z_2$ contains 8 free parameters in the potential.
One can trade them for $v$, $\tan\beta$, the masses of the physical Higgses $m_h$, $m_H$, $m_A$, $m_{H^\pm}$,
plus the $CP$-even mixing angle $\alpha$ and the remaining $\Z_2$-breaking parameter $m_{12}^2$.
The quadratic coefficients $m_{11}^2$ and $m_{22}^2$ remain fixed by the extremization (``tadpole'') conditions,
while the quartic parameters can be expressed as \cite{Gunion:2002zf,Ginzburg:2004vp}
\bea
&&\lambda_1=
\frac{1}{v^2 c_\beta^2}\left(c_\alpha^2 m_H^2 + s_\alpha^2 m_h^2
- m^2_{12}\frac{s_\beta}{c_\beta}\right), \quad
\lambda_2=
\frac{1}{v^2 s_\beta^2}\left( s_\alpha^2 m_H^2 + c_\alpha^2 m_h^2
- m^2_{12}\frac{c_\beta}{s_\beta}\right), \label{eq:coup}\\
&& \lambda_3
=
\frac{2 m_{H^\pm}^2}{v^2} +
\frac{s_{2 \alpha} (m_H^2 - m_h^2)}{v^2 s_{2 \beta}} - \frac{m^2_{12}}{v^2 s_\beta c_\beta},
\quad \lambda_4
=
\frac{m_A^2 - 2 m_{H^\pm}^2}{v^2} + \frac{m^2_{12}}{v^2 s_\beta c_\beta}, \quad
\lambda_5
= \frac{m^2_{12}}{v^2 s_\beta c_\beta} - \frac{m_A^2}{v^2}.\nonumber
\eea
It turns out however that a random scan over this parameter space often produces points
which violate perturbativity constraints. To optimize this computationally costly
procedure, Haber and Stal \cite{Haber:2015pua} proposed a ``hybrid'' basis of parameters for the
$CP$-conserving model with softly broken $\Z_2$:
$v$, $m_h$, $m_H$, $\beta-\alpha$, $\tan\beta$, and three quartic couplings
defined in the Higgs basis. In this basis, they produced several carefully selected benchmark scenarios
to be checked in detail at the LHC.

For $CP$-violating model with softly broken $\Z_2$, the expressions become more
cumbersome \cite{Arhrib:2010ju,Fontes:2014xva}.
A study of the optimal choice of observable parameters in $CP$-violating
2HDM with softly broken $\Z_2$ symmetry was undertaken in \cite{Grzadkowski:2013rza}. 
The input parameters were chosen to be a single quadratic parameter $\mu^2$, $\tan\beta$,
three angles $\alpha_i$ of the neutral Higgs mixing matrix, and the masses of two neutral Higgses
and the charged ones. A strategy of determining these quantities via experimental measurements
was presented. This choice facilitated exploration of various regimes of $CP$-violation 
in the scalar sector of 2HDM.

For the most general 2HDM the problem is further complicated.
Due to a large number of vertices which could in principle be measured,
the best choice of free parameters is far from obvious.
The approach of \cite{Grzadkowski:2013rza} was extended in \cite{Grzadkowski:2014ada,Grzadkowski:2016szj}
to the general 2HDM bosonic sector, with a special focus on detecting $CP$-violating features of the scalar potential.
Another work \cite{Ginzburg:2015cxa} suggests to use, apart from masses of all Higgs bosons and $v$,
the couplings of two neutral Higgses to gauge bosons,
three triple couplings $h_i H^+H^-$ and one quartic coupling $H^+H^-H^+H^-$.
The eleven physical free parameters of the potential
can be expressed in terms of those observables, but of course their experimental determination
is very challenging.

\subsection{Building 2HDMs: the Yukawa sector}

As mentioned earlier, the completely general Yukawa matrices in (\ref{NHDM-Yukawa}) 
lead to wild fermion-family mixing transitions via Higgs exchanges, which must be harnessed
in the light of strong experimental constraints from meson properties.
This requirement has led to two large families of 2HDM Yukawa sectors: 
those which completely forbid the tree-level flavor-changing neutral currents (FCNC)
by implementing the symmetry-protected natural flavor conservation (NFC)
and those which do not, allowing for FCNC but keeping them under control.

\subsubsection{Natural flavor conservation}

In 1976, Glashow and Weinberg \cite{Glashow:1976nt} and Paschos \cite{Paschos:1976ay}
proved that a necessary and sufficient condition for complete absence of tree-level FCNC 
is that all fermions with the same charges couple only to one Higgs doublet. 
It can be achieved by imposing an extra global $\Z_2$ symmetry for each Higgs doublet 
which flips its sign together with the signs of the corresponding right-handed fermions.
The mass matrices (\ref{NHDM-fermion-masses}) for any fermion sector, leptons, up, and down-type quarks,
are proportional to the single Yukawa matrix. 
As in SM, diagonalizing the fermion masses automatically diagonalizes the scalar-fermion interactions, and no FCNC
transitions occur at tree level.

In 2HDM, one imposes a single symmetry $\Z_2$, which selects one doublet and the fermions
it couples to.
Connecting all fermions to doublets, one can construct four different types of 2HDM Yukawa sectors 
with NFC \cite{Gunion:1989we,Barger:1989fj,2HDM-review}, which are listed in Table~\ref{table:2HDM}.
This traditional nomenclature is based on the convention that right-handed up-quarks always couple to $\phi_2$.
\begin{table}[h]
\begin{center}
\begin{tabular}{lcccccccc}
\hline
model & $\quad$ & $u_R$ & $d_R$ & $\ell_R$ & $\quad$ & $\xi^h_u$ & $\xi^h_d$ & $\xi^h_\ell$ \\[1mm] 
\hline
Type I & & $\phi_2$ & $\phi_2$ & $\phi_2$ && $\cos\alpha/\sin\beta$ & $\cos\alpha/\sin\beta$ & $\cos\alpha/\sin\beta$ \\
Type II & & $\phi_2$ & $\phi_1$ & $\phi_1$  && $\cos\alpha/\sin\beta$ & $-\sin\alpha/\cos\beta$ & $-\sin\alpha/\cos\beta$ \\
Lepton-specific& & $\phi_2$ & $\phi_2$ & $\phi_1$  && $\cos\alpha/\sin\beta$ & $\cos\alpha/\sin\beta$ & $-\sin\alpha/\cos\beta$ \\
Flipped & & $\phi_2$ & $\phi_1$ & $\phi_2$  && $\cos\alpha/\sin\beta$ & $-\sin\alpha/\cos\beta$ & $\cos\alpha/\sin\beta$ \\
\hline
\end{tabular}
\end{center}
\caption{Yukawa sectors in the four types of 2HDM models with natural flavor conservation and couplings 
of the lighest $CP$-even Higgs to fermions relative to the SM.}\label{table:2HDM}
\end{table}
\begin{itemize}
\item
Type I: all fermions couple to $\phi_2$, while $\phi_1$ remains inert in the fermion sector.
The $\Z_2$ symmetry consists in flipping only the sign of $\phi_1$.
Since the fermions interact with neutral scalars only via $y_f \bar f f \eta_2/\sqrt{2} = (m_f/v_2) \bar f f \eta_2$,
and since the lightest $CP$-even Higgs boson $h$ has projection $\cos\alpha$ on $\eta_2$, 
its coupling to the fermions relative to the SM is $\xi^{h}_{f} = \cos\alpha/\sin\beta$ and is the same across all fermions.
For the heavy neutral $H$, the relative coupling is $\xi^{H}_{f} = \sin\alpha/\sin\beta$. 
Notice that in the alignment limit, $\xi^{h}_{f} \to 1$ and $\xi^{H}_{f} \to -\cot \beta$.
\item 
Type II: all down-type fermions, $d_R$ and $e_R$, get their masses from $\phi_1$
while the up-type quarks $u_R$ couple to $\phi_2$. 
This Yukawa structure arises in supersymmetric models; in particular, the Higgs sector of MSSM
is exactly Type II 2HDM with extra constraints on the parameters of the potential 
\cite{Gunion:1984yn,Gunion:1989we,Carena:2002es,Djouadi:2005gj}.
The $h\bar f f$ couplings now differ for up quarks $\xi^{h}_{u} = \cos\alpha/\sin\beta$ and 
down quarks $\xi^{h}_{d} = -\sin\alpha/\cos\beta$.
This model leads to a richer phenomenology as the hierarchy of the 125 GeV Higgs couplings
to fermions can strongly deviate from their mass pattern.
As the result, it can be constrained stronger than the Type I model.
Type I and II are the most explored versions of the 2HDM with NFC. 
\item
The other two cases are less developed. 
Lepton-specific, or Type X: the quarks couple only to $\phi_2$
while the leptons get their masses from $\phi_1$. Suggested first in \cite{Barnett:1983mm,Barnett:1984zy} in 1984,
it has recently received attention \cite{Aoki:2009ha} as a means of explaining several lepton-relation
deviations such as muon $g-2$ discrepancy or the cosmic positron excess.
The last case is the so-called flipped, or Type Y, model which is similar to Type X but with $\phi_1$ coupling to $d_R$ rather than $e_R$.
Type X and Type Y 2HDMs were denoted Type III and Type IV in the early literature;
the present day convention is that Type III labels the most-general Yukawa sector with FCNC.
\end{itemize}
The Higgs sector does not need to be exactly invariant under the $\Z_2$ symmetry.
In fact, imposing the exact $\Z_2$ on the Higgs potential
simplifies it so much that there remains no $CP$-violation in the scalar sector.
To relax this constraint, one introduces the soft $\Z_2$-breaking term $m_{12}^2 \fd_1 \f_2 + h.c.$ 
with either real or complex $m_{12}^2$ in order to incorporate explicit and spontaneous $CP$-violation.
The expressions for the Higgs-fermion couplings will, of course, differ from those given in Table~\ref{table:2HDM};
they can be found in \cite{Fontes:2015mea}. However the flavor-diagonal structure of these couplings is preserved,
and no FCNC appear at tree level.

A separate class of models is represented by the so-called aligned 2HDM (a2HDM) \cite{Pich:2009sp}.
In this model, the fermions are allowed to couple to both Higgs doublets but their
Yukawa matrices are proportional to each other: $\Gamma_1 \propto \Gamma_2$ and 
$\Delta_1 \propto \Delta_2$.
Diagonalization of the mass matrices $M_d = (\Gamma_1 v_1 + \Gamma_2 v_2)/\sqrt{2}$
and $M_u = (\Delta_1 v^*_1 + \Delta_2 v^*_2)/\sqrt{2}$ automatically diagonalizes
each Higgs-fermion interaction, preventing FCNC. 
Notice that the word ``aligned'' in the name of this model refers to the Yukawa structures
as aligned vectors in the space of $3\times 3$ matrices, which is not to be confused 
with alignment of the Higgs states.
In fact, the proportionality of the matrices $\Gamma_1$ and $\Gamma_2$ implies 
that it is possible to perform a basis change in the space of Higgs doublets 
$\Gamma_1 \phi_1 + \Gamma_2 \phi_2 = \Gamma'_1 \phi'_1 + \Gamma'_2 \phi'_2$
such that $\Gamma'_2 = 0$, which restores hidden NFC in this sector.
The same can be achieved, separately, for the up-quarks with a different basis rotation.
Thus, the up and down-type quark sectors also follow the NFC principle
but in two misaligned bases.
This misalignment implies that there is no exact $\Z_2$ symmetry, which would protect the entire lagrangian
and be stable under renormalization \cite{Ferreira:2010xe,Bijnens:2011gd}.
Nevertheless, a2HDM provides a convenient Ansatz for phenomenological studies.

\subsubsection{Models with tree-level FCNC}
The general Yukawa sector in 2HDM leads, for non-zero $v_1$ and $v_2$, to large tree-level FCNC 
interactions via neutral Higgs exchanges. This is most clearly seen in the Higgs basis,
where only the first doublet $H_1$ gets vev and gives masses to the fermions.
The neutral field $h'_1$ residing in $H_1$ does not induce FCNC transitions:
diagonalization of the fermion mass matrix also diagonalizes the $h'_1\bar f f$ interactions.
However the neutrals from the second doublet can couple to fermion bilinears in an arbitrary way.
The FCNC-inducing interactions can be written as
\be
{\cal L}_{\mathrm{FCNC}} = \bar u_{Li} (Y_u)_{ij} u_{Rj} H_2^{0*} + \bar d_{Li} (Y_d)_{ji} d_{Rj} H_2 
+ \bar \ell_{Li} (Y_\ell)_{ij} \ell_{Rj} H_2 + h.c.
\ee
The fermion fields already correspond to the mass eigenstates, so that the off-diagonal elements
in matrices $Y$ induce tree-level FCNC transitions.
If unsuppressed, these transitions can lead to dramatic effects, such as the $\mu \to e\gamma$
decay \cite{Bjorken:1977vt} and the $K-\bar K$ oscillations \cite{McWilliams:1980kj}.
Assuming that all entries in each $Y$ are of the same order and are given by the mass of the heaviest
fermion in that sector ($m_b$ for $Y_d$, which is relevant for kaon mixing), 
one could establish a lower bound on the mass of the extra neutral scalars of the order of 100 TeV \cite{McWilliams:1980kj}.
In early 80's, this was seen by many as a strong argument in favor of complete absence of tree-level FCNC.

Not willing to give up this attractive phenomenological opportunity,
Cheng and Sher put forth in \cite{Cheng:1987rs} a proposal to make FCNC effects naturally small.
They argued that since the fermion masses display a hierarchical structure,
the non-diagonal Higgs couplings with fermions could as well. 
They proposed what is now known as the Cheng–Sher Ansatz: 
$Y_{ij} \propto \sqrt{m_i m_j}$ up to coefficients of order one.
This hierarchical structure of non-diagonal interactions was not a pure guess;
it was backed up by concrete examples such as the Fritzsch Ansatz \cite{Fritzsch:1977za}
for the quark mass matrices,
which at that time was well compatible with experimental data on the CKM matrix.
The Cheng-Sher Ansatz suppresses the non-diagonal Yukawa couplings for the first two fermion generations
not through very heavy Higgses but through the small masses of the quarks involved.

Another class of models in which the Higgs-induced FCNCs are naturally suppressed 
was proposed by Branco, Grimus, and Lavoura \cite{Branco:1996bq} and is known as BGL models.
In these models, the tree-level flavor-changing couplings of the neutral scalars are related in an exact way 
to elements of the CKM matrix $V$. 
To achieve that, they started with the generic Yukawa structures, diagonalized the fermion masses,
and then expressed the Higgs field interactions with physical quarks
with two matrices $N_u$ and $N_d$.
For example, 
\be
N_d = \tan\beta D_d - {v e^{i\beta}\over \sqrt{2}\cos\beta} U_{dL}^\dagger \Gamma_2 U_{dR}\,.
\ee
where $D_d$ is the diagonal mass matrix and the second term induces FCNC.
Imposing invariance under rephasing symmetry $Q_{L3} \to e^{i\psi}Q_{L3}$,
$u_{R3}\to e^{2i\psi}u_{R3}$, $\phi_2 \to e^{i\psi} \phi_2$, they found that 
the second term in $N_d$ can be expressed via elements of the CKM matrix 
$V_{i3}^\dagger V_{3j} D_{jj}$. For up-quarks, $N_u$ is just diagonal.
Thus, FCNC exist in this model but they are suppressed by the combination of small
elements $V_{i3}^\dagger V_{3j}$.

Some variants of these models bear much in common with 
the hypothesis of Minimal Flavor Violation (MFV) \cite{Buras:2000dm,D'Ambrosio:2002ex}.
In this framework, one forbids New Physics to bring its own flavor-violating effects.
The only source of flavor violation in the entire theory, at least at low energies, is the Yukawa sector.
Tecnically, one promotes the Yukawa couplings to scalar fields, called flavons, which transform 
in certain way under the global flavor symmetry group $G_F = U(3)^5$,
so that the entire lagrangian, together with the would-be Yukawa sector, is now $G_F$-symmetric. 
When flavons acquire their own vevs, the conventional Yukawa interactions arise, 
and their pattern reflects the initial flavor symmetry.
Then, after electroweak symmetry breaking, flavor-changing interactions of neutral Higgses
appear, but they become related in an exact way with the quark mixing matrix.
The explicit demonstration of how this idea work in multi-doublet models, thus, generalizing 
the original BGL models, was presented in \cite{Botella:2009pq}.

Yukawa sectors can also be shaped with discrete symmetry
groups such as $\Z_3$ \cite{Ferreira:2010ir,Aranda:2014lna}, $S_3$ \cite{Kajiyama:2013sza,Cogollo:2016dsd},
or the ``nearest-neighbour interaction'' model \cite{Branco:2010tx}.
When imposed on the Higgs potential, they produce an accidental continuous symmetry,
which needs either to remain unbroken or be explicitly broken by soft terms to avoid massless scalars,
and leads to characteristic phenomenological predictions.

\subsection{$CP$-violation in 2HDM}\label{section-CPV-2HDM}

The experimental fact that weak interactions are not invariant under $C$ and $P$ transformations
is incorporated in theory via the gauge multiplet choices for the fermions. 
But the gauge structure does not specify whether the theory is invariant under the combined transformation $CP$.
It is again an experimental fact that the world is not $CP$-invariant \cite{CPV-book}.
Within SM, the $CP$-violation (CPV) is put by hand:
it arises from the fact that the CKM matrix is not real 
and originates from the complex Yukawa matrices.
The weak-basis invariant formulation of this result is that the mass matrices $M_d$ and $M_u$
arising from Yukawa couplings (\ref{SM-mass-matrices})
are such that the weak basis invariant $I_{CP}$ defined as 
\be
I_{CP} = \Tr [H_u, H_d]^3\,, \quad H_u = M_u M_u^\dagger, \quad H_d = M_d M_d^\dagger \label{ICP}
\ee
is non-zero \cite{Bernabeu:1986fc}.

Models with extended scalar sector including NHDMs 
are attractive in this respect in that they can provide a rationale for the emergence of $CP$-violation
and can also accommodate additional CPV interactions beyond the charged weak currents.
The initial proposals of the 2HDM by T.~D.~Lee \cite{Lee:1973iz} and 3HDM by Weinberg \cite{Weinberg:1976hu}
were motivated exactly by the search for scalar sector explanation of CPV.
In fact, when T.~D.~Lee proposed in 1973 spontaneous $CP$-violating two-Higgs-doublet model, 
the three-generation paper by Kobayashi and Maskawa was just published \cite{Kobayashi:1973fv} 
and had not yet become the widely accepted CKM paradigm.

In spontaneous CPV, we start with a perfectly $CP$-invariant
lagrangian with two Higgs doublets $\phi_i$ \cite{Lee:1973iz}. 
With the standard definition of how $CP$-transformation
acts on the scalar fields
\be
\phi_i(\vec r, t) \toCP (CP) \phi_i(\vec r, t) (CP)^{-1} = \phi_i^*(-\vec r, t)\,,\label{CP-standard}
\ee
one can guarantee explicit $CP$-conservation by setting all parameters
in the scalar potential real. However the minimum
can be such that the two vevs have a non-zero relative phase (\ref{vacuum-2HDM}). 
The mass matrices $M_d$ and $M_u$ in (\ref{NHDM-fermion-masses}) 
become complex even if the Yukawa couplings are chosen real, which can lead to the complex CKM matrix. 
In addition, the model provides $CP$-violating effects via Higgs boson exchanges. 
Notice that this form of CPV does not require three fermion generations: 
if the same physical neutral scalar couples to fermions via both $\bar \psi \psi$ and $\bar \psi \gamma_5 \psi$,
then CPV can take place even for single generation.

The explicit CPV originating from the scalar sector as formulated in 1976 by Weinberg 
\cite{Weinberg:1976hu} postulates that the scalar potential itself violates $CP$-invariance.
Whatever the vacuum, the Higgs boson exchanges, both charged and neutral ones, 
lead to CPV even for two fermion generations.
The essential difference of Weinberg's proposal with Lee's 2HDM is that
he managed to combine this effect with NFC at the expense of 
increasing the number of Higgs doublets from two to three.

Even this short recap demonstrates that this phenomenon has several facets. 
Further developments uncovered many subtle issues in this subject, 
and the entire literature on CPV in extended scalar sectors is a perfect illustration that 
these subtleties should not be disregarded.
Many of them were described in the classical textbooks and reviews \cite{CPV-book,2HDM-review}. 
We briefly summarize them and add results
published in the last few years, both for 2HDM and for NHDM in section~\ref{section-NHDM-CPV}. 

When building a model with desired $CP$ properties, one must keep in mind 
that quantum field theory by itself does not uniquely specify how discrete symmetry transformations
act on fields \cite{Feinberg1959,Lee:1966ik,CPV-book}.
This action must be assigned, and it can happen that this assignment is not unique. 
The standard convention (\ref{CP-standard}) is basis-dependent: 
the same transformation viewed in another basis with rephased scalar field 
contains a phase factor accompanying the conjugation.  
With several Higgs doublets with identical quantum numbers, 
the freedom of defining $CP$-transformation is even larger \cite{Feinberg1959,Lee:1966ik,Grimus:1995zi,CPV-book}:
\be
\phi_i(\vec r, t) \toCP X_{ij} \phi_j^*(-\vec r, t)\,,\label{GCP}
\ee
with any unitary $X \in U(N)$. 
These are often called generalized $CP$-transformations (GCP) \cite{Ecker:1987qp} 
as if contrasted to the ``standard'' $CP$. 
Upon a basis change $\phi'_i = U_{ij} \phi_j$, the new fields $\phi'$ 
transform under the same transformation (\ref{GCP})
with the matrix $X' = U X U^T$. So, what looks standard in one basis 
becomes generalized in another and a generalized one can become standard. 
The fact that this rule involves $U^T$ rather than $U^\dagger$ has an important consequence:
not every unitary $X$ in (\ref{GCP}) can be diagonalized
by a basis change. 
The simplest form of $X$ one can achieve 
is the block-diagonal form \cite{Ecker:1987qp,Weinberg:1995mt,Ferreira:2009wh}, with the blocks
being either phases or $2\times 2$ matrices
\be
\mmatrix{\cos\alpha}{\sin\alpha}{-\sin\alpha}{\cos\alpha}\quad \mbox{as in \cite{Ecker:1987qp},}\quad \mbox{or}\quad
\mmatrix{0}{e^{i\alpha}}{e^{-i\alpha}}{0}\quad \mbox{as in appendix 2C of \cite{Weinberg:1995mt}.}\label{block}
\ee
Each block contains its own parameter $\alpha$ which can be arbitrary.
Notice that applying GCP twice, one gets a usual family transformation,
$\phi_i \to (XX^*)_{ij} \phi_j$, with $XX^*$ not necessarily being identify.
This opens up the possibility of $CP$-transformations of higher order \cite{Ferreira:2009wh}:
if $\alpha = \pi/p$ with integer $p$, then one needs to apply it $2p$ times
to obtain identity. By $CPT$-invariance of the standard interaction terms,
this implies the $T$-transformation of higher order. Although such a possibility sounds exotic,
it is well consistent with all requirements of local causal quantum field theory.
Conversely, if one assumes that a GCP symmetry is an inversion, that is, it has order two,
then there exists a basis change which makes it diagonal, as in the left form in Eq.~(\ref{block})
with $\alpha = 0$ (the case of $\alpha=\pi$ reduces to it via an additional rephasing).
Thus, order two implies possibility to bring the GCP transformation to the standard form.

When building a multi-Higgs-doublet model, one can identify its $CP$-properties 
with the following procedure \cite{Lee:1966ik}.
One starts with kinetic terms and gauge interactions, with the lagrangian invariant under (\ref{GCP}) with any unitary $X$.
Then one adds the scalar potential, which breaks this large symmetry group. 
If there remains no $CP$-symmetry of the form (\ref{GCP}) which would leave the potential invariant, 
the model is said to have explicitly $CP$-violating scalar sector. If there is at least one such symmetry 
with whatever fancy $X$, the model is explicitly $CP$-conserving. 
There may be several GCP transformations with different $X$'s which leave the potential
invariant, and any of them can play the role of ``the'' $CP$-symmetry of the model
if one needs to assign it.
Finally, if the vacuum of an explicitly $CP$-conserving model violates all $CP$-symmetries of the potential, 
we have spontaneous CPV. 
If at least one among $CP$ symmetries of the potential leaves the vacuum invariant, the model continues to be $CP$-conserving.

\subsubsection{Explicit $CP$-violation}

For the ``standard'' $CP$ transformation (\ref{CP-standard}), 
explicit $CP$-conservation implies that all coefficients in the potential are real.
The basis-independent form of necessary and sufficient condition for the scalar potential
to be explicitly $CP$-conserving under an order-2 GCP is that {\em there exists a real basis},
the basis in which all coefficients are real \cite{Gunion:2005ja}.
Relaxing the requirement of order-2 transformation invalidates this link:
there exist models with higher-order $CP$-symmetries but without real basis \cite{Ivanov:2015mwl},
though such models require at least three doublets, see section~\ref{section-NHDM-CPV}. 

In simple cases, the presence of a $CP$-symmetry can be easily detected.
But when building more sophisticated 2HDMs and checking their $CP$-properties, 
one encounters a serious technical challenge: 
even if the coefficients are complex, a hidden, non-obvious symmetry of type (\ref{GCP}) may still exist.
To check that, one needs to construct all $CP$-odd basis-invariants $I_i$ 
and verify that all $I_i = 0$.
They can be built as imaginary parts of various fully contracted products 
of tensors $Y_{ij}$ and $Z_{ijkl}$ defined in Eq.~(\ref{V-NHDM});
the challenge is to identify the full set of algebraically independent ones. 
In \cite{Branco:2005em,Davidson:2005cw,Gunion:2005ja},
such invariants were explicitly constructed for 2HDM scalar sector. In \cite{Gunion:2005ja}, they were labeled
as $I_{Y3Z}$, $I_{2Y2Z}$, $I_{3Y3Z}$, $I_{6Z}$, where the indices show the powers of $Y$ and $Z$ tensors
used in their construction. 
The 2HDM scalars sector is explicitly $CP$-conserving if and only if all of them are zero --- 
and this can now be checked in any basis. 
This formalism was recently extended in \cite{Varzielas:2016zjc} to NHDM as well as to more general sectors.
However, the problem of finding the minimal set of algebraically independent invariants ---
as consequently the task of writing down all necessary and sufficient conditions for explicit $CP$-conservation ---
for more than two doublets still remained open.

Back to 2HDM, proving within this formalism that all higher-order invariants 
are not independent anymore turned out to be a highly non-trivial task and required explicit check of 
millions of invariants with Mathematica \cite{Gunion:2005ja}.
A much shorter and more intuitive derivation of these invariant was achieved 
with the geometric bilinear formalism \cite{Ivanov:2005hg,Nishi:2006tg,Maniatis:2007vn}. 
Here, the basic objects are $M_\mu$ and $\Lambda_{\mu\nu}$ in (\ref{V-bilinears-2HDM}),
or more concretely their three-dimensional components: 
$M_a$ in (\ref{M-2HDM}) and $\Lambda_{ab}$ and $L_a \equiv \Lambda_{0a}$ in (\ref{Lambda-2HDM}).
Any basis change corresponds to a proper $SO(3)$ rotation in the space of bilinears,
and any GCP transformation corresponds to a rotary reflection.
$CP$-odd invariants must be constructed as triple products of certain vectors.
Introducing in addition to $L_a$ and $M_a$ 
the extra vectors $L^{(p)}_a = (\Lambda^p)_{ab} L_b$ and $M^{(p)}_a = (\Lambda^p)_{ab}M_b$
and denoting the triple products as $(A,B,C)  \equiv \epsilon_{abc} A_a B_b C_c$, 
one can construct the four $CP$-odd invariants
\be
{\cal I}_1 = (M, M^{(1)}, M^{(2)})\,, \quad 
{\cal I}_2 = (L, L^{(1)}, L^{(2)})\,, \quad 
{\cal I}_3 = (M, M^{(1)}, L^{(2)})\,, \quad 
{\cal I}_4 = (M, L^{(1)}, L^{(2)})\,.\label{triple-products}
\ee
Any GCP symmetry is equivalent to all four invariants ${\cal I}_i=0$.
Geometrically, it means that there exists an eigenvector of $\Lambda_{ab}$ orthogonal
to both $L_a$ and $M_a$.
The relation between these ${\cal I}_i$ and $I_{Y3Z}$, $I_{2Y2Z}$, $I_{3Y3Z}$, and $I_{6Z}$ of \cite{Gunion:2005ja} 
was given in \cite{Nishi:2006tg}.
We note also that $I_{6Z} \propto {\cal I}_2$ is special in the sense that 
it involves only quartic coupling constants. If it deviates from zero,
the model features hard $CP$-violation; if it vanishes but other invariants are non-zero,
the explicit CPV is soft.

Although the general 2HDM potential (\ref{V-2HDM}) contains four complex
parameters, one can always make a basis transformation to set $m_{12}^2 = 0$.
Thus, there are only {\em two} independent relative phases:
$\arg\lambda_6-\arg\lambda_7$ and $\arg\lambda_5-2\arg\lambda_6$ 
\cite{Davidson:2005cw}. In that basis, it would be sufficient to set these two relative phases to zero
to impose explicit $CP$-conservation. It is then surprising
that we needed four basis-invariant conditions for the same task.
The explanation is that in each particular case it is sufficient to use
a subset of these four but, depending on the case, this subset can be different.
For example, if $\lambda_6 + \lambda_7 = 0$ and $\lambda_1 = \lambda_2$,
which translates into $L_a = 0$, then ${\cal I}_2 = {\cal I}_3 = {\cal I}_4 = 0$
automatically, and one needs to check only ${\cal I}_1$. Similarly, when $M_a = 0$,
one needs only ${\cal I}_2$. 
If $L_a$ and $M_a$ are non-zero, 
it can happen than ${\cal I}_1 = {\cal I}_2 = 0$, that is, both vectors are orthogonal to an eigenvector 
of $\Lambda_{ab}$, but these are {\em distinct} eigenvectors.
In that case, one needs ${\cal I}_3$ and ${\cal I}_4$ to guarantee that this eigenvector is the same.
The task of identifying these special points in the parameter space
and disentangling different regimes of CPV was investigated in \cite{Grzadkowski:2013rza}.

So far, the discussion was on such GCP transformations which can be brought
to the ``standard'' form by a basis change. 
As we mentioned above, these correspond to Eq.~(\ref{GCP}) with $\alpha=0$
and were labeled in \cite{Ferreira:2009wh} as CP1.
In the bilinear formalism, they correspond to plane reflections. 
In the notation of \cite{Maniatis:2007de}, these are $\mathrm{CP}^{(ii)}_g$ transformations,
whose explicit form depends on the specific reflection plane choice.
Setting $\alpha = \pi/2$ in Eq.~(\ref{GCP}) leads to a GCP transformation of order 4,
labeled in \cite{Ferreira:2009wh} as CP2. 
In the bilinear space, this is a point reflection $r_a \to - r_a$. 
In the notation of \cite{Maniatis:2007de}, it is labeled as $\mathrm{CP}^{(i)}_g$ transformations.
The third class of GCP transformations,
labeled as CP3, corresponds to any other $\alpha$,
corresponding in the bilinear space to a plane reflection followed by a generic rotation.

CP2 and CP3 can also be imposed on the scalar potential, strongly restricting 
its free parameters, so that resulting model contains accidental symmetries.
Refs.~\cite{Ferreira:2009wh,Ferreira:2010yh} proved that these three classes of GCP symmetries exhaust the
GCP-based model building possibilities within the scalar sector of 2HDM 
and investigated their parameter ranges. 
A special feature of CP3 is that it is so restricting that, even when softly broken, 
it does not allow neither for explicit nor spontaneous CPV \cite{Ferreira:2010hy}.
CP2 can be extended to the Yukawa sector. 
Maniatis et al \cite{Maniatis:2007de} constructed a model with ``maximal CP symmetry''
in which the scalar and Yukawa sectors were required to be invariant under three orthogonal usual 
CP symmetries and CP2. These conditions produced a very non-trivial 
Yukawa interaction patterns and predicted a rather unique phenomenology with various
flavor-violating processes, which was explored in detail in 
\cite{Maniatis:2009vp,Maniatis:2009by,Brehmer:2012hh}. 
Finally, CP3 can be extended as well but only for the special value of $\alpha = \pi/3$ \cite{Ferreira:2010bm}, 
see details below.

\subsubsection{Spontaneous $CP$-violation}
For the ``standard'' definition of $CP$ symmetry (\ref{CP-standard}),
spontaneous CPV usually manifests itself via the presence of complex vevs.
The word ``usually'' signals that complex vevs do not always imply spontaneous CPV.
The simplest, almost trivial example is the SM itself, see (\ref{vev}). 
A more interesting example is the 2HDM potential (\ref{V-2HDM}) 
with the exact $\Z_2$-symmetry (\ref{Z2-2HDM}), which implies $m_{12}^2 = 0$ and $\lambda_6=\lambda_7=0$.
The only complex parameter $\lambda_5$ can be set real and positive by global rephasing of one of the doublets,
so that the ``standard'' $CP$-symmetry is present. If $v_1 v_2 \not = 0$, then
the minimum corresponds to the vev alignment
with the relative phase $\pi/2$: $\lr{\phi_1^0} = v_1/\sqrt{2}$, $\lr{\phi_2^0} = \pm i v_2/\sqrt{2}$,
which is clearly non-invariant under the ``standard'' $CP$.
Nevertheless, the model is not $CP$-violating \cite{Branco:1980sz}.
One way to see it is to remark that the potential contains yet {\em another} $CP$-symmetry,
the GCP with $X = \mathrm{diag}(1, -1)$, which is left unbroken. 
The $\Z_2$-symmetric model is ``double-protected'' against explicit CPV,
and minimization removes only one of the two symmetries.
Alternatively, by performing the basis change $\phi_2 \to i\phi_2$,
one obtains the same potential with $\lambda_5 < 0$ and, consequently, with the vev alignment $(v_1, \pm v_2)$.
The two $CP$-symmetries swap their appearance:
the ``standard'' one becomes GCP, and the GCP is now standard.
In general, the impossibility to break both $CP$-symmetries
simultaneously follows from the fact that the tree-level potential in 2HDM
can have at most two minima \cite{Ivanov:2007de}.

In the most general 2HDM, if the explicit $CP$-conservation is verified with the help of basis invariants,
one can ask if it is possible to detect whether spontaneous CPV happens
without having to actually find the minimum.
This question was answered in the geometric bilinear approach \cite{Ivanov:2006yq}. 
These conditions are especially simple in the $\Lambda_{\mu\nu}$-diagonal basis of 
Eq.~(\ref{V-bilinears-2HDM}): spontaneous CPV takes place if the vector $M_a$ lies inside
an ellipse whose semiaxes depend in a known and simple way on the eigenvalues of $\Lambda_{\mu\nu}$.

In a more phenomenology-oriented analysis, we first select vevs and then build the 2HDM potential around it.
One may ask for a basis-invariant criterion to detect the presence of $CP$-violation in the scalar sector
without caring if $CP$-violation is explicit or spontaneous.
A natural path to such conditions is to switch to the Higgs basis
and check whether the three quantities
\be
J_1 \propto \Im (\bar \lambda_5^* \bar \lambda_6^2)\,, \quad 
J_2 \propto \Im (\bar \lambda_5^* \bar \lambda_7^2)\,, \quad 
J_3 \propto \Im (\bar \lambda_6 \bar \lambda_7^*)\,, 
\ee
are zero or not \cite{Mendez:1991gp,Lavoura:1994fv}. The parameters $\bar \lambda_5$, $\bar \lambda_6$, $\bar \lambda_7$
refer to the Higgs basis, which makes these three $J_i$ basis invariant quantities.
Again, in each particular case, two invariants out of three suffice, but the selections of this pair depends
on the case, therefore all three are required.
The technical task of actually constructing $J_i$ in terms of tensors $Y_{ij}$ and $Z_{ijkl}$
as well as complex $v_i$ was solved in \cite{Botella:1994cs}. If fact, $Y$'s can be eliminated via
extremization condition, giving these invariants in terms of vevs and the quartic tensor $Z$'s \cite{Gunion:2005ja}.
The same conditions can also be derived in the bilinear approach \cite{Nishi:2006tg,Maniatis:2007vn}.
Determining invariants $J_i$ via physical observables and disentangling
forms of CPV in the general 2HDM, in the softly broken $\Z_2$, and in the alignment limit 
was studied in \cite{Grzadkowski:2014ada,Grzadkowski:2016szj}.
The latter is also a useful resource of various relations among parameters, physical observables, 
and $CP$-sensitive invariants.

A remarkable example of CPV arising from cooperation between two sectors 
was found by Ferreira and Silva in \cite{Ferreira:2010bm}.
They started with a GCP symmetry of type CP3 parametrized with angle $\alpha$ (\ref{block})
and extended it to the fermion sector. In all but one cases it led to unphysical quark sector.
The exceptional case corresponds to $\alpha = \pi/3$; 
it fixes uniquely the transformation properties of the fermions and their Yukawa matrices
and allows for a good fit to the CKM matrix.
The scalar potential has a continuous symmetry as the result of CP3, 
and in order to prevent a massless scalar, a real soft breaking term was introduced (thus, no explicit CPV is present).
The vevs get a relative phase but there is still no spontaneous CPV within scalar sector, 
as all $J_i = 0$. In the Yukawa sector, all coefficients are real, thus CPV is not put by hand.
Yet, the Jarlskog invariant $I_{CP}$ (\ref{ICP}) is non-zero, which indicates CPV.

\subsection{Testing 2HDM}

2HDM can be experimentally checked in several ways.
One can directly search for phenomena which definitely go beyond SM, such as direct
production of additional scalars or observation of the 125 GeV Higgs decay channels impossible in SM.
One can also measure the rates of SM-allowed processes and see them differ in a way favored by 2HDM.
Or at least, if all checks give negative results, one can limit the 2HDM parameter space
or even close certain versions of the model.

Before going into details,
let us make several remarks.
First, although the literature on 2HDM phenomenology is vast, 
the majority of phenomenological studies are limited to simple versions of 2HDM,
especially to softly broken $\Z_2$ models with or without $CP$-violation and often accompanied by NFC in the Yukawa sector.
This is a reasonable approach given the lack of definitive experimental indications 
in support of non-minimal Higgs sectors.
Trying to experimentally check if we are dealing with the 2HDM that strongly differs from
these simple versions would require access to triple and quartic
Higgs couplings \cite{Grzadkowski:2014ada,Ginzburg:2015cxa,Grzadkowski:2016szj} 
and can certainly be postponed.
Second, the 2HDM has a decoupling limit with heavy extra Higgses \cite{Ginzburg:2001ss,Gunion:2002zf}.
If accompanied with Type I Yukawa sector model, it produces a very SM-like picture,
and it will be extremely difficult to reveal its non-SM properties. 
Thus, it is impossible to completely rule out 2HDM without first ruling out the SM itself.
Third, a good feature of 2HDM is that the same interactions can affect the observed Higgs
properties and the signals involving new scalars. 
The measurement of the observed Higgs couplings 
and direct searches for additional scalars are complementary and correlated probes of the model \cite{Craig:2013hca}.
Together with astroparticle features that the 2HDMs possess, one can speak of a ``multi-messenger approach''
to testing 2HDM and other similar sectors.

The pre-LHC expectations on the production and decay channels
of neutral and charged Higgs bosons focused mostly on Type I and Type II 
Yukawa sectors and especially on the MSSM Higgs sector. 
They are extensively covered in the Higgs Hunter's Guide \cite{Gunion:1989we},
by Djouadi in the MSSM part of his review \cite{Djouadi:2005gj}, and in the recent review on 2HDM \cite{2HDM-review}.

Despite a large variety of 2HDMs, some experimental bounds are rather generic. 
New Higgs bosons participate in gauge interactions, and the vertices such as $\gamma H^+H^-$ and $ZHA$ 
are proportional to gauge coupling constants.
As a result, the charged scalars cannot be too light, otherwise one would copiously produce them at colliders.
The recent combined LEP result \cite{Abbiendi:2013hk} excluded
charged Higgs bosons lighter than 80 GeV (Type II) and 72.5 GeV (Type I).
Within Type II, the flavor observables and in particular $b\to s\gamma$ place a much stronger 
lower limit on the charged Higgs mass $m_{H^\pm} > 480$ GeV at 95\% C.L. \cite{Misiak:2015xwa}.
A light pseudoscalar Higgs is possible \cite{Bernon:2014nxa} if it is not accompanied by a light second $CP$-even Higgs $h$,
because otherwise $Z\to hA$ decay would have been detected.
Curiously, it is still possible for $m_h + m_A$ to be slightly below $m_Z$ in Type I 2HDM without violating
all experimental constraints \cite{Enberg:2016ygw}; this regime can be soon closed by LHC Run 2 data.
The values of $\tan\beta$ can also be limited on general grounds, though these limits depend on the Yukawa sector.
For Type I, $\lr{\phi_2} = 0$ meaning $\tan\beta = 0$ is still viable; the Inert doublet model
featuring a DM candidate which is described in section~\ref{section-IDM} is of this kind.
For other models including the popular Type II, one typically starts with $\tan\beta \gsim 0.3$, 
which comes from the requirement that the top-quark coupling be not too large,
and goes up to $\tan\beta \sim 100$ in order to preserve the $b$-quark Yukawa coupling 
going non-perturbative at a rather low energy scale.

\begin{figure}[!htb]
   \centering
\includegraphics[width=0.45\textwidth]{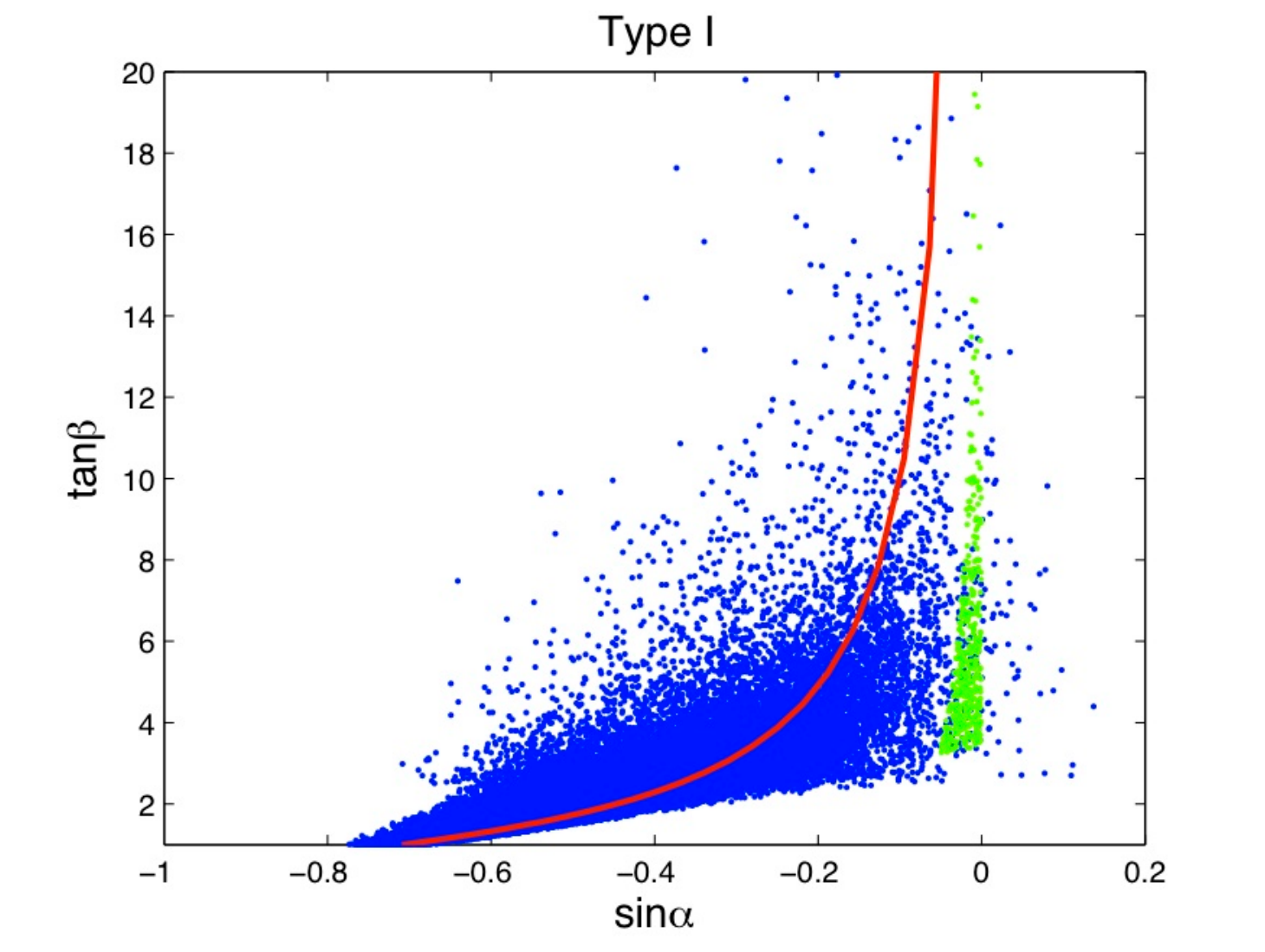}
\includegraphics[width=0.45\textwidth]{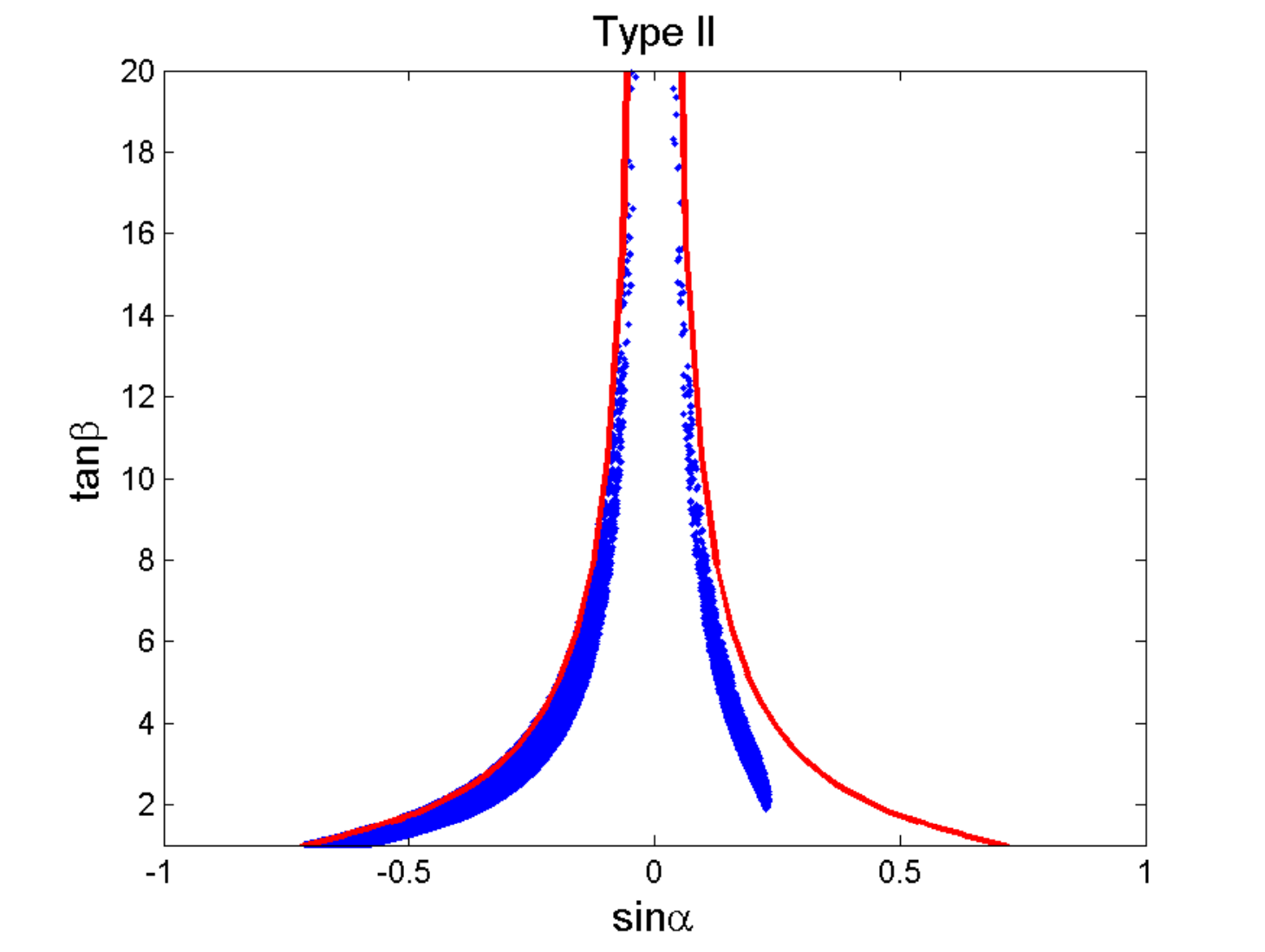}
\caption{Points in the ($\sin \alpha$, $\tan \beta$) plane in Type I (left) and Type II (right) 2HDMs 
which passed all the constraints as of early 2013 at 1$\sigma$ in green (light grey) and 2$\sigma$ in blue (dark grey). 
Also shown are the lines for the SM-like limit, that is $\sin(\beta - \alpha) =1$ (negative $\sin \alpha$), and for the limit
$\sin(\beta + \alpha) =1$ (positive $\sin \alpha$). Reproduced with permission from \cite{Barroso:2013zxa}.}
   \label{fig-2HDM-cabtb}
\end{figure}

As the ATLAS and CMS collaborations were finalizing their Run 1 Higgs results,
many analyzes started to appear focusing on heavy Higgs searches and the resulting constraints 
on the 2HDM parameter space 
\cite{Barroso:2013zxa,Eberhardt:2013uba,Dawson:2013bba,Celis:2013rcs,Chen:2013rba,Craig:2013hca,Celis:2013ixa,%
Coleppa:2013dya,Dumont:2014wha,Bernon:2015qea,Bernon:2014nxa,Bernon:2015wef,Han:2017pfo}.
Some studies tested the LHC Higgs data against not only 2HDM but also other examples of extended Higgs sectors
\cite{Belanger:2013xza,Lopez-Val:2013yba,Kanemura:2014bqa}.

As an illustration of the early analyses, Fig.~\ref{fig-2HDM-cabtb} 
shows one of the first results of the $CP$-conserving 2HDM parameter space scan 
for Yukawa models Type I (left) and Type II (right) \cite{Barroso:2013zxa}. 
The points here represent models from a wide parameter scan
which passed all the experimental constraints available at the beginning of 2013 at $1\sigma$ and $2\sigma$ levels.
We remind that in early 2013, the LHC data showed a $\sim 2\sigma$ excess in $H \to \gamma\gamma$
decay channel with respect to the SM \cite{Arbey:2012bp}, which was also at odds with many 2HDM models. 
It is this $H \to \gamma\gamma$ excess that explains why all the green points in Fig.~\ref{fig-2HDM-cabtb} 
fall on the right of the solid red line describing the SM limit. 
When this excess evaporated in the later data, the parameter space scans yielded broader regions compatible with experiment.

In later publications, Craig et al \cite{Craig:2013hca} present a detailed analysis 
of numerous suitable heavy Higgs production and decay processes.
Coleppa et al \cite{Coleppa:2013dya} focus on the parameter space of Type II 2HDM surviving 
after the LHC Run 1 results. 
One of the most detailed analyses of 2HDM Type I and II parameter space 
was given by Dumont et al \cite{Dumont:2014wha}, together with implications for future measurements.
Bernon et al published an extensive two-part analysis of the 2HDM Type I and II in the vicinity of the alignment limit,
corresponding to the 125 GeV Higgs being the lighter $h$ \cite{Bernon:2015qea} 
or the heavier $H$ \cite{Bernon:2015wef} $CP$-even scalar.
In a related work \cite{Bernon:2014nxa} they studied the case of the light second Higgs: $m_h < 62$ GeV, without assuming alignment.
A very recent review of experimental results on light additional bosons can be found in \cite{Aggleton:2016tdd}.

\begin{figure}[!htb]
   \centering
\includegraphics[width=0.49\textwidth]{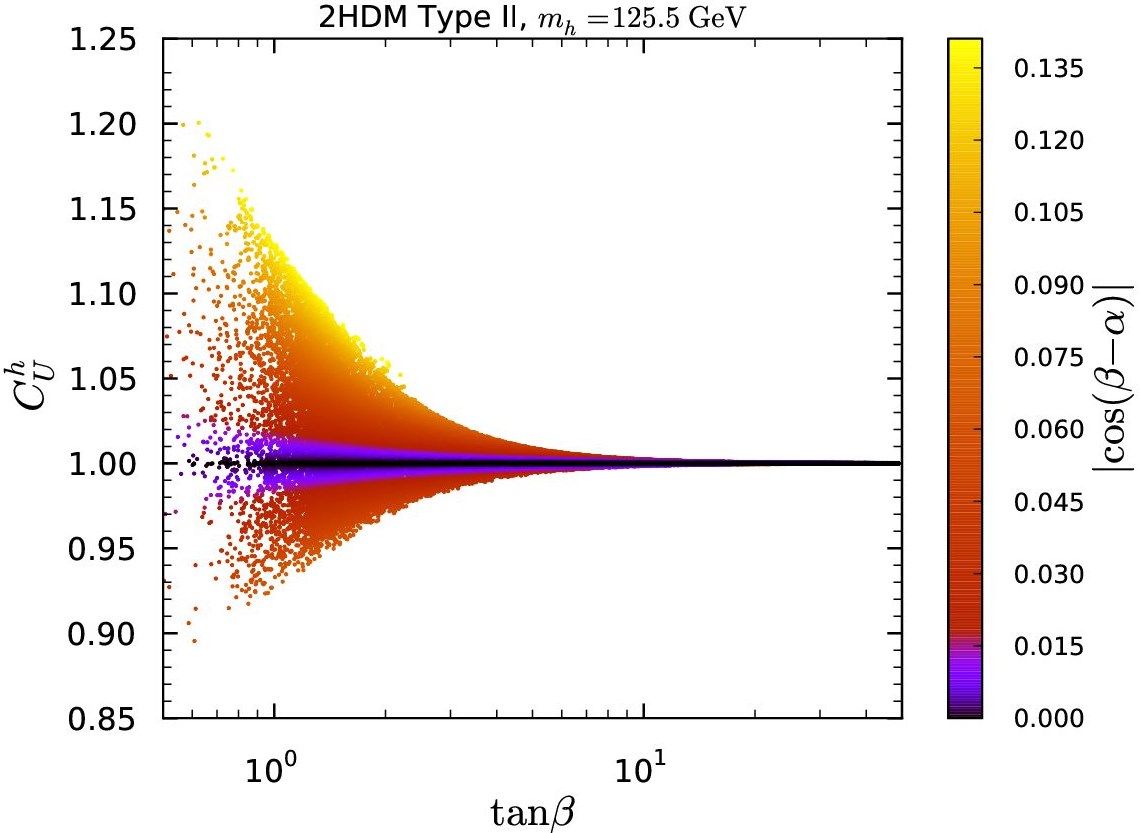}\hfill
\includegraphics[width=0.49\textwidth]{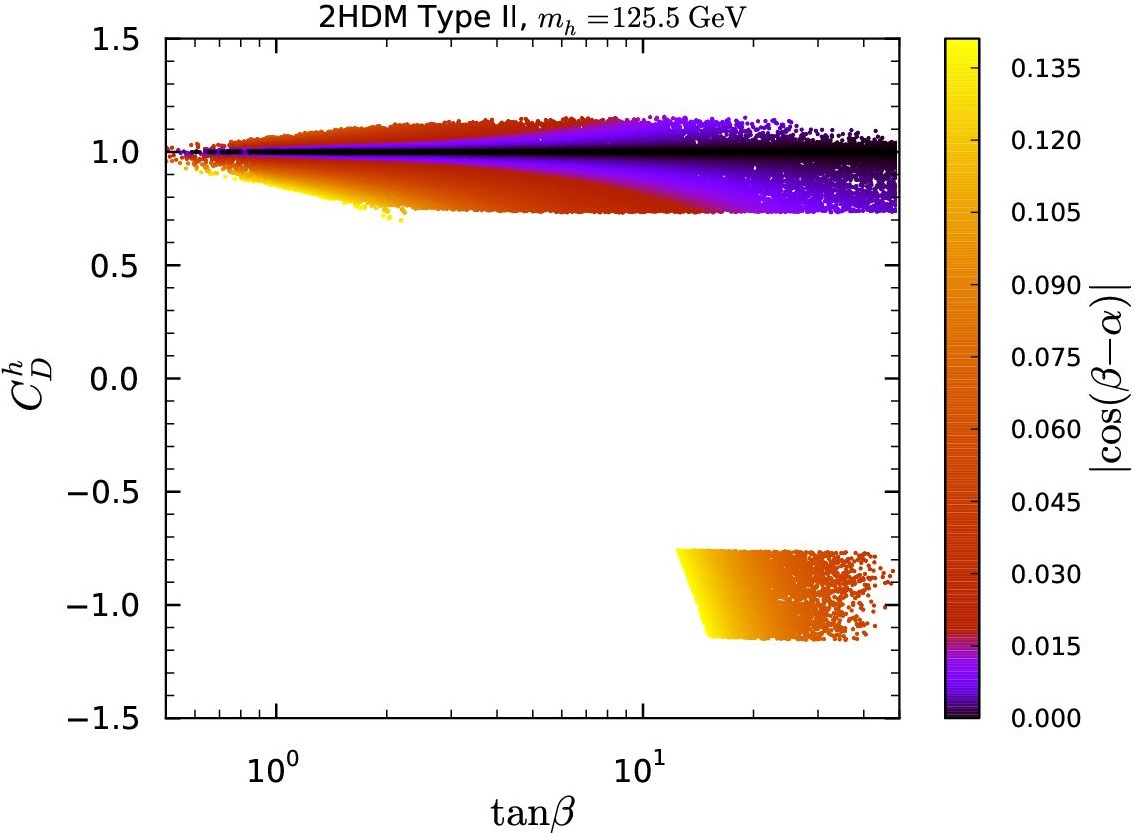}
\caption{Up-quark (left) and down-quark (right) couplings relative to the SM 
versus $\tan\beta$ in Type II 2HDM with $h_{125}$ identified with the lighter $CP$-even Higgs. 
The color encodes the values of misalignment quantity $|\cos(\beta-\alpha)|$.
Reproduced with permission from \cite{Bernon:2015qea}.}
   \label{fig-2HDM-fermioncouplings}
\end{figure}

One curious possibility of 2HDM is that Yukawa coupling of the SM-like Higgs 
can have right magnitude but ``wrong'' sign with respect
to the SM expectations \cite{Ginzburg:2001ss}. 
It turns out that among various options, only down-quark Yukawa coupling of $h_{125}$
could have wrong-sign solutions for $\tan\beta > 1$ and only in Type II and Flipped 2HDM 
\cite{Celis:2013ixa,Ferreira:2014naa,Ferreira:2014dya,Bernon:2015qea}, see Fig.~\ref{fig-2HDM-fermioncouplings}.
This wrong sign limit corresponds to $\sin(\beta+\alpha)$
rather than $\sin(\beta-\alpha)$ close to 1 (see Fig.~\ref{fig-2HDM-cabtb}, right), 
so that the model is not in the decoupling regime 
and is strongly constrained by the data. Still, there are regions of the parameter space which
survive these constraints. Trying to close this option via $gg \to h$ or $h\to \gamma\gamma$ processes
would be difficult as the $b$-quark loops are subdominant.
However, as Modak et al showed \cite{Modak:2016cdm}, 
once the exclusive decay $h \to \Upsilon\gamma$ is measured, it will be able to immediately probe
the wrong-sign solution.

\begin{figure}[!htb]
   \centering
\includegraphics[height=5cm]{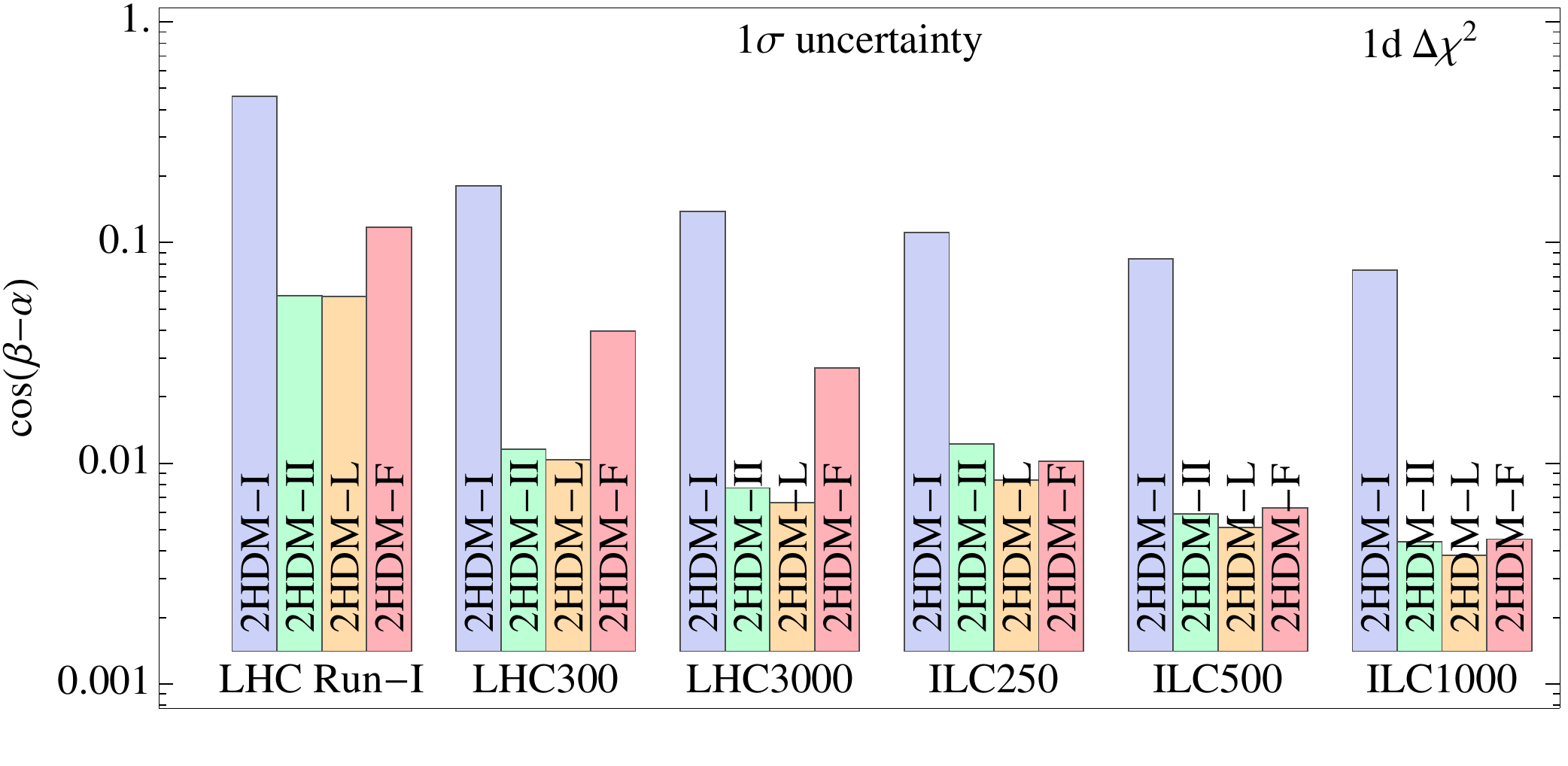}
\caption{The largest $\cos(\beta-\alpha)$ which pass machine benchmarks for different 2HDM Yukawa sectors with NFC
and at different colliders: LHC Run-1, 300 fb$^{-1}$, 3 ab$^{-1}$, as well as several options for ILC. 
Reproduced with permission from \cite{Barger:2013ofa}.}
   \label{fig-2HDM-cab}
\end{figure}

Anticipating future experiments, Barger et al \cite{Barger:2013ofa} developed strategies of experimental investigation of 2HDM
through the 125 GeV Higgs at future colliders, see Fig.~\ref{fig-2HDM-cab}.
Kanemura et al \cite{Kanemura:2014bqa} presented a systematic analysis of the properties
of all heavy Higgses for all four types of flavor conserving Yukawa sectors. 
Expected sensitivity limits for signal strengths were calculated for the LHC and future colliders.
The recent Snowmass report of the Higgs working group \cite{Dawson:2013bba} also summarizes 
future collider expectations for the Higgs physics, including 2HDM.

Several works focused specifically on ways to detect and measure $CP$-violation in 2HDM.
The key objects proposed included $ZZZ$ and $ZH^+H^-$ vertices \cite{Grzadkowski:2014ada,Grzadkowski:2016lpv}, 
$h \to Z\gamma$ decay \cite{Fontes:2014xva}, difference between $H^+$ and $H^-$
production and decays \cite{Arhrib:2010ju}, 
as well as the patterns of $h_i \to h_j Z$ and $h_i \to ZZ$ decays \cite{Fontes:2015xva}.
Specifically, Fontes et al \cite{Fontes:2015xva} presented five combinations of decays
whose simultaneous detection will be an ``undoubtable'' sign of $CP$-violation 
in the 2HDM scalar sector, such as $h_3 \to h_2 Z$, $h_2 \to h_1Z$, and $h_3\to h_1Z$. 

Pair production of the SM-like Higgs has also been proposed as a means of detecting the presence
of extra scalars. For example, in \cite{Baglio:2014nea,Hespel:2014sla}, this process was
analyzed for several benchmark scenarios in 2HDM, with the conclusion that the resonance production
of heavy Higgs can boost $hh$ production by up to two orders of magnitude above the SM expectations.
For non-resonant production, the effect is significantly more moderate and the rate can even be below the SM.
Subtle effects specific to Yukawa types can be seen in differential distributions.

Quite a number of publications \cite{Celis:2012dk,Crivellin:2012ye,Cheng:2014ova,Broggio:2014mna,Li:2014fea,%
Crivellin:2015hha,Crivellin:2015mga,Benbrik:2015evd,Enomoto:2015wbn}
were devoted to possible 2HDM explanations of several flavor physics
discrepancies, especially those involving $B$-decays.
In particular, Crivellin et al showed in \cite{Crivellin:2015mga} that it is well possible
to accommodate $B \to K^*\mu\mu$, $R(K) = Br(B\to K\mu\mu)/Br(B\to K ee)$,
and $h \to \tau\mu$ anomalies within a 2HDM with the gauged $L_\mu - L_\tau$ symmetry.
Another departure from the SM, the lack of leptonic universality in 
$R(D^{(*)}) = Br(B \to D^{(*)}\tau\nu)/Br(B \to D^{(*)}\mu\nu)$,
was interpreted in \cite{Celis:2012dk,Crivellin:2012ye} as the charged-Higgs induced enhancement
of $\tau\nu$ production. Although Type II 2HDM cannot accommodate both $R(D)$
and $R(D^*)$ simultaneously, a less constrained Yukawa sector, for example, the aligned 2HDM,
is capable of doing that without much fine-tuning \cite{Enomoto:2015wbn}.

In 2015, CMS reported a hint of the lepton-flavor-violating Higgs decay $h \to \tau\mu$ 
at about 1\% level \cite{Khachatryan:2015kon}. 
The latest ATLAS results on this decay \cite{Aad:2016blu} do not support this deviation from SM, 
although they do not rule out the CMS hint either.
This decay, if it is real, can be well accommodated within 2HDM without violating bounds on other LFV processes, 
but of course in the Yukawa sector beyond NFC \cite{Sierra:2014nqa}. 
One aspect here is that 
it is exceedingly difficult to organize for such a decay within MSSM. If future data confirms
$h \to \tau\mu$ decay at the present level, it will basically rule out MSSM \cite{Aloni:2015wvn}.
It is also interesting that, when describing this excess within 2HDM, one can obtain models
which predict much stronger lepton-flavor-violating signals for the {\em heavier} Higgs decays $H/A \to \tau\mu$
\cite{Buschmann:2016uzg}, with branching ratios which can go as high as 60\% \cite{Sher:2016rhh}.

The magnitude of electric dipole moments achievable within 2HDM was analyzed in \cite{Jung:2013hka,Ipek:2013iba}.
The neutrinophilic 2HDM, a model in which the second doublet gets vev $v_2 \sim 1$ eV and gives neutrinos their masses \cite{Gabriel:2006ns,Davidson:2009ha}, was tested in \cite{Machado:2015sha,Bertuzzo:2015ada} 
against the LHC and lepton physics data.


\section{More than two doublets}\label{section-NHDM}

\subsection{Two examples of NHDM flavor models}\label{section-NHDM-examples}

There are hundreds of publications on models with more than two Higgs doublets,
often accompanied with other fields. Instead of grazing superficially over many of them,
we will go into some detail with a few historical examples from late 1970's. 
These examples of three-Higgs-doublet models (3HDM)
nicely illustrate new possibilities and new challenges which come into play with more than two Higgs doublets.
The driving idea behind them was to find global symmetries 
which would shape the Yukawa sector just in the right way. Namely, one expects the global symmetries 
to strongly reduce the number of free parameters, to naturally lead to the patterns in masses
and mixing angles we observe, and hopefully to provide symmetry-imposed relations
among masses and mixing, as well as to lead to new beyond the SM signals.
We will close this subsection with a general theorem
which severely restricts the use of large global symmetry groups for flavor model-building within NHDM.

\subsubsection{3HDMs with permutation symmetries}

Building a symmetry-based NHDM flavor sector starts with a choice of the global symmetry group $G$ 
and an assignment of fermions and Higgs doublets to irreducible representations of $G$.
For an introduction to finite groups and formulas useful for model building, 
see for example reviews \cite{Altarelli:2010gt,Ishimori:2010au}.

One natural choice for $G$ is the permutation symmetry group of three elements $S_3$,
which was first considered by Derman \cite{Derman:1977wq,Derman:1978nz,Derman:1978rx}.
The irreducible representations of $S_3$ are singlets and doublets, not triplets.
If the three left-handed fermion electroweak doublets, three right-handed singlets, and three Higgs generations
are all assumed to be in the singlet plus doublet $1\oplus 2$ of $S_3$,
then the Yukawa sector is represented as the product 
$(1+2)_L\otimes(1+2)_R\otimes(1+2)_\phi$. 
This product contains five singlets, each coming with its own free parameter. 
The parameters across the sectors --- up-quarks, down-quarks, and leptons --- do not need to be correlated.
The fermion mass matrices are obtained by multiplying Yukawa matrices with vevs $v_i$
as described in section~\ref{section-NHDM-generic-features}.
The question is which vev alignment one is allowed to choose.
Building the $S_3$-symmetry potential and working in the region of free parameters 
in which there is no $CP$-violation, neither explicit nor spontaneous, Derman
found that only two vev alignments were possible: $(v_1,\, v_1,\, v_1)$
and $(v_1,\, v_1,\, v_2)$.
In the former case, the $S_3$ symmetry is conserved, in the latter case it is spontaneously broken
to the $\Z_2$ subgroup generated by the family permutation.
The former option leads to two fermions being mass-degenerate,
which is clearly ruled out experimentally,
while the latter allows the masses to be all distinct.
However the presence of a residual symmetry strongly restricts the mixing among the quark generations.
The quark mixing matrix contains only one mixing angle $\theta_C$, whose value 
cannot be predicted from quark mass ratios. 
The just-discovered $b$-quark, although not mixing with other quarks in this model, 
was nevertheless predicted to be metastable due to the semileptonic decay 
$b \to d e^+\mu^-$ mediated by the $\Z_2$-odd Higgs boson. 
Of course, this remarkable prediction was quickly checked and ruled out.
Pakvasa and Sugawara \cite{Pakvasa:1977in} used the same permutation symmetry group $S_3$
in a somewhat different setting and managed to obtain the Cabibbo angle 
$\tan\theta_C = m_d/m_s$.
At that time the current masses of light quarks were poorly known;
they were compatible with $m_d/m_s \sim 1/5$, which roughly agreed with the Cabibbo angle.

One can go further with permutation groups and unite three generations in one irreducible representation
of $S_4$. The Yukawa sector becomes more economic.
In one example of extreme simplicity, one can assume that only one group singlet is present in each sector,
$1 = 3_L\otimes 3_R\otimes 3_\phi$,
making all three Yukawa sectors governed by a single coupling constant.
In this case, one fermion in each sector, $u$, $d$, and $e$, remains massless,
so that the origin of their masses is delegated to yet another complementary mechanism.
This was seen not as a bug but as a feature, an appealing argument in favor of one fermion generation being very light.
In this approximation one gets $M_\ell \propto M_d \propto M_u^*$,
which implies $m_\mu/m_\tau = m_c/m_t = m_s/m_b$ and predicts the top-quark mass
to be around $m_t \sim 20$--$30$ GeV.
Later, the negative results of top quark searches in this mass range discouraged the simplistic intuition
of the sequential family mass ordering.

Works \cite{Pakvasa:1978tx,Yamanaka:1981pa,Brown:1984mq} 
put more sophistication to this $S_4$ construction by assigning $Q_L,\, L,\, \phi \sim\, 3$ 
and $\ell_R,\, u_R,\, d_R \sim \, 1\oplus 2$.
The $S_4$-symmetric 3HDM scalar potential has only five free parameters, its minimum can be found
explicitly. In a certain range of the parameters, the vev alignment becomes
\be
v_i \propto (e^{i \xi},\, e^{-i\xi},\, r)\,,\label{S4-minimum}
\ee
with the phase $\xi$ calculated via the parameters of the potential.
For Yukawas, one gets only two singlet terms per sector, coming from the $S_4$ tensor decomposition 
$3_L \otimes 3_\phi = 3+3'+2+1$ coupled to $(1+2)_R$.
These two singlet combinations come with free parameters $f$ and $g$, different for each sector.
Taking into account that one fermion generation is much lighter than the other two,
one assumes $f_i / g_i \equiv \epsilon_i \ll 1$ for $i = \ell, u, d$.
The mass matrices become 
\be
M_\ell = {g_\ell v_1 \over\sqrt{2}} \mmmatrix{\epsilon_\ell e^{i\xi}}{e^{-i\xi}}{e^{i\xi}/\sqrt{3}}
{\epsilon_\ell e^{-i\xi}}{e^{i\xi}}{e^{-i\xi}/\sqrt{3}}
{r \epsilon_\ell}{0}{-2r/\sqrt{3}}\,,\label{yamanaka-Ml}
\ee
with $M_d$ and $M_u^*$ being of the same form with their own $\epsilon$'s.
The parameters $r$ and $\xi$ are universal across the Yukawa sector, while $\epsilon$'s are different.
It is only due to $\epsilon_u \not = \epsilon_d$ that quarks have a non-trivial CKM matrix.

The small parameter simplifies the analysis allowing one to investigate mass matrices 
in the first non-trivial order of $\epsilon$.
Since one has eight free parameters and nine masses,
there must exist one relation among them. In this approximation, it has the form
\be
\left|\! \begin{array}{ccc} m_e^2 & m_\mu^2 & m_\tau^2\\[1mm]
m_u^2 & m_c^2 & m_t^2\\[1mm]
m_d^2 & m_s^2 & m_b^2\\ \end{array}\!\right|=0\,.\label{yamanaka-det}
\ee
This equation can be used to predict the mass of the top quark between $25$ and 35 GeV 
\cite{Pakvasa:1978tx,Brown:1984mq}.
Next, one can express the vev ratio $r$ and phase $\xi$
fully in terms of masses, and, as a consequence, reconstruct the {\em entire} CKM matrix 
in terms of quark and lepton masses.
For the Cabibbo angles one obtains
\be
\sin\theta_C \approx {m_d \over m_s}\left(1 - {m_\mu^2 m_b^2 \over m_\tau^2 m_s^2}\right)^{-1/2}\,.
\ee
Due to strong sensitivity to quark masses, it was not possible
to accurately estimate $\theta_C$ from this relation, but at least
it did not contradict the experimental value $\tan\theta_C \approx 0.22$.
Similarly, one can express other mixing angles as well as the $CP$-violating phase.
The elements of the predicted CKM were perfectly compatible with the data available in the late 70's,
and even if compared with the present day measurements,
they are off by at most the factor of 2. 
The amount of $CP$-violation turned out to be too small in this model, though.

\subsubsection{3HDM with $\Delta(54)$ symmetry}

A somewhat different way of using global discrete symmetries 
to link quark mixing and masses was suggested in 1978 by Serge, Weldon, and Weyers \cite{Segre:1978ji}. 
They start with a natural idea that the very different fermion masses across generations
may arise not from very different Yukawa couplings, but from highly hierarchical vev alignment $v_1 \ll v_2 \ll v_3$
of the three Higgs doublets.
The simplest implementation of this idea is when each Higgs doublet couples only to one fermion generation
with a universal coupling,
\be
f \sum_i \bar Q_{Li} d_{R i} \phi_i + f' \sum_i \bar Q_{Li} u_{R i} \tilde \phi_i\,.\label{serge-dominant}
\ee
Unfortunately, as it stands, this structure leads to unphysical CKM matrix. One needs to add non-diagonal terms, 
still assuming that (\ref{serge-dominant}) are dominant. 
A very economic way is to impose a discrete symmetry group with permutations
and discrete rephasing: 
\be
Q_{Lk} \mapsto \omega^{-k} Q_{Lk}\,, \quad \phi_{k} \mapsto \omega^k \phi_{k}\,, \quad 
d_{Rk} \mapsto \omega^k d_{Rk}\,, \quad u_{Rk} \mapsto u_{Rk}\,, 
\ee
where $\omega^3 = 1$. This rather minimalistic choice leads to a predictive model.
The full symmetry group generated by these transformations and arbitrary permutations 
is known as $\Delta(54)$. The quark Yukawa sector compatible with it is defined by 
\be
\Gamma_1 = \mmmatrix{f}{0}{0}{0}{0}{g}{0}{g}{0}\,, \quad
\Delta_1 = \mmmatrix{f'}{g'}{g'}{0}{0}{0}{0}{0}{0}\,,  
\ee
and $\Gamma_2, \Gamma_3$ and $\Delta_2, \Delta_3$ obtained from those by cyclic permutations.
The resulting mass matrices $M_d$ and $M_u$, despite their rather simple form,
are quite different, and it is this mismatch that drives the quark mixing pattern.

The challenge comes, however, from the scalar sector.
It is possible to write down and explicitly minimize the $\Delta(54)$-symmetric renormalizable 3HDM potential.
The problem is that it does not generate the desired vev alignment $v_1 \ll v_2 \ll v_3$ for any choice of parameters.
Colloquially, highly symmetric potentials refuse to completely lose their symmetry upon minimization.
The authors of \cite{Segre:1978ji} admitted that unpleasant fact, and 
without providing further detail, assumed that appropriate soft symmetry breaking terms 
could solve this problem.
This has become the standard philosophy behind vev engineering:
if the exactly symmetric potential cannot provide the desired vacuum, one breaks it softly.

Nevertheless, this rather simple setting allows for a number of phenomenological predictions.
First, determining vevs via lepton mass ratios and substituting them for quarks,
one arrives again at an approximate relation $m_\mu/m_\tau \approx m_s/m_b \approx m_c/m_t$.
Using the $s$-quark mass $m_s \approx 270$ MeV from the current algebra evaluations,
Serge et al obtained $m_b \approx 4.7$ GeV, close to the experimental observation.
For the top-quark, the prediction was $m_t \approx 20$ GeV, similarly to the previous models.

The CKM matrix is also reconstructed in terms of mass ratios.
In particular, the Cabibbo angle becomes 
$\theta_C \approx \sqrt{m_d/m_s}$, quite distinct from models based on permutation symmetries,
but similar to the left-right symmetric and texture-based predictions \cite{Fritzsch:1977za}.
Another prediction was that the $b$-quark decays 
predominantly into $c$ with the lifetime below $10^{-11}$ s.
As we now know, this upper bound is correct,
but at that time it was still four orders of magnitude below then existing experimental bounds.
The amount of $CP$-violation could not be cleanly predicted within this model,
as it relied in the relative phase of the two largest vevs, 
but it was noted that no additional smallness was needed. 

\subsubsection{Quark sector against residual symmetries}

The above examples teach us two important lessons.
First, residual global symmetries can easily lead to non-physical quark sector,
for example, by predicting degenerate quarks or by generating too few mixing angles.
This was formulated and proved as a general theorem in \cite{Leurer:1992wg},
and recently it was further refined in \cite{Felipe:2014zka}
by including the overlooked role of the residual symmetries in the right-handed quark space.
The final statement of the theorem reads: given a group $G$ acting on the flavor space, 
the only way to obtain a non-block-diagonal CKM mixing matrix and, simultaneously, non-degenerate
and non-zero quark masses, is that the vev alignment 
breaks the group $G$ {\em completely}, except possibly for some symmetry
belonging to the baryon number or a symmetry located purely within the inert sector 
(i.e. for the doublets which do not couple to quarks).

Thus, when imposing a discrete symmetry group on Higgs and fermion fields,
one should make sure that this symmetry is completely broken. However, it is not easy to force
the symmetric potentials to give up all of their symmetries upon minimization.
In all the 3HDM examples listed above, $CP$-conserving $S_3$, $S_4$, or $\Delta(54)$-symmetric potentials,
any neutral minimum for any choice of parameters contained a residual symmetry,
which would immediately render the quark sector non-physical.

An excellent and pedagogical illustration of what exactly goes wrong with $S_4$ and $A_4$ 3HDMs,
at each possible minimum and for each allowed representation assignment,
was given by González Felipe, Serôdio, and Silva in \cite{Felipe:2013ie}. 
They checked the full list of representation assignments and constructed the Yukawa sectors
for each combination.
As was found in \cite{Degee:2012sk}, the 3HDM renormalizable
potential with the exact $A_4$ or $S_4$ symmetry can have only four vev alignment patterns at the global minimum.
Remarkably, the vev alignment described in (\ref{S4-minimum}) was {\em not} among them: this minimum can only be local, not global.
Merging these vev alignments with Yukawa sectors, the authors of \cite{Felipe:2013ie} obtained mass matrices,
and deduced from them the quark masses and the CKM matrix. 
In all cases, there was at least one quark-sector parameter that went non-physical. 
The case closest to reality was the model 
with the representation assignment $Q_L \sim (1,1',1'')$, the three singlets of $A_4$, with
the Higgs doublets and the right-handed quarks transforming as triplets, and with vev alignment $(1, e^{i\alpha}, 0)$.
The residual symmetry is the one which flips the sign of the third Higgs doublet and third right-handed quarks.
This produces, after the up- and down-quark mass matrix diagonalisation, three mixing angles
and three non-degenerate quark masses, but with one massless quark in each sector.

All in all, the exactly symmetric situations are often overly restrictive.
One can now ask: which discrete symmetry groups can be completely broken upon minimization
of the potential and which cannot?
In the case of 3HDM, it was answered by algebraic {\em tour de force} in \cite{Degee:2012sk,Ivanov:2014doa}, 
see the next subsection; beyond 3HDM, no general criterion is known.

\subsection{Scalar sector and its symmetries}

The most general renormalizable Higgs potential (\ref{V-NHDM}) contains a huge number 
of free parameters. 
Its minima cannot be expressed in terms of elementary functions of free parameters.
One can try to apply the bilinear formalism, which was very successful in 2HDM, 
to the general NHDM potential. 
Similarly to (\ref{bilinears-2HDM}), one introduces the real-valued vector $(r_0, r_a)$, 
where $r_0 = \fd_i \f_i$ and $r_a = \fd_i (\lambda^a)_{ij} \f_j$,
$a=1, \dots N^2-1$, $\lambda^a$ being the $SU(N)$ generators,
and express the potential exactly as (\ref{V-bilinears-2HDM}).
However in contrast to the 2HDM, the orbit space inside $\R^{N^2}$ has a more complicated shape. 
It can be characterized algebraically via a set of inequalities \cite{Ivanov:2010ww,Maniatis:2014oza,Maniatis:2015gma}
and the minimization conditions can also be written explicitly \cite{Maniatis:2014oza,Maniatis:2015gma}.
One can proceed further and apply the homotopy continuation method \cite{Maniatis:2012ex}
to numerically locate the global minimum. However, this approach has not yet produced as powerful results
as in 2HDM. 
On the other hand, studying the general NHDM potential seems to be of little practical use.
Arguably the only way to keep the situation under control is to impose global symmetries.
They can dramatically simplify the potential, guarantee that its form is stable against corrections,
and hopefully render the model tractable analytically.

This brings up several symmetry-related questions. 
Which global symmetries can be imposed on the NHDM scalar potential?
What is the most general potential for each symmetry group $G$? 
What minima can a $G$-symmetric potential have and how do they break $G$?
Do minima of different nature coexist? 
What is the relation between these horizontal symmetries and the $CP$-violation
in the scalar sector? 
In 2HDM, all these questions were answered within the bilinear formalism, see section~\ref{section-2HDM}.
With three doublets, even the first examples constructed back in 1970s showed that 
allowed discrete symmetry groups can be non-Abelian, they do not necessarily imply explicit $CP$-conservation,
and, depending on free parameters, the minima can either conserve or (partially) break
these symmetry groups.

The full systematic study of symmetry-related opportunities within 3HDM 
was undertaked recently in \cite{Ivanov:2011ae,Ivanov:2012ry,Ivanov:2012fp,Degee:2012sk,Ivanov:2014doa}.
In \cite{Ivanov:2011ae}, Ivanov, Keus, and Vdovin developed a linear-algebraic approach 
based on the so-called Smith normal forms to identify all possible Abelian symmetry groups
which can be imposed on the NHDM scalar sector without producing accidental symmetries.
For 3HDM, the list is rather short: $\Z_2$, $\Z_3$, $\Z_4$, $\Z_2 \times \Z_2$, $\Z_3 \times \Z_3$,
and continuous groups.
In fact, the method itself is universal and can be applied to an arbitrary model with any number
of complex fields, see examples for NHDM Yukawa sectors \cite{Ivanov:2013bka,Nishi:2014zla} and 
for $SO(10)$ GUT Yukawa sectors \cite{Ivanov:2015xss}.
The next step is to find all allowed non-Abelian groups, whose all Abelian subgroups belong to the list
obtained at the previous step.
This classification for 3HDM was completed by Ivanov and Vdovin in \cite{Ivanov:2012ry,Ivanov:2012fp}.
The key observation was that the order of each discrete Abelian $A$ is a power of 2 or 3. By Cauchy's lemma,
it implies that the order of the non-Abelian group $G$ can also contain only these two primes:
$|G| = 2^a 3^b$. Then, using Burnside's theorem, the requirement that $G \subset PSU(3)$,
and the absence of accidental symmetries, one can express $G$ as extension of $A$ by its automorphisms:
$G = A.K$, $K \subseteq Aut(A)$. 
Checking one by one all five $A$'s, one can find all possible $G$. The final list of discrete family-symmetry groups
of the 3HDM scalar potential is
\be
\Z_2, \quad \Z_3,\quad \Z_4,\quad \Z_2\times\Z_2,\quad S_3,\quad D_4, \quad A_4,\quad S_4\,, \quad
\Delta(54)/\Z_3,\quad \Sigma(36)\,.
\label{groups-3HDM}
\ee
This list is complete: trying to impose any other finite symmetry group of Higgs-family transformations
leads to a potential with a continuous symmetry.
Most of these groups have been used previously for model-building
with three doublets, with the exception of $\Sigma(36)$.
To avoid confusion, we stress that all finite groups in (\ref{groups-3HDM}) are considered as subgroups of 
$PSU(3)\simeq SU(3)/\Z_3$, not $SU(3)$ itself, see explanations in \cite{Ivanov:2011ae,Ivanov:2012fp}. 
Their full preimages inside $SU(3)$ contain elements of the center of $SU(3)$.

The third step is to minimize the potential for each $G$. 
For high symmetry groups the number of free parameters is small, and minimization
can be done exactly. Most of these minima were well known and used by model builder
since early papers, but the full classification of all minima possible for every $G$ was
done recently in \cite{Degee:2012sk,Ivanov:2014doa}.
For small groups, the straightforward minimization usually becomes
cumbersome if not impossible. The standard procedure then is to select the 
vevs as input parameters, which leads to constraints among coefficients. 
Similarly to 2HDM, there is a hidden danger in this approach: 
the chosen vacuum might represent a local, not the global minimum of the potential.
An example of a situation when the metastability issue was overlooked is given by the $A_4$-symmetric 3HDM
with minimum (\ref{S4-minimum}), which was used in several publications 
\cite{Lavoura:2007dw,Morisi:2009sc,Toorop:2010ex}.
It was found only in \cite{Degee:2012sk} that this alignment cannot represent the global minimum.

Beyond three doublets, not much is known. The list of Abelian symmetry groups
was constructed for each $N$ \cite{Ivanov:2011ae} and includes essentially all finite groups
with order $|A| \leq 2^{N-1}$ plus one special case for every $N$. 
However combining them into non-Abelian groups is, apparently, an arduous task.
Not much is known as well about possible symmetry breaking of these discrete groups upon minimization.
One insight may come from the observation that a renormalizable Higgs potential in NHDM,
because of its algebraic structure, can have at most $p$ distinct minima. 
For 2HDM, $p=2$; for 3HDM, most likely $p=8$, 
for $N>3$ the exact value of $p$ is unknown \cite{Ivanov:2014doa}. 
A discrete group $G$, if broken completely, produces at least $|G|$ distinct 
degenerate minima, which are related by symmetry transformations. 
If $|G|>p$, it leads to a contradiction; therefore, such groups cannot be broken completely.

\subsection{Novel forms of $CP$-violation}\label{section-NHDM-CPV}

With more than two Higgs doublets, one has more room for various forms of $CP$-violation
in the scalar sector. Below we list several types of CPV which were impossible
in 2HDM.

The first unusual form comes under the name of geometrical $CP$-violation.
In the first multi-Higgs-doublet models with spontaneous CPV \cite{Lee:1973iz,Branco:1980sz},
the $CP$-violating relative phase between vevs is a smooth function of the free parameters of the potential.
Its numerical value depends on the input parameters,
and as they evolve upon renormalization group flow, the phase evolves accordingly.
In 1984 Branco, Gerard, and Grimus found a 3HDM model \cite{Branco:1983tn} in which
the phase was {\em calculable}: it does not depend on the numerical values of the free parameters
but is fixed by the structure of the symmetry-constrained potential.
The model is based on the exact symmetry group $\Delta(54)$ mentioned earlier,
which is generated by arbitrary permutation of the three doublets $\phi_k$
as well as order-3 rephasing transformation: $\phi_k \to \omega^k \phi_k$, where $\omega = \exp(2\pi i/3)$.
The potential contains permutation-symmetric phase-insensitive part and a single phase-sensitive combination
\be
V_{\rm phase} = \lambda \left[
(\fd_1\f_2)(\fd_1\f_3) + (\fd_2\f_3)(\fd_2\f_1) + (\fd_3\f_1)(\fd_3\f_2) + h.c.\right].\label{geometricCPV}
\ee
If all vevs are non-zero and $\lambda$ is positive, 
the vev alignment must be of the form $v_i \propto (1,\, 1,\, \omega)$ or its conjugate,
up to permutations. Indeed, 
the bracket in (\ref{geometricCPV}) can be rewritten as
\be
|\fd_1\f_2 + \fd_2\f_3 + \fd_3 \f_1|^2 - |\fd_1\f_2|^2 - |\fd_2\f_3|^2  - |\fd_3\f_1|^2\,,
\ee
so that minimizing it amounts to setting $\fd_1\f_2 + \fd_2\f_3 + \fd_3 \f_1$ to zero,
from which one obtains this $CP$-violating alignment.
Since the relative phase $2\pi/3$ arises in a purely geometric way,
this phenomenon was baptized in \cite{Branco:1983tn} the geometrical $CP$-violation.

The interest in this possibility was revived recently,
both within the original model \cite{deMedeirosVarzielas:2011zw,Varzielas:2012nn}
including the full fermion sector \cite{Varzielas:2013eta,Varzielas:2013sla}
and accompanied with other new fields,
and with larger multi-Higgs extensions of the same idea \cite{Varzielas:2012pd,Ivanov:2013nla}.
Remarkably, the calculability of the vev phase survives even in the explicitly $CP$-violating model
with $\Delta(27)$ symmetry, which motivated the authors of \cite{Branco:2015hea}
to speak of explicit geometrical CPV. 

It tuns out that there is an intimate relation between Higgs family symmetry groups 
and $CP$-violation in the scalar sector.
In 2HDM, imposing an exact discrete symmetry $\Z_2$ immediately prevents both explicit and spontaneous CPV.
It is precisely because of that feature that Weinberg moved on to three Higgs doublets:
imposing the exact NFC via $\Z_2 \times \Z_2$ symmetry, one can still have explicit \cite{Weinberg:1976hu} 
or spontaneous \cite{Branco:1980sz} CPV.
On the other hand, it was known that imposing a certain Higgs-family symmetry groups $G$
on the scalar potential often precludes any form of CPV in the scalar sector.
This group does not have to be very large. Again, staying within 3HDM, 
the exact $Z_4$ symmetry forbids CPV \cite{Ivanov:2011ae}.
At present, no simple rule is known that could universally tell which symmetry group $G$
for which number of doublets prevents explicit or spontaneous CPV.
For 3HDM, the situation is now resolved \cite{Ivanov:2014doa}
only because the full list of all discrete symmetry groups and all their symmetry
breaking patterns have been derived.
For arbitrary number of doublets and for Abelian groups, there is a general
result which says colloquially that ``CPV comes in pairs'':
if $G$ prevents explicit CPV within the neutral scalar sector, 
then it prevents spontaneous as well, and vice versa \cite{Branco:2015bfb}. 
Generalization of this statement to non-Abelian groups has not been proved, although it holds for all cases studied so far.
Finally, we remark that there is a deeper mathematical relation between $CP$-symmetries
and the automorphism group of $G$, 
the transformations which can be viewed as ``symmetries of symmetries'' 
\cite{Grimus:1995zi,Holthausen:2012dk,Chen:2014tpa,Trautner:2016ezn}.

3HDMs can accommodate not only new forms of $CP$-violation but also an exotic form of $CP$-conservation:
a $CP$-symmetry of order 4. It is the GCP symmetry of type (\ref{GCP})
with the angle $\alpha = \pi/2$ (\ref{block}); in order to get identity transformation,
one must apply it not twice but four times.
Within 2HDM, we already encountered such symmetry, dubbed CP2. But when imposing it on 2HDM,
we automatically had extra accidental symmetries including usual $CP$.
In 3HDM, one can construct an example for which this order-4 GCP is the only symmetry of the model.
Remarkably, its scalar potential does not have real basis, and it serves as an explicit
counterexample to the attempts to extend to NHDM the theorem \cite{Gunion:2005ja} 
relating the real basis existence with explicit $CP$-conservation.
This model was briefly mentioned already in \cite{Ivanov:2011ae} and was elaborated recently in
\cite{Ivanov:2015mwl,Aranda:2016qmp}. It leads to degenerate real scalars,
which can be combined into complex scalar fields $\varPhi$ with exotic $CP$-property: 
$\varPhi(\vec x, t) \toCP i \varPhi(-\vec x, t)$, without complex conjugation.
Phenomenologically, it looks like a curious generalization of the Inert doublet model, see section~\ref{section-IDM},
with a different quantum number protection against DM candidate decay.

This symmetry can also be extended to fermion sector \cite{Aranda:2016qmp} leading to degenerate fermions.
In that aspects, it resembles the 2HDM with maximal $CP$ symmetry suggested in \cite{Maniatis:2007de,Maniatis:2009vp}.
The difference is that with two doublets, this symmetry is effectively of order two not four,
and that it must be spontaneously broken by minimization of the potential.
Here, an additional doublet which is unaffected by the family transformations,
makes this symmetry genuinely of order four and, by acquiring a vev instead of the other two doublets, 
it can leave the symmetry unbroken.
It would be interesting to see whether a realistic model with spontaneous breaking of this symmetry
differs in any essential way from spontaneous CPV of the usual $CP$-symmetry.
Two questions concerning the possibility of distinguish this models with a usual $CP$-conserving one
which remain open are 
(i) whether it can be done experimentally, at least in principle,
and (ii) how one can tell these two cases apart with the basis-invariant approach,
which was recently constructed for NHDM in \cite{Varzielas:2016zjc}.


\section{Beyond doublets}\label{section-beyond-doublets}

\subsection{Singlet extensions}\label{section-singlets}

\subsubsection{Motivations}

An even simpler extension of the SM scalar sector consists in introducing just one extra scalar field, 
either real $S$ or complex $\S$, which transforms trivially 
under all the gauge transformations of the Standard model. Being electroweak singlet with zero hypercharge,
this scalar does not couple to the SM fermions nor does it contribute to the masses of the $W$ and $Z$ bosons.
The only way it can couple to the SM fields in renormalizable way is via the unique dimension-2 
SM gauge-invariant operator $\fd \f$. 
Even more, the new singlet can be the tip of an iceberg. A whole new sector
could in principle exist, with a potentially rich field content and new gauge interactions 
to which the SM fields are completely blind;
the mirror world \cite{Lee:1956qn,Kobzarev:1966qya} and the twin Higgs models \cite{Chacko:2005pe} are two popular frameworks
which fully use this idea. 
The lagrangian of the full theory can be written as ${\cal L}_{SM} + V_{int} + {\cal L}_{new}$,
where $V_{int}$ describes the interaction between the SM with the hidden world.
Staying with renormalizable interactions only, one can introduce
$V_{int} = \lambda_{hs} \fd\f S^2$ for real $S$ or $\lambda_{hs}\fd\f |\S|^2$ for complex $\S$.
The SM Higgs boson/gauge singlet scalar combination represents in these models the only link 
which couples the two sectors of the theory.
In this way, a rich hidden sector even with light fields could remain hidden until now.
Once the Higgs bosons are copiously produced at colliders, 
this Higgs portal opens, allowing us to explore the hidden sector.

As a side remark, one can notice that, within the pure SM, the Higgs bilinear term enters 
the Higgs potential (\ref{singlePhi}) and is accompanied by the dimensional coefficient $\mu^2$.
It is the only place in the entire SM lagrangian where a mass scale is introduced by hand.
A new gauge singlet scalar field interacting with the Higgs doublet can offer a dynamical origin of this mass scale.
Starting from the quartic interaction $V_{int}$, the self-interaction of the singlet $S$ can drive it to acquire 
a non-zero vev, $\lr{S} \not = 0$, which reproduces the $\mu^2 \fd\f$ term in the SM Higgs potential.

This simple and far-reaching idea of the Higgs plus gauge singlet scalar combination as a portal to a new hidden world
was explicitly articulated in \cite{Patt:2006fw}, although models exploiting
such interaction appeared earlier \cite{Silveira:1985rk,McDonald:1993ex,Schabinger:2005ei}.
Although other variants of the Higgs portal have been proposed and explored in the past few years, 
as described in the recent review \cite{Assamagan:2016azc}, the models with gauge singlets
remain the most studied examples.

Another source of inspiration for gauge singlets is that they can naturally lead to scalar dark matter candidates.
They are inert by construction, and if, in addition, their interaction with the SM Higgs fields is protected
by a conserved global symmetry, they will remain stable against decay \cite{McDonald:1993ex,Bento:2000ah,Burgess:2000yq}. 
Additionally, with the complex singlet field one can simultaneously achieve the correct
DM relic abundance and a sufficiently strong electroweak phase transition \cite{Barger:2008jx},
accompanied with interesting collider signals.
Finally, an extra scalar which mixes with the neutral component of the Higgs doublet 
emerges as a perfect remedy against the Higgs potential metastability, an issue which raises concerns within the SM.

In short, models with gauge singlets are a convenient bSM playground, equally well suited for analytical
calculations and incorporation in computer tools, for numerical scans of the parameter space, 
for experimental tests, and also serve as simple ultraviolet-complete toy models for exploration of various subtle
quantum field theoretical issues, see e.g. \cite{Buchalla:2016bse}.

\subsubsection{Real singlet extension}

In models with a SM Higgs doublet $\phi$ and a gauge singlet scalar $S$, the new scalar links
the SM fields to the unknown hidden sector. 
In spite of our ignorance of this hidden sector, we can simply assume a generic renormalizable 
self-interaction for the scalar $S$ and investigate the joint $(\phi, S)$ potential.

A simple, easily tractable, and phenomenologically interesting version of this model arises when 
we assume that the lagrangian is invariant under the $\Z_2$ symmetry $S \to -S$. 
The general renormalizable potential takes the following form:
\be
V(\phi, S) = - m^2 \fd\f - {1 \over 2}\mu^2 S^2 + \lambda_1 (\fd\f)^2 + {1 \over 4}\lambda_2 S^4 + {1 \over 2}\lambda_3 \fd\f S^2\,.
\label{singlet1}
\ee
The potential must be bounded from below, which implies $\lambda_1 > 0$, $\lambda_2 > 0$, 
and $\sqrt{4\lambda_1 \lambda_2} + \lambda_3 > 0$. 
If one wishes to investigate the model with $\lr{\f} \not = 0$ and $\lr{S} \not = 0$, 
one needs to place a stronger bound $4\lambda_1 \lambda_2 > \lambda_3^2$,
which additionally forbids $\lambda_3$ to be too large \cite{Robens:2015gla}.
Minimization of the potential (\ref{singlet1}) generates vevs for both the doublet $\lr{\phi^0} = v/\sqrt{2}$ 
and singlet $\lr{S} = v_s$. Similarly to the 2HDM notation,
we define $\tan\beta \equiv v/v_s$.
The electroweak Goldstone fields are located inside the doublet. 
In the unitary gauge the fields can be expanded near the vacuum point as $\phi = (0,\, v+h_\phi)^T/\sqrt{2}$
and $S = v_s + h_s$, and the physical Higgses $h$, $H$ arises as their mixtures:
\be
\doublet{h}{H} = \mmatrix{\cos\alpha}{-\sin\alpha}{\sin\alpha}{\cos\alpha}\doublet{h_\phi}{h_s}\,,
\quad m^2_{h,H} = \lambda_1 v^2 + \lambda_2 v_s^2 \mp \sqrt{(\lambda_1 v^2 - \lambda_2 v_s^2)^2 + (\lambda_3v v_s)^2}\,.
\label{singlet-mixing}
\ee 
The mixing angle $\alpha$ is defined in the range from $-\pi/2$ to $\pi/2$.
With these masses, one can replace the five free parameters of the original potential (\ref{singlet1}) 
with the five phenomenologically relevant quantities $v$, $m_h$, $m_H$, $\alpha$, $\tan\beta$.

Notice that the fermions and gauge bosons couple only to $h_\phi$ but not to $h_s$.
Therefore, the couplings of $h$ and $H$ follow the same pattern as in the SM 
but with an extra suppressing factor of $\cos\alpha$ and $\sin\alpha$, respectively.
This leads, in particular, to the $\cos^2\alpha$ and $\sin^2\alpha$ suppression with respect to the SM 
of the production rates of the light and the heavy Higgses.
The decay widths pattern exhibits a similar suppression with one modification: the heavy Higgs
can decay into a pair of light ones if this channel is open kinematically.
 
So far, we have not specified which of the two scalars is the 125 GeV Higgs boson $h_{125}$.
Both scenarios are possible. In the high-mass scenario, $h = h_{125}$,
and the alignment limit corresponds to $\sin\alpha = 0$. In the low-mass scenario, $H = h_{125}$,
the alignment limit is $\cos\alpha = 0$, and an extra invisible decay channel for the SM-like Higgs is potentially open.
When parametrizing the model in terms of the five phenomenological parameters and placing limits
on them, one should not be surprised to see jumps at the second Higgs mass of 125 GeV.

Lifting the $\Z_2$ symmetry allows cubic and linear terms to appear in the potential,
$a S \fd \f + b S^3 + c S$,
which was impossible in multi-doublet models.
The linear term can be removed by a shift of the field $S$, 
so that one adds only two more free parameters to the model, the total counting rising to seven.
But then whatever vev the field $S$ acquires, be it positive or negative, it cannot be manipulated anymore.
Alternatively, one can perform such a shift that sets $\lr{S}=0$, but then the linear term will enter the tadpole conditions.
Notice also that the extra terms do not change the stability conditions for the potential
but do modify the vacuum structure and can lead to a metastable minimum, 
see \cite{Ghosh:2015apa} for a relevant analysis at the tree level and at one loop.
The presence of cubic terms makes the minimization of the potential and the resulting interactions
between the scalars $h$ and $H$ more involved.
One useful relation which holds for any scalar potential is
\be
\sin^2 \alpha = ({\cal M}_{\phi\phi}^2 - m_h^2)/(m_H^2 - m_h^2)\,,\label{singletmixing}
\ee
where ${\cal M}_{\phi\phi}$ is the coefficient in front of the $h_\phi^2/2$ term.
This relation can be used to trade angle $\alpha$ for ${\cal M}_{\phi\phi}$ when defining parameter space.
It also makes evident that if $\lr{S}=0$ after removal of the linear term, then there is no mixing between the two scalars: 
${\cal M}_{\phi\phi}^2 = m_h^2$ and $\alpha = 0$.
Explicit expressions for the $hhh$ and $Hhh$ couplings in terms of seven parameters of the model
were given in \cite{Buttazzo:2015bka}.

\subsubsection{Complex singlet extension}

An extra complex scalar $\S = S + i A$ brings in two new degrees of freedom.
Depending on the singlet vev, the three neutral scalars can mix, producing a richer
collider phenomenology and complicating its analysis.
On the other hand, one or both new scalars can be symmetry protected against decay,
yielding simple models of two-component dark matter or models with one DM candidate
and a strong electroweak phase transition \cite{Barger:2008jx}.

For complex $\S$, the number of free parameters in the scalar sector rises sharply, 
reaching 21 in the most general renormalizable potential \cite{Barger:2008jx}. 
In the quartic sector alone, a single term $S^4$ for the real scalar
is now replaced with $\S^4$, $(\S^4)^*$, $|\S|^2 \S^2$, $|\S^2| (\S^2)^*$, and $|\S|^4$, 
all coming with their own coefficients.
An exhaustive phenomenological scan of the entire parameter space is of questionable use, 
taken into account the minimalistic attitude one takes in singlet models.
It makes more sense to simplify the problem by imposing extra global symmetries such as a discrete $\Z_2$ 
(generated by $\S \to - \S$), a $\Z_2 \times \Z_2'$ (generated by $\S \to - \S$ and $\S \to \S^*$, 
or alternatively by $S \to - S$ and $A \to - A$), or a continuous $U(1) \rtimes \Z_2$ 
(generated by phase rotations and $\S \to \S^*$, or alternatively by $O(2)$ transformations in the $(S,A)$ space).

Before moving on, let us make a remark on the (absence of) $CP$-violation in the singlet extension of SM. 
Although the potential contains many complex coefficients, 
it does not produce $CP$-violating effects in the scalar sector, see Chapter 23.6 in \cite{CPV-book}.
The fundamental reason is that in the presence of gauge-singlet scalars there exists even larger freedom 
of defining $CP$-symmetry. In particular, $\phi \toCP \phi^*$, $\S \toCP \S$ is an equally acceptable
definition as the traditional one $\phi \toCP \phi^*$, $\S \toCP \S^*$.
Alternatively, in terms of real fields, when building the model, we are free to arbitrarily assign the $CP$-properties
of both $S$ and $A$. In particular, we can assume that both $S$ and $A$ 
are $CP$-even: in the absence of fermion and gauge interactions, 
there is no justification to prefer the $CP$-odd $A$ over 
the $CP$-even one --- at least, as long as only the SM fermions are considered.
Thus, whatever the doublet plus singlet potential is, there is no $CP$ violation in the scalar sector.
Adding vector-like quarks will, however, open up the possibility of $CP$-violation 
with this scalar content \cite{Bento:1990wv,Branco:2003rt,Darvishi:2016tni}.
It must be added that not all authors agree that this extended freedom leads to a legitimate $CP$-transformations.
For example, in \cite{Barger:2008jx}, the two real scalars $S$ and $A$ are explicitly labeled as the $CP$-even and $CP$-odd one,
although the authors mention that they do not consider $CP$-violating variant of the model. 
The authors of the recent work \cite{Darvishi:2016gvm} also claim that the model with cubic terms 
formally exhibits $CP$-violation, although in order to observe $CP$-violating effects in physical processes 
one needs to complement the content of the model with vector-like quarks.

A relatively simple and physically motivated model can be obtained by assuming a $U(1)$ global symmetry
which is softly broken by the quadratic and linear terms \cite{Barger:2008jx,Coimbra:2013qq,Costa:2015llh}. 
The scalar potential takes the form
\be
V = {m^2 \over 2} \fd\f + {\lambda \over 4}(\fd\f)^2 + {\delta_2 \over 2}\fd\f |\S|^2 + {b_2 \over 2}|\S|^2 + {d_2 \over 4}|\S|^4
 + \left(a_1 \S + {b_1\over 4} \S^2 + h.c.\right)\,,\label{cxSM}
 \ee
with five real and two complex parameters.
The $U(1)$ breaking quadratic term is needed to avoid the massless Goldstone field at $\lr{\S} \not = 0$,
while the linear term breaks the residual $\Z_2$ symmetry and prevents formation of domain walls
in early Universe, which would normally appear when an exact discrete symmetry is spontaneously broken.
Cubic terms can also be added \cite{Darvishi:2016gvm}.

\begin{table}[h]
\begin{center}
\begin{tabular}{|c|c|c|l|}
\hline
model & phase & vevs & symmetry\\[1mm] \hline 
$U(1)$ & Higgs + 2 degenerate dark & $\lr{\S}=0$ &  conserved $U(1)$\\[1mm]
 & 2 mixed + 1 Goldstone & $\lr{A} = 0$ & $U(1) \to \Z'_2$ \\ \hline
$\Z_2\times\Z'_2$ & Higgs + 2 dark & $\lr{\S}=0$ &  conserved $\Z_2\times\Z'_2$\\[1mm]
 & 2 mixed + 1 dark & $\lr{A} = 0$ & $\Z_2\times\Z'_2 \to \Z'_2$ \\ \hline
$\Z'_2$ & 2 mixed + 1 dark & $\lr{A}=0$ &  conserved $\Z'_2$\\[1mm]
 & 3 mixed & $\lr{\S} \not = 0$ & broken $\Z'_2$ \\
\hline
\end{tabular}
\end{center}
\caption{Phase classification for possible models with one complex singlet \cite{Coimbra:2013qq}}\label{table-singlets}
\end{table}

Depending on whether $a_1$ and $b_1$ are zero or not, and whether the singlet gets a non-zero vev,
several classes of models are possible. 
Table~\ref{table-singlets} summarizes the global symmetries, the vev patterns, 
and the main phenomenological consequences \cite{Coimbra:2013qq}.
In the $U(1)$ symmetric model, with $a_1 = b_1 = 0$, the singlet-unbroken phase with $\lr{\S} = 0$ produces
two degenerate DM candidates, while broken phase contains a massless scalar, --- one can choose it to be $A$, ---
which is phenomenologically ruled out.
The $\Z_2\times \Z_2'$ symmetry 
is obtained by setting $a_1=0$ and $b_1$ real. Again, for $\lr{S}=0$, we have no mixing and two DM candidates,
which are not degenerate anymore. In the broken phase, the minimization removes only one of the two $\Z_2$ factors,
so that one scalar remains inert and is a DM candidate, while the other mixes with the doublet neutral.
Again, without the loss of generality, one can choose $\lr{A}=0$, as the other possibility is obtained
just by the $S\leftrightarrow A$ swap and the corresponding sign flip of the real $b_1$.
The $\Z_2$-symmetric model with real $a_1$ and $b_1$, or to be precise, with a real $b_1^* a_1^2$, 
can either have one DM candidate and two-scalar mixing, or no DM candidates at all and the full three-scalar mixing.
Finally, the generic potential (\ref{cxSM}) with a complex $b_1^* a_1^2$ does not possess any global symmetry
and does not lead to DM candidates.

In any of these cases, the stability and perturbative unitarity conditions are calculated straightforwardly,
as the quartic part of the potential is simple \cite{Coimbra:2013qq}.
In the full-mixing case, the $3\times 3$ real rotation matrix $R_{ij}$ with angles $\alpha_1$, $\alpha_2$, $\alpha_3$ 
brings the initial fields $h_\phi$, $S$, and $A$ to the three mass eigenstates $h_1$, $h_2$, and $h_3$.
Each of them couples to the SM fields via their $h_\phi$ component,
so that the coupling coefficients are proportional to their SM values times the $R_{i1}$ matrix elements.
Expressions for all triple and quartic couplings among physical scalars in terms of the parameters of the potential 
and the entries of the rotation matrix $R$ are given in \cite{Costa:2015llh}.

Another way to cope with the many interaction terms is 
to use the exact $\Z_2 \times \Z_2'$ symmetry instead of softly broken $U(1)$ \cite{Ferreira:2016tcu}.
Analysis of this potential is more complicated but still manageable, especially
if one works with the real fields $S$ and $A$. Since the potential is now expressed in terms of quadratic 
quantities $\fd\f$, $S^2$, $A^2$, the stability conditions can be derived exactly with the copositivity criteria \cite{Kannike:2012pe}.
The free parameter counting rises to nine, and vacuum structure of the model is more elaborate
than in the softly broken $U(1)$ case \cite{Ferreira:2016tcu}.
The potential can possess simultaneous minima at different depths, which differ by their symmetry breaking patterns.
This brings again the issue of the tree-level stability vs metastability of a chosen vacuum.
Ref.~\cite{Ferreira:2016tcu} presents a detailed analysis of these stability issues.

\subsubsection{Phenomenological studies}

Collider phenomenology of the gauge singlet extensions is less rich than in multi-Higgs-doublet models.
Singlets interact neither to the gauge bosons nor to fermions, and the physical scalars couple 
to the SM fields with the same pattern as in the SM just downscaled by the elements of the rotation matrix. 
There are no charged Higgses, the neutral ones cannot induce FCNC or LFV transitions, 
there is no $CP$-violation coming from the scalar sector.
When checking whether a specific model is compatible with existing data, 
one just needs to consider electroweak precision tests, the LEP and LHC bounds on direct production of new scalars,
both light and heavy, 
and to satisfy the relic DM abundance and the constraints coming from the direct DM detection experiments.
On the other hand, singlets do have important astroparticle and cosmological consequences, see section~\ref{section-astro}.
At colliders, apart from the reduced couplings to the SM fields and existence of new states, 
the potential signals include a strongly enhanced $hh$ production around the mass of the heavier scalar,
cascade decays within the scalar sector, and displaced vertices characteristic of light extra scalars.

Several detailed phenomenological studies have been published over the last decade, 
both in the pre-LHC era \cite{Barger:2007im,Barger:2008jx} and in the recent years 
\cite{Pruna:2013bma,Coimbra:2013qq,Robens:2015gla,Buttazzo:2015bka,Costa:2015llh,Robens:2016xkb}. 
In the last couple of years, as the LHC is gearing up for the Run 2 data taking,
several works came up with carefully selected benchmark models with real \cite{Robens:2016xkb,Lewis:2017dme} and complex 
\cite{Costa:2015llh} singlets, which can be tested in the years to come.
The LHC Higgs cross section Working Group reports \cite{Dawson:2013bba,deFlorian:2016spz} provide extra details
on what parameter space regions can be probed at future colliders.

\begin{figure}[!htb]
   \centering
\includegraphics[height=7.1cm]{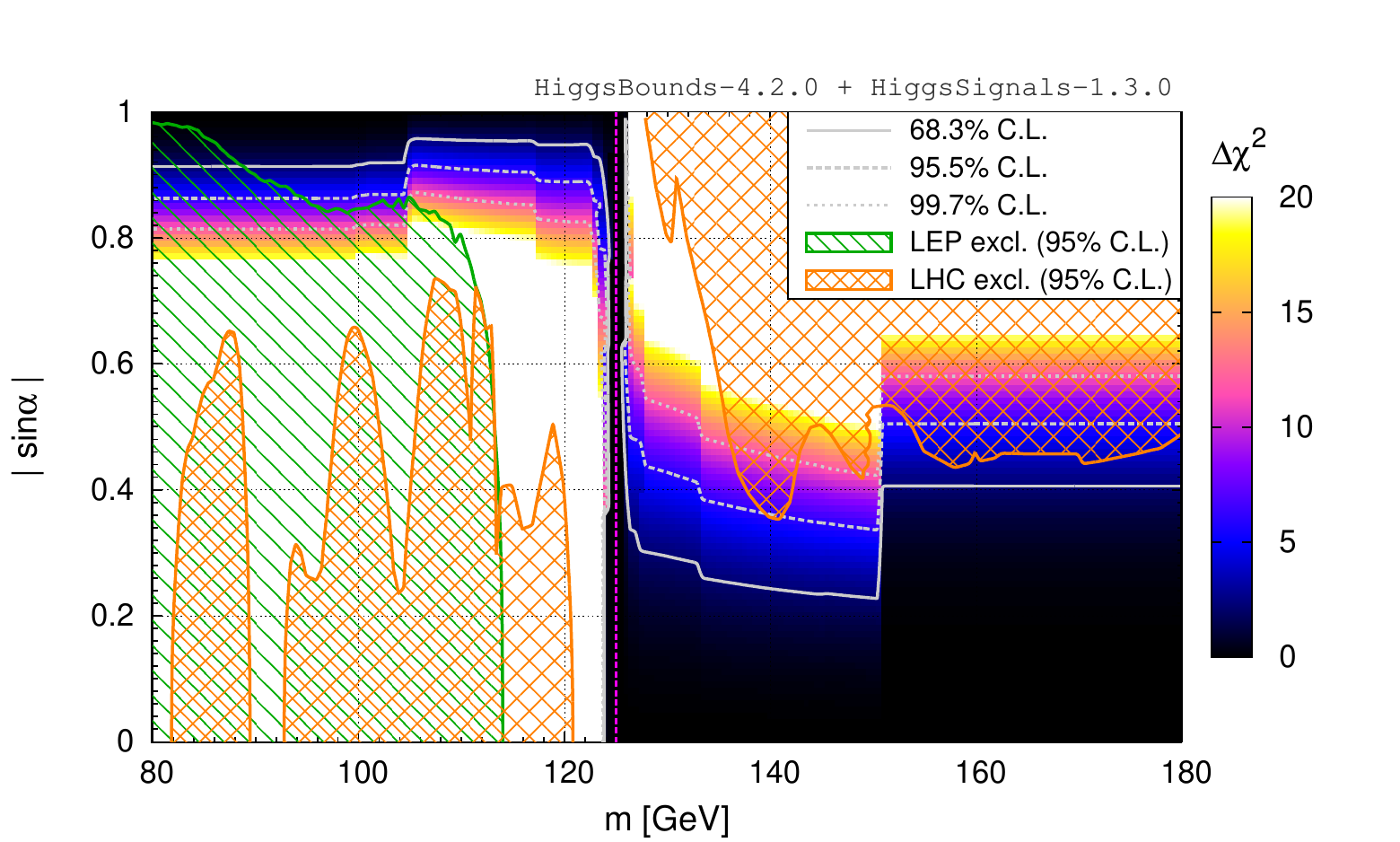}\hfill
\includegraphics[height=6.5cm]{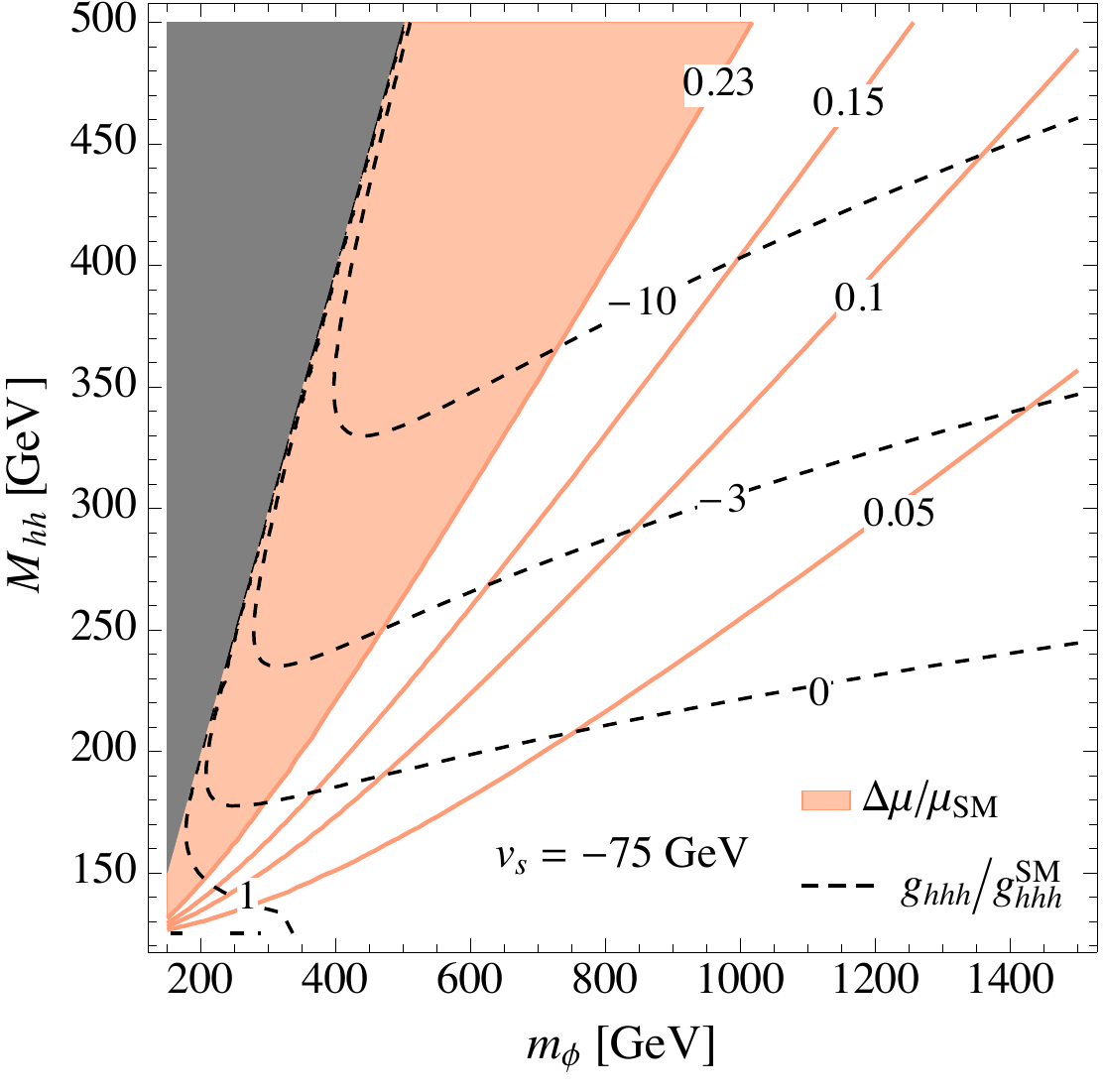}
\caption{Exclusion regions for the real singlet extension.
Right: Constraints on the $(|\sin\alpha|, m_H)$-plane arising from the LEP and LHC exclusion limits
as well as from the HiggsSignal $\Delta\chi^2$ distribution; reproduced with permission from \cite{Robens:2015gla}. 
Left: Exclusion at 95\% C.L.\ from current Higgs couplings measurements at the LHC8 (light region), and deviation in the Higgs signal strengths $\Delta \mu/\mu_{SM} \equiv \sin^2\alpha$ (light solid lines). The ratio of the trilinear Higgs coupling to its SM value is drawn for $v_s = -75$ GeV (dashed black lines). Dark gray: unphysical parameters. Reproduced with permission from from \cite{Buttazzo:2015bka}.}
   \label{fig-singlet-plane}
\end{figure}

For the real singlet extension, a global fit to the Higgs signal strengths at the 8 TeV LHC leads to $\sin^2\alpha \le 0.23$ 
at 95\% C.L. \cite{Buttazzo:2015bka}.
The study \cite{Robens:2015gla} focusing on the real singlet extension gave a more detailed picture.
Fig.~\ref{fig-singlet-plane}, left, reproduced from that work shows the regions in the plane of mixing parameter $|\sin\alpha|$
and the mass of the second boson $m_H$, which are excluded by LEP and the LHC,
as well as the regions disfavored by the $\Delta \chi^2$ analysis obtained with the HiggsSignals package on the basis of Run 1 data.
The black regions are still allowed. 
As shown by the vertical black stripe in Fig.~\ref{fig-singlet-plane}, left,
the sensitivity to $\sin\alpha$ is limited in the vicinity of the degenerate situation $m_h \approx m_H$,
where the alignment jump from the light to the heavy-mass scenario takes place.
For the exact degeneracy the sensitivity vanishes since the mixing angle is undefined here. 
In the near-degenerate region, depending on the exact mass splitting, 
one can hope to get a better constraint from the high-resolution LHC channels, 
$\gamma\gamma$ and $ZZ^*\to 4\ell$, or at the future colliders \cite{Dawson:2013bba}.
In the high mass region the strongest upper limit on $|\sin\alpha|$ arise
from the NLO corrections to the $W$ mass and the perturbativity constraints, 
pushing this limit to below $0.2$ at $m_H = 1$ TeV \cite{Robens:2015gla,Robens:2016xkb}.

Relaxing the $\Z_2$-symmetry assumption in the real singlet model 
leads to minor complication of the analysis \cite{Buttazzo:2015bka}.
Fig.~\ref{fig-singlet-plane}, right, reproduces the plot from \cite{Buttazzo:2015bka}
which shows, for a particular choice of quartic coefficients, the values of $\Delta \mu/\mu_{SM} \equiv \sin^2\alpha$ 
and the triple Higgs coupling relative to the SM $g_{hhh}/g_{hhh}^{SM}$
as functions of the entry $M_{hh}^2 \equiv {\cal M}_{\phi\phi}^2$ in Eq.~(\ref{singletmixing}) 
and the mass of the heavy scalar denoted as $m_\phi$.
The dark region is unphysical, the light region is excluded by the LHC Run 1 Higgs data.
The white region is allowed and the lines show the predicted parameters $\sin^2\alpha$ and 
$g_{hhh}/g_{hhh}^{SM}$.
Future colliders will further constrain these two parameters.
Citing \cite{Dawson:2013bba,Buttazzo:2015bka} for rough estimates, 
one expects that the HL-LHC with 3 ab$^{-1}$ of data will probe $\sin^2\alpha$ down to $\sim$ 4--8\% 
and yield the first indication of the triple Higgs coupling.
The future 100 TeV hadron collider should help detect deviations of $g_{hhh}$ from its SM value below 10\%.
The linear $e^+e^-$ colliders aim at $\sin^2\alpha$ of about 2\% level (ILC) or 0.3\% level (CLIC).
Using this plot, one can visualize the parameter range accessible to the future colliders.

The case of a light second scalar, with mass in the GeV range, 
is severly constrained by the heavy meson decays and by non-observation of 
new long-lived particles which could produce displaced vertices \cite{O'Connell:2006wi,Dermisek:2006py}. 
A moderately heavy scalar, but still lighter than 62.5 GeV, would be well visible in the decays of the SM-like Higgs boson.
A review from 2008 of the possible collider signals in this regime can be found in \cite{Chang:2008cw}.
Many of those opportunities are now of historic interest after the discovery of the 125 GeV Higgs bosons
and placing an upper bounds on its invisible decay branching fraction.

\begin{figure}[!htb]
   \centering
\includegraphics[height=7.7cm]{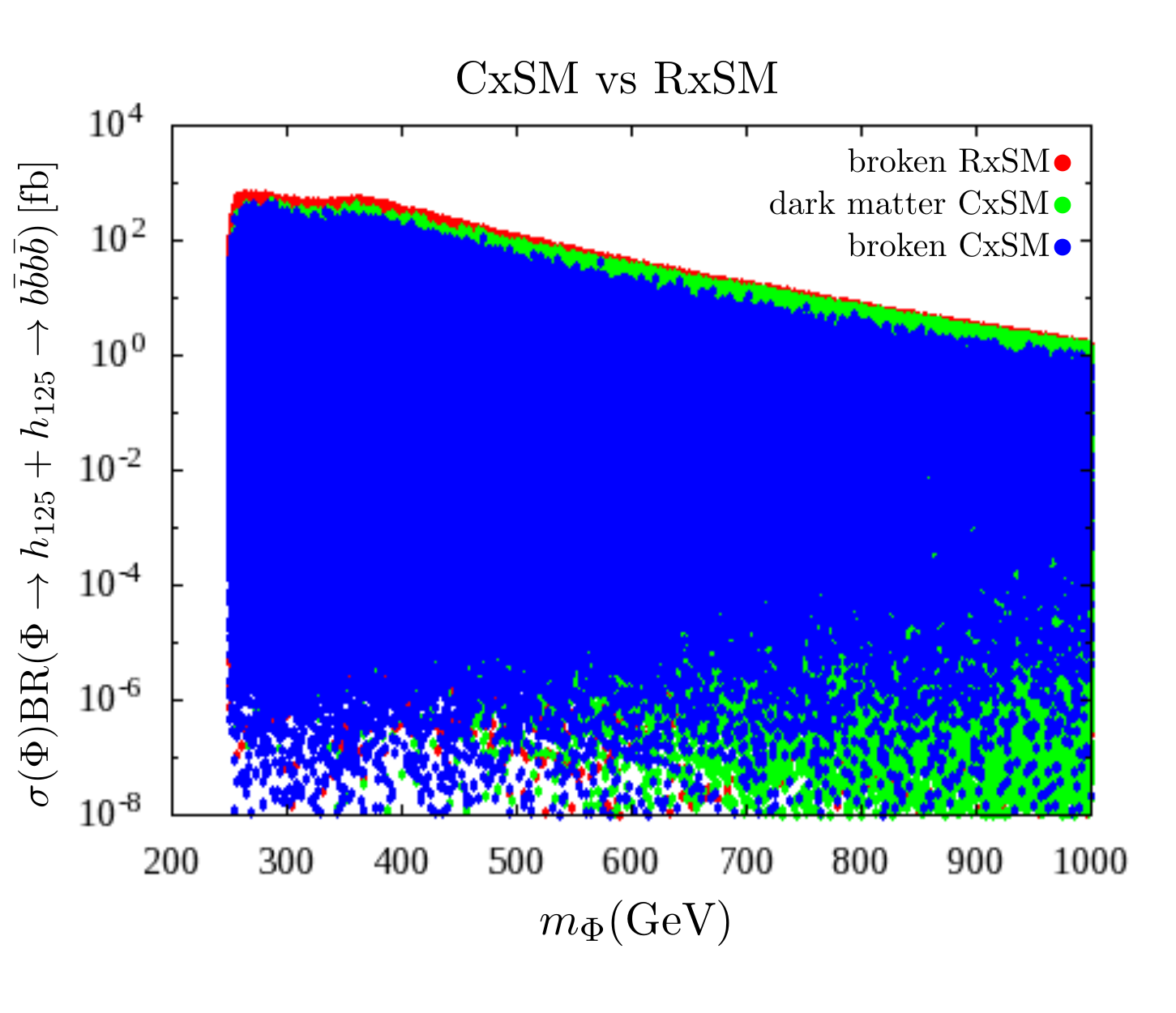}\hfill
\includegraphics[height=7.7cm]{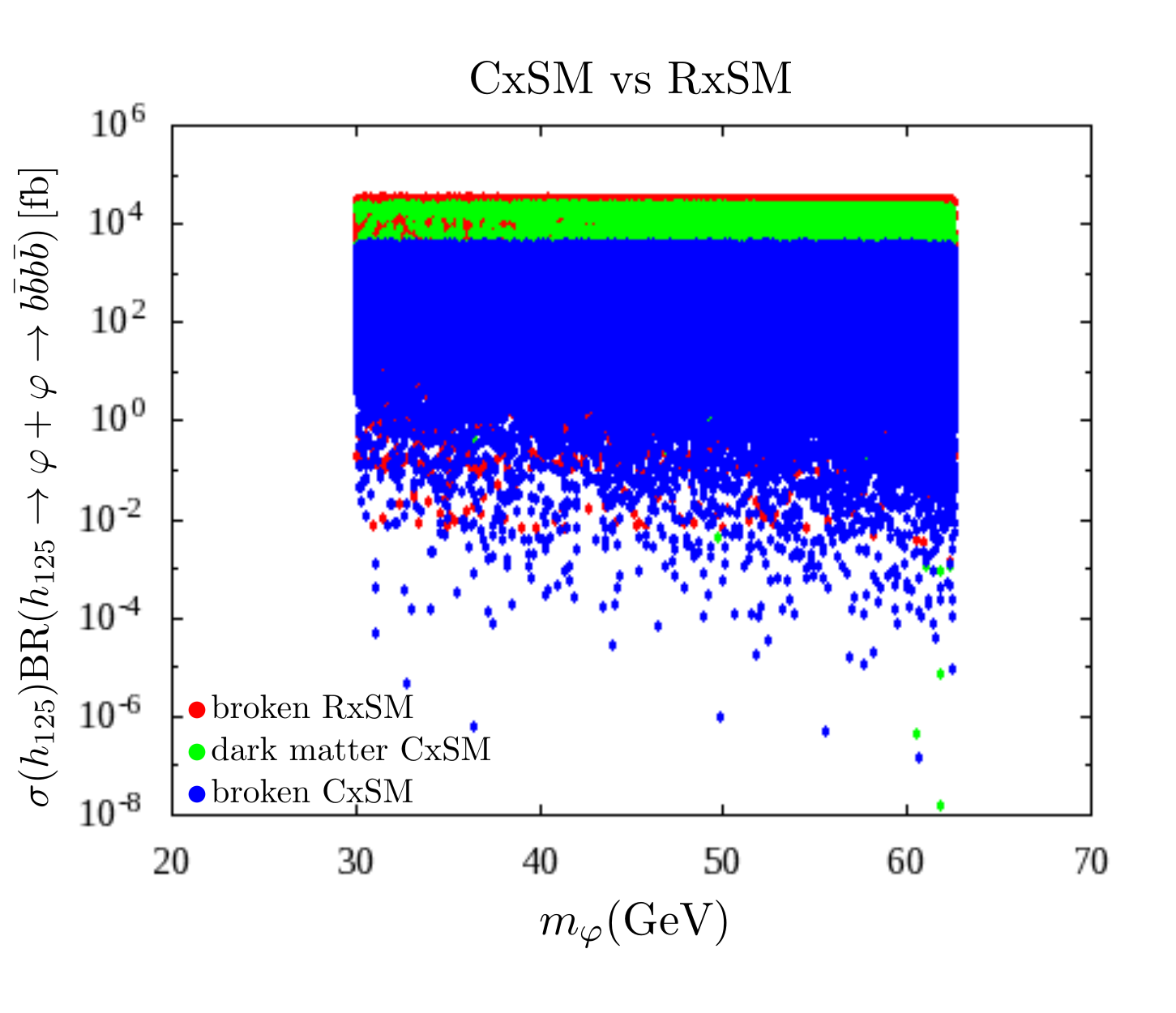}
\caption{The $4b$ final state rates for a heavier Higgs decaying into the $h_{125}h_{125}$ pair
(left) and for the case where $h_{125}$ decays into a pair of lighter bosons. 
The production process is gluon fusion at 13 TeV. The black (blue) and light gray (green) points
correspond to the fully mixing and DM phases of the complex singlet extensions,
the dark grey (red) points correpond to the mixing phase of the real singlet model.
Reproduced with permission from \cite{Costa:2015llh}.}
   \label{fig-cx-4b-rate}
\end{figure}

The collider phenomenology of the complex singlet extension depends on the scenario chosen.
For two DM candidates, either mass-degenerate or not, the phenomenology is not dramatically
different from the real singlet case: one expects missing $E_T$ events and, for light DM candidates,
an extra contribution to the invisible decay of $h_{125}$. 
For a more attractive variant with one DM candidate and two mixing scalars,
the second Higgs can be seen directly or via its $H \to hh$ decay \cite{Barger:2008jx}.
In the fully mixed case, the three scalars can give rise to additional collider signals.
Any final state $X$ that one studies at colliders searching for the Higgs signal
can be reached not only by the one-step production and decay of the SM-like Higgs
but also through intermediate production of $h_i$ decaying into two lighter scalars $h_j, h_k$,
which may be different or not, to finally produce the state $X$.
This channel can be competitive to the direct channel
in the case of resonant production of the heavy scalar.
The recent work \cite{Costa:2015llh} investigated in detail this issue
via numerical scans of the parameter space in the model given by (\ref{cxSM})
and for two different assignments for $h_{125}$: either the lightest or next-to-lightest.
A typical contribution of the chain production channel stays within a few percent,
reaching about 1/6 at $m_2 = 250$ GeV.

Another result of that study is given in Fig.~\ref{fig-cx-4b-rate}.
Shown on these plots is the production rate of the $4b$ final state,
either via heavy scalar $\Phi \to h_{125}h_{125}\to 4b$
or via the SM-like Higgs and its decay into two ligher scalars $h_{125} \to \varphi\varphi \to 4 b$.
Since the scan takes into account all experimental and theoretical constraints,
these plots offer an estimate of typical signals one can expect at the LHC Run 2.
The same plots also show the difference between the real and complex singlet extensions 
for this particular collider signature.

In the last couple of years, several authors went beyond tree-level analysis of this model.
One-loop calculation of the $H\to hh$ decay was reported in \cite{Bojarski:2015kra}.
Another recent work \cite{Kanemura:2016lkz} found that one-loop corrections to the $hhh$ coupling 
can be comparable and even larger than the tree-level values, 
and studied its effect on the double Higgs production at the LHC, 
which in the high-mass scenario can dramatically differ from the SM. 
On the other hand, if this enhancement is too high, the electroweak minimum ceases to be the global minimum,
which places extra constraints on the parameters \cite{Chen:2014ask}.

An extra scalar field additionally stabilizes
the Higgs potential at very large quasiclassical values.
Theoretical analyses of the Higgs potential stability, assuming that no New Physics
is present up to Planck scale, indicate that the present low-energy vacuum is marginally metastable \cite{Buttazzo:2013uya}.
Although it does not run into immediate problems for the present-day physics, it does raise questions about 
how the Universe ended up in such a metastable state \cite{Lebedev:2012sy}. 
The singlet scalar mixing with the Higgs boson dispels such doubts. 
Even with small mixing angles, the new scalar manages to stabilize the potential at large field values 
\cite{Lebedev:2012zw,EliasMiro:2012ay}.
This stabilization is well compatible with other experimental constraints.
Costa et al \cite{Costa:2014qga} studied the high-scale stability of the scalar potential 
in the softly broken $U(1)$ model (\ref{cxSM}) at the two-loop level. 
All relevant collider data, dark matter direct detection limits, and the relic abundance results 
were found to be consistent with stability up to the Planck scale, with the sharp lower bound 
on the second Higgs mass at about 140 GeV, with a sizable mixing angle $\alpha$,
with the dark matter particle heavier than about 50 GeV. If one wants the relic density measured by Planck 
to be exactly saturated, still keeping the stability at least up to the GUT scale, the lower limit on the second Higgs
shifts to 170 GeV. Stability can also be achieved in the broken phase, with all three mixing scalars.
All these scenarios can be explored at the LHC.

A similar conclusion for the real singlet extension and at the one-loop level only was 
obtained later in \cite{Falkowski:2015iwa}.
Stability up to the Planck scale was well feasible, with the second scalar being moderately heavy, well below 1 TeV,
and the mixing angle being sizable, leading to $\sigma(pp \to H \to hh) \sim$ 1 pb.
A similar prediction was also obtained in \cite{Godunov:2015nea}.
We also mention here the detailed study \cite{Cheung:2012nb} which investigated, via the two-loop analysis,
the stability and perturbativity of the Higgs potential in a variety of models with scalars and fermions in various representations.
Although it appeared before the official Higgs boson discovery, it assumes the Higgs mass to be at 125 GeV,
which the December 2011 LHC presentations were pointing to.

We end the discussion of singlet extensions with a remark which is hard to avoid. 
Recently, the singlet extension models received a turbo boost 
from the 750 GeV diphoton excess reported by the ATLAS and CMS collaborations after the analysis
of the first Run 2 data collected in 2015 \cite{Aaboud:2016tru,Khachatryan:2016hje}.
The enhancement appeared to originate from production and decay of a neutral scalar,
and models with singlet extensions were among the first summoned to explain the anomaly, 
see for instance \cite{Falkowski:2015swt,Chakraborty:2015jvs} to get the impression.
Clearly, the very fact that the hypothesized neutral scalar couples to photons and gluons
and, at the same time, stays invisible in other production channels, 
requires introduction of yet more new fields.
The resulting phenomenological analyses cannot be attributed to the singlet extension of the scalar sector itself
but rather refer to synthetic models with several ingredients  \cite{Strumia:2016wys}.
Although it can be debatable whether this activity produced new insights into the model itself, 
it was at the very least a rare ``real fight'' test of this model and 
it remains a valuable experience from the historical point of view.



\subsection{Higher multiplet extensions}\label{section-triplets}

Apart from electroweak doublets and singlets, one can also think of adding
scalar multiplets in higher $SU(2)$ representations. 
Historically, a scalar triplet was especially attractive because, if coupled to a pair of lepton doublets, 
it can generate neutrino masses without introducing right-handed neutrinos
\cite{Konetschny:1977bn,Magg:1980ut,Schechter:1980gr,Cheng:1980qt,Gelmini:1980re}.
In the simplest setting with one additional complex triplet $X = (\chi^{++}, \chi^+, \chi^0)$,
this interaction has form
\be
f_{ij}\left(\chi^0 \nu_i \nu_j + \chi^+ {\nu_i \ell_j + \ell_i \nu_j \over \sqrt{2}} + \chi^{++}\ell_i \ell_j\right)\,.
\ee
If $\chi^0$ gets a tiny vev, thus spontaneously breaking the lepton number, the Majorana neutrino mass matrix arises.
This simple approach leads, however, to a massless Goldstone boson (Majoron), which would be detected via its contribution
to the $Z$ width. Hence, the construction must be modified.
In the modern models of tree-level neutrino mass generation with triplets, known as the seesaw Type II models,
the triplet interacts with leptons and with the Higgs doublet and just connects the Higgs vev with the neutrinos,
see e.g. \cite{Melfo:2011nx,Boucenna:2014zba} for the models themselves and their LHC-related signals.

If one wants the triplet or higher-dimensional multiplet to develop a sizable vev, then apart from special cases mentioned later
the value of the $\rho$-parameter strongly constrains its value.
Consider a scalar multiplet $X$ with weak isospin $T > 1/2$ and hypercharge $Y$. 
The gauge-kinetic part of the new multiplet $X$ is $|D_\mu X|^2$ (with the factor $1/2$ for real fields), 
where $D_\mu$ is still given by the same Eq.~(\ref{gauge-SM}). 
The weak isospin generators $T_i$ still satisfy the $SU(2)$ commutation relations $[T_i, T_j] = i \epsilon_{ijk} T_k$,
but they now act in the $2T+1$-dimensional space.
Defining raising and lowering generators $T_{\pm} = T_1 \pm i T_2$, we 
rewrite the product $T_iW^i_\mu$ as $(W^+_\mu T_+ + W^-_\mu T_-)/\sqrt{2} + W^3_\mu T_3$.
The scalar multiplet $X$ must acquire vev $v_\chi$ only in its neutral component with $T_3 = -Y/2$. 
This leads to the following contribution to the $W$ and $Z$ mass terms:
\be
|D_\mu X|^2 \to \lr{X}^\dagger {1\over 4}\left[ g^2 W^+_\mu W^{-\mu}\cdot 2 (T_+T_- + T^-T^+) 
+ \bar g^2 Z_\mu Z_\mu Y^2\right]\lr{X}\,.
\ee
Using $T_+T_- + T_-T_+ = 2[T(T+1) - T_3^2]$, we find a contribution to the $W$ mass 
$\Delta m_W^2 = g^2 [T(T+1)-Y^2/4] v_\chi^2$ and to the $Z$ mass $\Delta m_Z^2 = {\bar g}^2v_\chi^2 \cdot Y^2/2$
from complex multiplet, and half these values for a real one.
For the Higgs doublet with $T=1/2$, $Y=1$, and vev $v/\sqrt{2}$, we recover (\ref{SM-WZmasses}),
and the $\rho$-parameter (\ref{rho}) is equal to 1. But this hinges upon relation $T(T+1)-Y^2/4 = Y^2/2$,
which is satisfied for doublets {\em accidentally}.
For a generic set of Higgs multiplets $X_i$ with neutral component
vevs $v_i$, one gets
\be
\rho = {\sum_i c_i [T_i(T_i+1) - Y^2_i/4] v_i^2 \over \sum_i Y^2_iv_i^2 /2 }\,,
\ee
where $c_i = 1$ and $c_i = 1/2$ for complex and real multiplets, respectively.
Thus, higher order multiplets lead generically to sizeable departures of $\rho$ from one.
With the present day global fit value $\rho_{\mathrm{exp}} = 1.00037(23)$ \cite{PDG},
one concludes that their vevs must be small, a few GeV at most.
One can trace this departure from unity to the lack of custodial symmetry with higher representations.
If $g'\to 0$, the three real fields $W^i_\mu$ get their masses proportional to 
$\lr{X}^\dagger T_i^2 \lr{X}$. For doublets, it turns out that $T_1^2 = T_2^2 = T_3^2$,
and it leads to the $SO(3)$ global symmetry, hence the custodial symmetry $SU(2)_V$.
For higher representations, $T_1^2 = T_2^2 \not = T_3^2$ even when inserted between $\lr{X}$,
which breaks the symmetry. 

Phenomenology of the generic triplet models was considered in several extensive studies,
both in the pre-LHC era \cite{Gunion:1989ci,Gunion:1990dt,Accomando:2006ga} 
and more recently \cite{Aoki:2011pz,Akeroyd:2011zza,Aoki:2012jj,Chiang:2012dk,Blunier:2016peh}.
One extra source of interest to this extension was an apparent excess in the early LHC data of the $H \to \gamma\gamma$
branching ratio, which could be accommodated in models with triplets \cite{Arhrib:2011vc,Akeroyd:2012ms,Wang:2012ts}.
To be fair to Higgs doublets, we hasten to add that 2HDM with nearly mass-degenerate
Higgses could also accomplish this task \cite{Gunion:2012he,Ferreira:2012nv} without showing a similar excess in other channels.

Before moving on, we mention the curious fact that the equation $T(T+1)-Y^2/4 = Y^2/2$
admits other solutions for higher-dimensional multiplets.
The next one, after the doublet, is septet (7-plet) with $T = 3$ and $Y=4$.
Its components are $\Phi_7 = (\phi_7^{5+},\, \phi_7^{4+},\, \dots,\, \phi_7^0,\, \phi_7^-)^T$,
with $\phi_7^0$ acquiring a non-zero vev, which by construction preserves $\rho=1$.
Collider phenomenology of this extension was explored in 
\cite{Hisano:2013sn,Kanemura:2013mc,Alvarado:2014jva,Geng:2014oea,Kanemura:2014bqa}.
There also exist radiative neutrino mass models based on such high scalar multiplets \cite{Aranda:2015xoa,Nomura:2016jnl}.
Dark matter candidates coming from large scalar multiplets were studied in \cite{Logan:2016ivc}.

\subsection{Georgi-Machacek model}

In models with scalar triplets, there exists an elegent way to keep $\rho$-parameter equal to one
at the tree level. It was suggested in 1985 by Georgi and Machacek \cite{Georgi:1985nv}
and Chanowitz and Golden \cite{Chanowitz:1985ug} and goes under the name of Georgi-Machacek model (GM).

Consider a model with two triplets, $T = 1$: a real $\Xi = (\xi^+,\, \xi^0,\, \xi^-)^T$ with $Y=0$ with 
vev $\lr{\xi} = (0,\, v_\xi,\, 0)^T$, and a complex $X = (\chi^{++},\, \chi^+,\, \chi^0)^T$ with $Y = 2$ 
and vev $\lr{\chi} = (0,\, 0,\, v_\chi)^T$, in addition to the usual doublet $\phi$ with vev $\lr{\phi^0} = v_\phi/\sqrt{2}$.
The $\rho$-parameter in this model becomes
\be
\rho = {v_\phi^2 + 4 v_\xi^2 + 4 v_\chi^2 \over v_\phi^2 + 8 v_\chi^2}\,,\label{rho-triplets} 
\ee
For $v_\xi \not = v_\chi$, we obtain a significant departure from unity.
The essence of the GM model is to make $v_\xi$ and $v_\chi$ equal via a natural construction.
To achieve that, we organize the two triplets into a single bi-triplet
\be
\Delta = (\tilde X, \Xi, X) = \left(
\begin{array}{ccc}
\chi^{0*} & \xi^+ & \chi^{++} \\
-\chi^{+*} & \xi^0 & \chi^{+} \\
\chi^{++*} & -\xi^{+*} & \chi^{0} 
\end{array}\right)\,,
\ee
and build a scalar sector with $\Delta$ and bidoublet $\hat\Phi$ (\ref{bidoublet}).
The kinetic term $\Tr[(D_\mu \Delta)^\dagger (D_\mu \Delta)]/2$ produces
$|D_\mu X|^2 + (D_\mu \Xi)^2/2$, while the general renormalizable potential contains 
several terms, including trilinear ones, which were impossible for multi-doublet models.
The presence of the trilinear terms in the scalar potential guarantees 
that the vev of the usual doublet triggers a non-zero vev of the bi-triplet.
The parameters of the potential must satisfy the usual stability bounds and 
perturbativity constraints \cite{Chiang:2012cn,Hartling:2014xma}.

By construction, the potential is invariant not only under the gauged left $U_L \in SU(2)_L$
but also under the global right $U_R \in SU(2)_R$ transformations, which act on the bi-triplet
$\Delta \to U_L \Delta U_R$ via their three-dimensional representations.
The key feature is that in a large part of the parameter space
the minimum of the scalar potential is achieved at 
\be
\lr{\Delta} = \mathrm{diag}(v_\chi,\, v_\chi,\, v_\chi)\,, \label{GM-vev}
\ee
which automatically leads to $v_\xi = v_\chi$ in (\ref{rho-triplets}) and $\rho=1$.
The origin of this relation is the restoration of approximate custodial symmetry: not only
the bidoublet vev, but also the bi-triplet vev (\ref{GM-vev}) is invariant under the diagonal
subgroup $SU(2)_V \subset SU(2)_L \times SU(2)_R$.

Counting bosonic degrees of freedom, 
we start with one (bi)doublet $\hat\Phi$ and one bi-triplet $\Delta$, which contain $4+9=13$ scalar fields.
Three of them become the would-be Goldstone modes, 
and the remaining 10 form multiplets under the custodial symmetry: one 5-plet 
$H_5 = (H_5^{++}, H_5^+, H_5^0, H_5^-, H_5^{--})$,
one triplet $H_3 = (H_3^+, H_3^0, H_3^-)$, and two neutral singlets, $H_1$ and $h$,
the last one identified with $h_{125}$.
Their explicit allocation in $\Phi$ and $\Delta$ is given in \cite{Chiang:2012cn,Hartling:2014xma}.
The 5-plet resides entirely in $\Delta$, while the Goldstone modes and the physical Higgs triplet come from 
both $\Phi$ and $\Delta$. Mixing between them can be parametrized with the angle $\beta$ in a way 
similar to 2HDM.
The two neutral singlets mix with the angle $\alpha$, also similarly to $CP$-conserving 2HDM.
Because of the custodial symmetry of the tree-level scalar potential, 
all states within the 5-plet and within the triplet are mass-degenerate.
Loop corrections lead to mass splitting, which stays small.

The phenomenology of the Georgi-Machacek models was studied in the last few years in quite a number of works.
Hartling et al \cite{Hartling:2014xma} produced a computer code GMCALC which allows one 
to conveniently explore the parameter space of the model.
Ref.~\cite{Hartling:2014xma} also contains a detailed list of vertices and basic processes.
Below, we will briefly list essential phenomenological features arising in the GM models.

Decay modes and widths for various scalars,
including the doubly charged $H^{++}$ which was absent in multi-doublet case,
are listed in \cite{Hartling:2014xma}.
Kanemura et al \cite{Kanemura:2014bqa} gave expectations for future colliders for various
extended scalar sectors, including the GM model.
One remark concerning the heavy neutral singlet $H$ is that its search can be tricky.
As it mixes with the doublet component, it interacts with fermions,
and can be produced in $gg$ fusion with the pb-level cross section \cite{Godunov:2015lea}. 
But the decay rates $H \to hh$ and $H \to VV$, although having the same leading dependence 
$\propto v_\chi^2 M_H^3/v_\phi^4$, contain significantly different numerical prefactors.
Estimates show \cite{Godunov:2015lea} that for $M_H = 300$ GeV and $v_\chi = 20$ GeV,
the decay channel $H \to hh$ will be strongly dominant, while Br$(H\to ZZ)$ stays below one percent.
A detailed analysis of the LHC reach of this process was recently presented in \cite{Chang:2017niy}.
It should be noted, however, that an enhancement of the double SM-like Higgs production 
without any accompanying $ZZ$ enhancement is not specific to the GM model;
the same conclusion was obtained for the real singlet extension \cite{Godunov:2015nea}.

Another feature of scalar triplets which cannot be mimicked by doublets and singlets
is the presence of the tree-level vertex $W^\pm Z H^\mp$ \cite{Grifols:1980uq}. 
In multi-doublet models is arises only at loop level, although in certain cases
is can be additionally enhanced \cite{Moretti:2015tva}.
Its absence at tree level is essentially due to orthogonality of charged Higgses to the Goldstone modes,
which is most clearly seen in the Higgs basis. Such a vertex would require that three factors $v$, $W^\pm$ and $Z$
come from the first doublet with vev, and one field $H^\mp$ comes from extra doublets without vev,
which is impossible to organize. In GM model, we have additional charged Higgses in the 5-plet $H^\pm_5$,
which reside inside $\Delta$ and are not orthogonal to the Goldstone modes.

Another firm prediction of multi-doublet models is that the $HWW$ and $HZZ$ couplings
for each neutral Higgs are less than one (\ref{2HDM-hVV}) and that their squares sum up to unity. 
This sum rule is required by the unitarization of the longitudinally polarized vector boson scattering 
at high energies, in particular $W^+W^- \to W^+ W^-$, 
in which the Higgs exchange cancels the $E^2/v^2$ growth of the cross section \cite{Gunion:1990kf}.
In triplet models, this sum rule is modified because the doubly charged scalar can now propagate in the $u$-channel
and compensate for the overshooting by the $HVV$ couplings \cite{Falkowski:2012vh}. 

Finally, we mention that the idea behind the GM model is not specific to triplets and can be extended to even higher multiplets.
One should however be careful not to run into troubles with perturbative unitarity. 
Logan and Rentala \cite{Logan:2015xpa} explored all such generalized GM models not violating these constraints, 
the largest one being GCM6 with three complex sextets.
Their general scalar potentials were analyzed and an overview of phenomenological features was presented.
One strong conclusion was an absolute upper bound
on the SM-like Higgs coupling to $VV$ possible in the GM framework: $\kappa_V^h < 2.36$.


\section{Astroparticle and cosmological implications}\label{section-astro}



\subsection{Scalar dark matter models}

\begin{figure}[htb]
\centering
\includegraphics[width=0.85\textwidth]{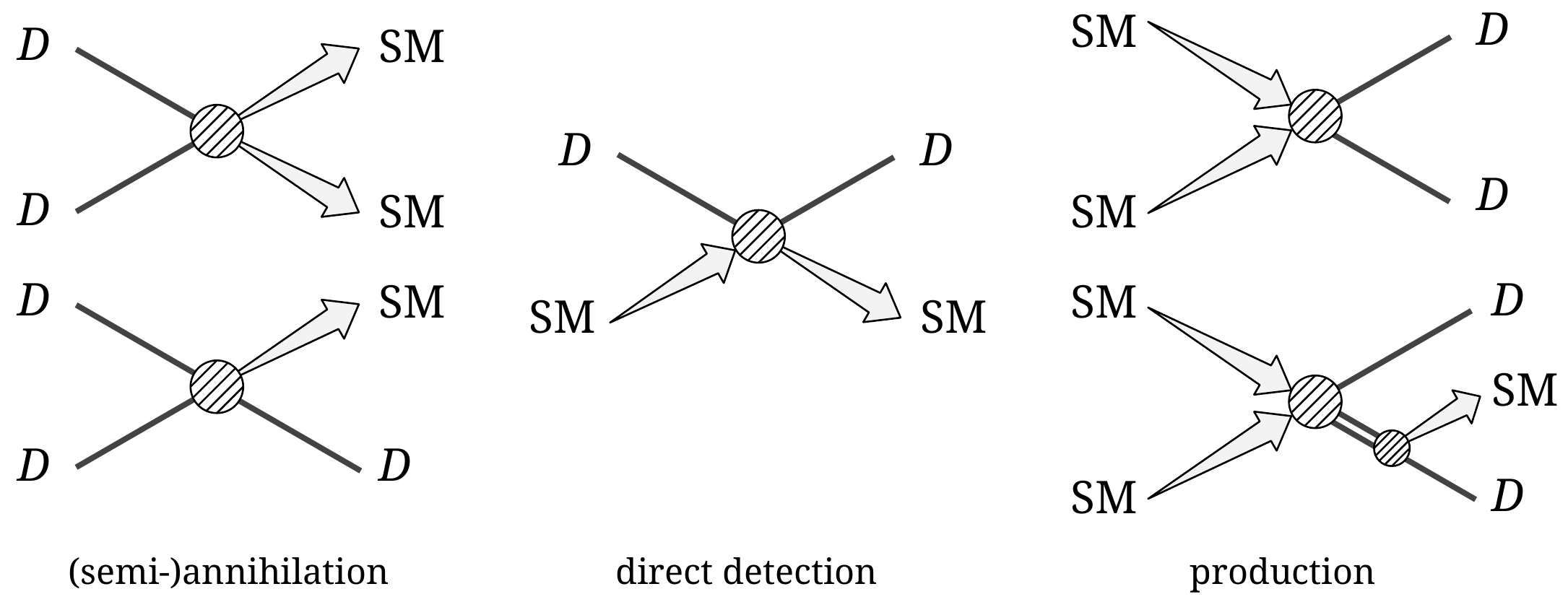}
\caption{Complementary ways to explore models with DM candidates. Left: annihilation or semiannihilation processes,
either in early Universe or at present, middle: DM scattering in DM direct detection experiments, 
right: pair production of new states possibly with a cascade decay inside the dark sector.}
\label{fig-DM-channels}
\end{figure}

Scalar dark matter (DM) candidates naturally arise in extended Higgs sectors 
equipped with global symmetries, 
under which the SM fields remain invariant while the extra scalars transform non-trivially.
In a large part of the parameter space these symmetries stay intact after EWSB 
and protect the lightest non-trivially transforming scalar against decay into SM fields.
An attractive feature of such models is that presence of DM candidates
can manifest itself in three complementary channels: annihilation, scattering, and production, Fig.~\ref{fig-DM-channels},
which can be probed via very different observables. 
Let us denote the DM candidates by $D$
and the heavier scalars from the same multiplet by $D'$.
Then annihilation $DD \to$ SM, coannihilation $DD' \to$ SM, or semi-annihilation $DD \to D +$SM
processes affect evolution of the early hot Universe and determine the DM relic abundance.
They also lead to present day signals such as $\gamma$-rays emission from galactic centers.
Scattering processes may produce signals in DM direct detection experiments.
The production channel if kinematically open will lead to large missing-$E_T$ signatures, 
often accompanied with additional objects from $DD' \to DD+$SM chain, which are observable at colliders
even if the DM particles are completely invisible. For light DM, it will also contribute to the invisible SM-like 
Higgs boson decay width. 
In fact, a decade ago the potentially dominant invisible Higgs decay was a source of worries that one could miss 
the sufficiently light Higgs boson at colliders \cite{Burgess:2000yq}; a one particular realization of this scenario
was described and investigated in \cite{He:2011de}.
All these signals are correlated and help constrain the parameter space.

Below, we highlight two simple scalar sectors which received attention
as viable sources of scalar DM candidates: the singlet extension models and the Inert doublet model.
Of course, several other scalar sectors have been studied as well.
One specific direction which received much attention goes under the name of ``minimal dark matter''
\cite{Cirelli:2005uq,Cirelli:2007xd,Cirelli:2009uv}. Here, one adds extra fields, which can be scalars, in a higher-dimensional 
representation of the $SU(2)_L$ group and demands that only renormalizable interactions be present in the lagrangian.
If the lightest component of the multiplet is neutral, it is automatically a DM candidate because
no renormalizable interaction can connect the new multiplet with the SM fields and induce its decay.
More sophisticated models include triplets \cite{FileviezPerez:2008bj,Araki:2011hm,Bahrami:2015mwa} 
or a triplet and a singlet \cite{Kanemura:2012rj,Wang:2012ts},
and large scalar multiplet DM models \cite{Logan:2016ivc}.
They have been invoked not only to introduce DM but also to explain small neutrino masses 
or some astrophysical observations, such as the 130 GeV $\gamma$ line which several groups seemed to observe 
in the Fermi-LAT data. A detailed overview of such models can be found in \cite{Assamagan:2016azc}.

\subsubsection{Singlets}

In models with one real gauge singlet, section~\ref{section-singlets}, 
the conserved $\Z_2$-symmetry, which flips the sign of the singlet, protects it from decay 
as well as from mixing with the neutral component of the doublet. 
Since the singlet is electroweak-blind, it interacts with SM fields only via the Higgs sector.
The three new parameters of the model (\ref{singlet1}) can be chosen 
as the DM candidate mass $m_D$, the SM-DM coupling $\lambda_3$, and the self-coupling
within the scalar sector $\lambda_2$, which is usually poorly constrained.

Already the first studies \cite{Silveira:1985rk,McDonald:1993ex,Bento:2000ah,Burgess:2000yq} showed
that the correct DM relic abundance can be reproduced within this minimalistic model
with the DM mass at the electroweak scale and a reasonably large SM-DM coupling $\lambda_3 \sim 0.1$.
When calculating DM relic abundance, one typically sees three different $m_D$ regions:
the small mass region, the funnel region $m_D \approx m_{h_{125}}/2$, and the large mass region.
As the annihilation proceeds via the $s$-channel Higgs diagram,
one can deduce the asymptotic behavior in the small and large mass regions \cite{Burgess:2000yq}:
\be
\lr{\sigma_{\mathrm{ann}}v} \propto {\lambda_3^2 m_D^2 \over m_{h_{125}}^4}\quad\mbox{for $m_D \ll m_{h_{125}}$}\,,
\quad 
\lr{\sigma_{\mathrm{ann}}v} \propto {\lambda_3^2 \over m_D^2 }\quad\mbox{for $m_D \gg m_{h_{125}}$}\,.
\ee
Deep into these two asymptotics, one needs a large $\lambda_3$ to prevent the cross section from becoming 
too small and the DM relic abundance too large.
Above a few TeV, one can run into troubles with perturbativity constraints.
Near the Higgs pole, the annihilation proceeds very efficiently, and if one wants to stay exactly at the measured relic abundance,
$\lambda_3 \sim 10^{-3}\div 10^{-4}$ are required. 
Away from the Higgs pole, the spin-independent elastic scattering cross section is typically 
of the order of $10^{-44}\div 10^{-43}$ cm$^2$ \cite{Burgess:2000yq}.
Two decades ago, it was below the direct detection limits but now this region is being probed by the present day direct detection experiments.
In the funnel region, if the coupling constant is indeed kept small, the elastic scattering cross section dives down by several orders of magnitude.

\begin{figure}[htb]
\centering
\includegraphics[width=0.75\textwidth]{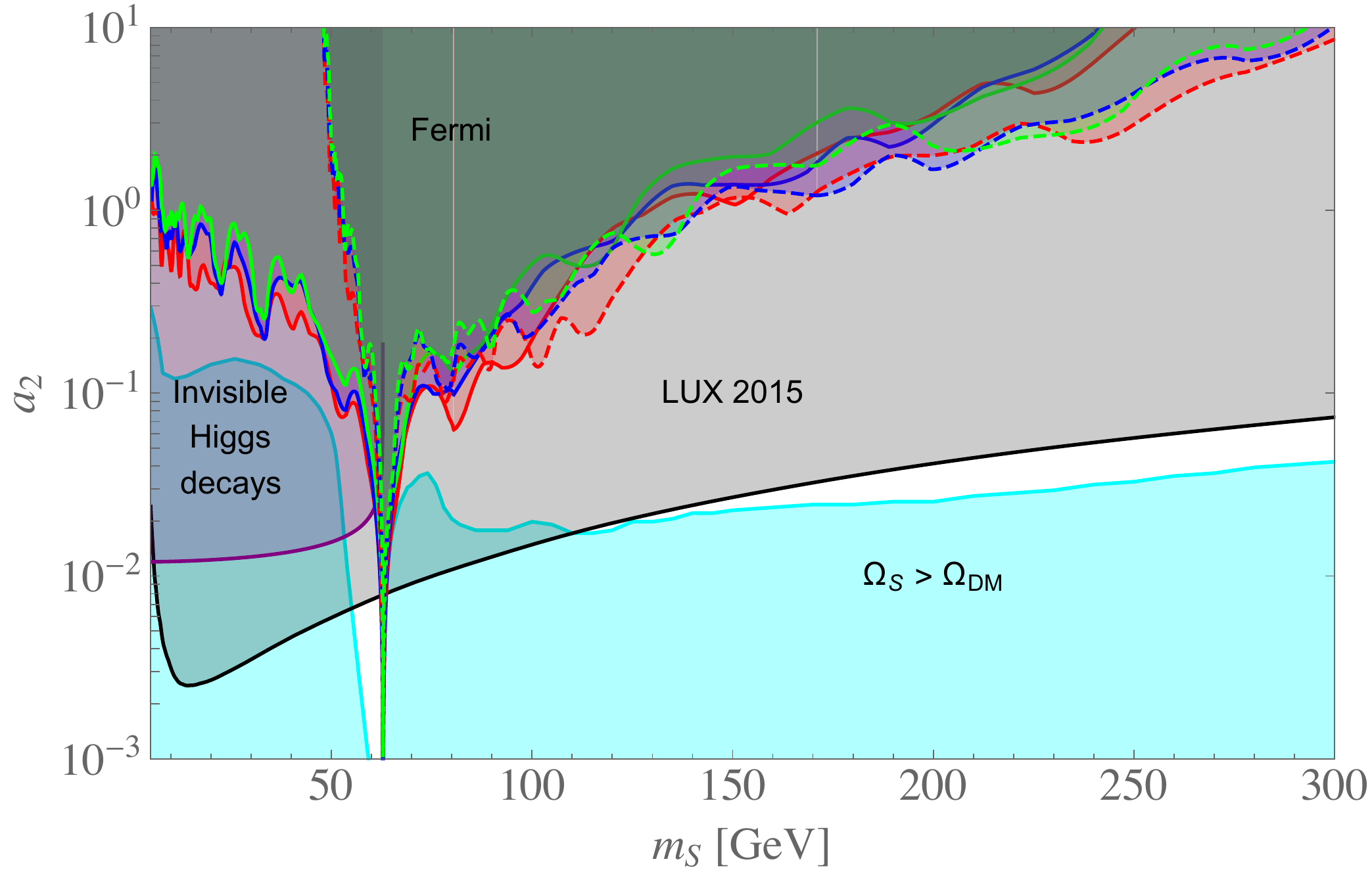}
\caption{Constraints on two parameters $(m_S \equiv m_D, a_2 \equiv \lambda_3) $ of the real singlet extension. 
Several shared regions are excluded by various observations. The white space is not yet excluded.
Reproduced with permission from the arXiv version of \cite{Feng:2014vea}.}
\label{fig-singlet-DM}
\end{figure}

The steady progress in DM direct detection experiments kept excluding more and more regions
in the parameter space \cite{He:2008qm}. The evidence for the Higgs boson at about 125 GeV which accumulated
towards the end of 2011 pinpointed the location of the funnel region \cite{He:2011gc}.
The current situation in the real singlet extension is shown in Fig.~\ref{fig-singlet-DM} 
from the arXiv version of \cite{Feng:2014vea} updated in 2016 (see also similar analyses in \cite{He:2016mls,Casas:2017jjg}). 
The combination of indirect detection limits from Fermi,
see a detailed analysis in \cite{Cuoco:2016jqt},
the Planck results on relic abundance \cite{Ade:2015xua}, 
the LHC Higgs data, and most importantly the LUX direct detection constraints \cite{Akerib:2015rjg}
close almost the entire parameter space of the model.
Only two small regions remain viable: (i) within the small mass range $55 < m_D < 63$ GeV
and within the interval of the DM-SM coupling $10^{-4}\lsim \lambda_3 \lsim 10^{-2}$, 
and (ii) for $m_D > 110$ GeV, again in a narrow range of coupling.
The former region can significantly shrink and eventually become excluded only after the future indirect detection observations
reach at least an order of magnitude better sensitivity than Fermi-LAT.
In the latter region, with $m_D > 110$ GeV and up to a few TeV, an order of magnitude improvement in the direct detection sensitivity 
will put this model to the conclusive test \cite{Feng:2014vea,He:2016mls}.
This will be a good illustration how minimalistic, highly predictive models can get constrained and eventually ruled out
by complementary measurements.

One real singlet can take care either of DM or of strong electroweak phase transition but not both. 
These two cosmological consequences can be achieved simultaneously in the complex singlet extension,
in which only one of the two extra scalars is stable. 
There have been several extensive analyses of the DM properties in complex singlet extensions \cite{Barger:2008jx,Barger:2010yn,Gonderinger:2012rd}; for more details see the overview \cite{Assamagan:2016azc}.
Models with two singlet scalars \cite{Abada:2011qb,Casas:2017jjg}
as well as combinations of the 2HDM with singlet scalar DM \cite{He:2011gc,vonBuddenbrock:2016rmr,He:2016mls}
have also been closely studied.

To close this section, we mention another interesting way how the singlet scalars 
can affect dark matter relic density: via a sequence of phase transitions.
Baker and Kopp \cite{Baker:2016xzo} proposed recently a model with ``vev flip-flop'', in which a sequence of two phase transitions
takes place as the Universe is cooling down. 
The DM candidates in this model are new fermions. They couple to the scalar singlet messenger $\S$ 
and freeze-out early and with a large relic abundance.
Then, at the first transition, $\S$ develops vev, which destabilizes DM and allows it to decay
until the second phase transition sets in, which restores $\lr{\S} = 0$ and generates the usual Higgs vev.
The DM relic abundance then stabilizes at the new level and it depends on the time interval between the two transitions
available for decay.

\subsubsection{Inert doublet model}\label{section-IDM}

2HDM can easily incorporate scalar DM candidates.
Taking Type I 2HDM and postulating the exact $\Z_2$ symmetry (\ref{Z2-2HDM})
in the scalar potential, one can obtain $\lr{\phi_2}=0$
in a large part of the parameter space without any fine-tuning \cite{Deshpande:1977rw}.
The scalars from the second, ``inert'' doublet are odd under $\Z_2$ and are protected against decay 
into the pure SM final states. 
Due to their mass splitting, the inert scalars can sequentially
decay into one another but the lightest scalar $h_1$ is stable and serves as the DM candidate.
Pairwise interactions between these scalars are still possible,
leading to DM annihilation and co-annihilation processes as well as to collider signatures.

This model known as the Inert doublet model (IDM) was resurrected a decade ago 
\cite{Ma:2006km,Barbieri:2006dq,LopezHonorez:2006gr} and its phenomenology has been explored in great detail.
It is easily doable with analytical calculations, it can be implemented in computer codes, 
and its parameter space is relatively small, allowing for efficient parameter scans.
It is predictive and can be strongly constrained by the present and future data.
Several comprehensive analyses of astroparticle and collider constraints on the IDM parameter space
have been published \cite{Goudelis:2013uca,Arhrib:2013ela,Ilnicka:2015jba,Datta:2016nfz,Belyaev:2016lok};
below we borrow some illustrative examples from \cite{Belyaev:2016lok}.

The scalar potential of IDM has the form (\ref{V-2HDM}) without the $m_{12}^2$, $\lambda_6$, and $\lambda_7$ terms.
All of its free parameters are real and satisfy the usual stability bounds (\ref{stability-2HDM}).
In addition, one assumes $|\lambda_5|-\lambda_4 > 0$ which guarantees that the vacuum
is neutral and that the charged inert scalars are heavier than the lightest neutral one.
The scalar spectrum in the inert sector is 
\be
M_{h^+}^2 = {1\over 2} \lambda_3 v^2 + m_{22}^2\,,\quad
M_{h_1}^2, M_{h_2}^2 = {1\over 2}(\lambda_3 +\lambda_4 \mp |\lambda_5|) v^2 + m_{22}^2\,.\label{IDM-masses}
\ee
In this way, $|\lambda_5| - \lambda_4$ defines the splitting between the charged and the lightest neutral scalars,
while $|\lambda_5|$ alone shows the splitting between the two neutrals $h_1$ and $h_2$.
The notation $(H, A)$ is often used instead of $(h_1, h_2)$ and it alludes to the specific $CP$-parity assignment
of the two neutral scalars. As we explained in section~\ref{section-CPV-2HDM}, this assignment is a matter of convention. 
Although $h_1$ and $h_2$ indeed have opposite $CP$-parities, which is manifest via the $Zh_1h_2$ vertex,
it is impossible to unambiguously assign 
which of them is $CP$-even and which is $CP$-odd.
In the absence of direct coupling to fermions, the model has two $CP$-symmetries, $h_1 \to h_1, h_2 \to -h_2$ and
$h_1 \to -h_1, h_2 \to h_2$, which get interchanged upon the basis change $\phi_2 \to i \phi_2$. 
Either can be used as ``the $CP$-symmetry'' of the model, 
making the specification of the $CP$ properties of $h_1$ and $h_2$ a basis dependent statement. 

The five IDM free parameters can be chosen, for example, as the three masses $M_{h_1}$, $M_{h_2} > M_{h_1}$, and $M_{h^+}$, 
the combination of quartic couplings $\lambda_{345} \equiv \lambda_3 +\lambda_4 - |\lambda_5|$, which determines the 
SM-DM interaction vertex $h_{125}h_1h_1$, and the parameter $\lambda_2 > 0$, which governs the self-interaction
inside the inert sector. Stability and perturbative unitarity places theoretical bounds
on $\lambda_{345}$ and $\lambda_2$. The values of $\lambda_{345}$ strongly affect phenomenology,
while $\lambda_2$ remains essentially unconstrained by data.

\begin{figure}[htb]
\centering
\includegraphics[width=0.49\textwidth]{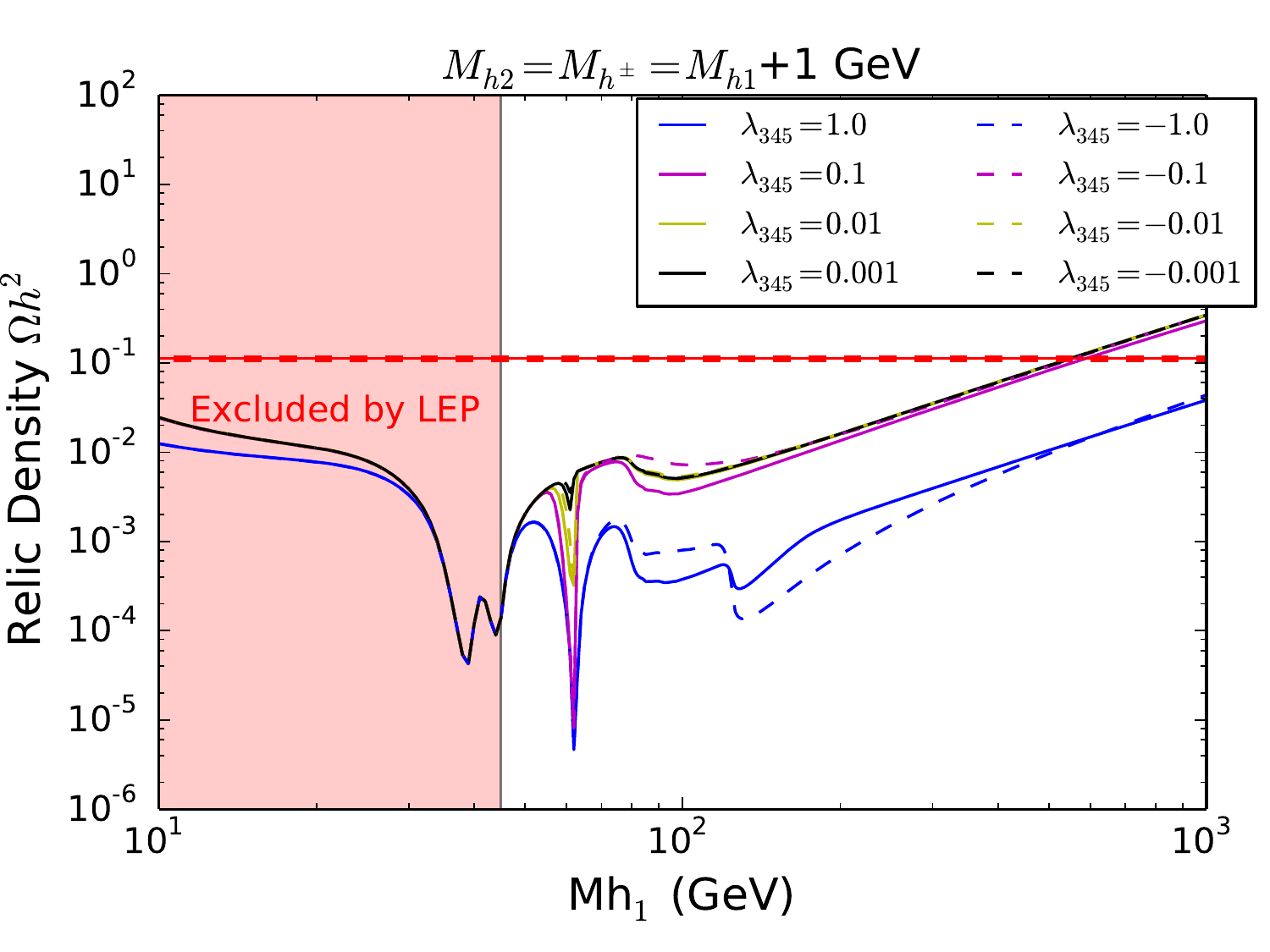}\hfill
\includegraphics[width=0.49\textwidth]{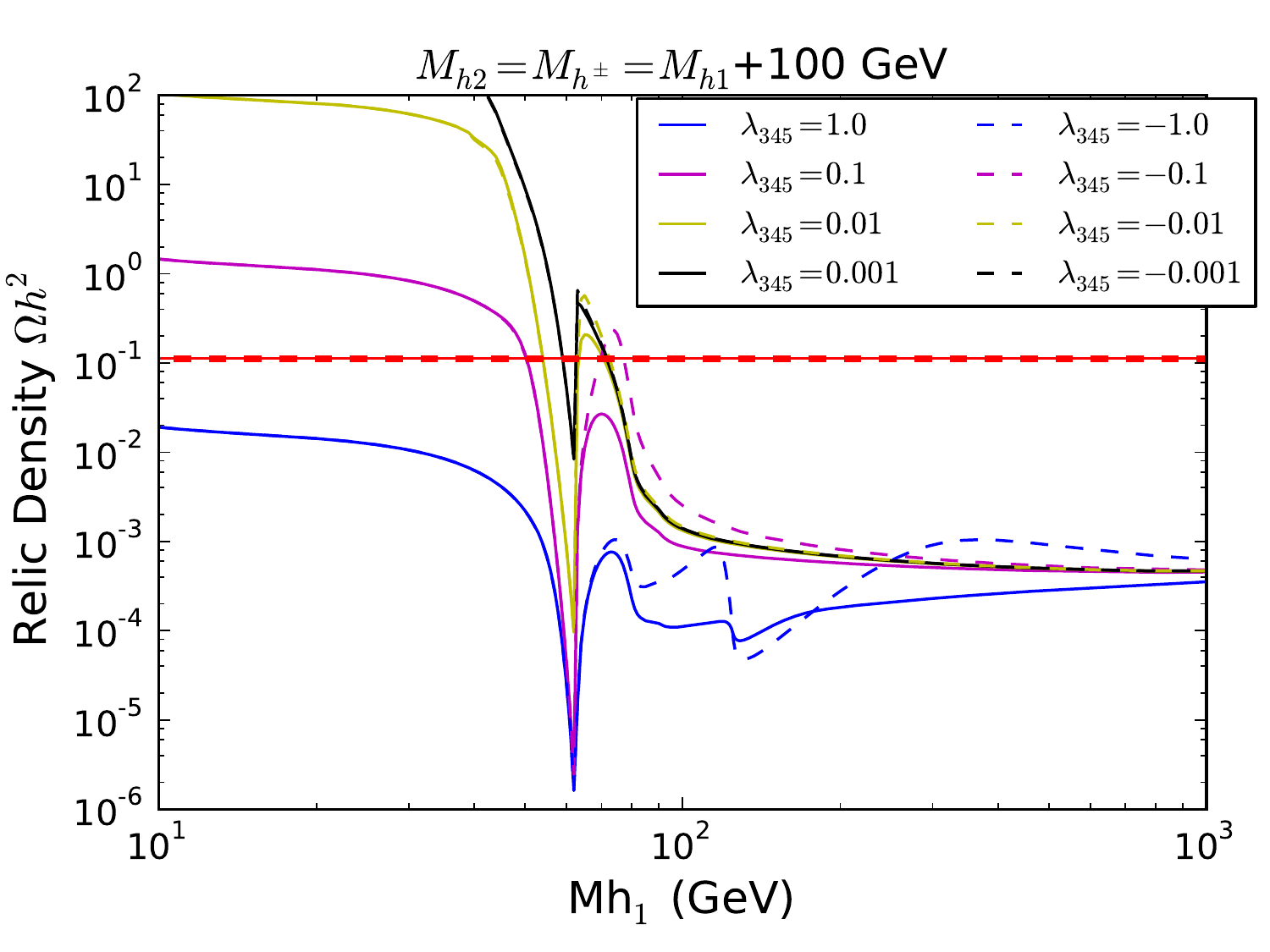}
\caption{The DM relic density as a function of $M_{h_1}$
for various $\lambda_{345}$ parameters for small (left) and large (right) mass splitting.
The horizontal line corresponds to the relic density obtained by Planck \cite{Ade:2015xua}.
Reproduced from \cite{Belyaev:2016lok}.}
\label{fig-IDM-Omega}
\end{figure}

To illustrate how the collider data and astrophysical observations constrain the parameter space of IDM,
we show in Fig.~\ref{fig-IDM-Omega} the results for the DM relic density $\Omega_{\rm DM} h^2$ \cite{Belyaev:2016lok} obtained with the
{\texttt{MicrOMEGAs 2.4.1}} package~\cite{Belanger:2013oya}.
The left and right plots correspond, respectively, to quasi-degenerate $h_1,h_2$ and $h^+$ masses, $M_{h_2}=M_{h^+}= M_{h_1}+$1~GeV,
and large mass splitting $M_{h_2}=M_{h^+}=M_{h_1}+$100~GeV;
on both plots, several scenarios with positive and negative $\lambda_{345}$ are shown. 
The relic density result obtained by Planck $\Omega_{\rm DM}^{\rm Planck} h^2=0.1184\pm0.0012$ \cite{Ade:2015xua} 
is shown by the horizontal line.
If one leaves room for other mechanisms which can produce additional DM candidates,
this value should be considered as an upper bound rather than the absolute value to be fitted within IDM.

Several features are visible on these plots.
For the light DM masses at small mass splitting, the curves go below the Planck value and display two dips 
at 40 and 45 GeV, corresponding to the resonant coannihilation $h_1 h_2 \to Z$ and $h_1 h^+ \to W^+$.
These processes are governed by the gauge couplings, and they efficiently reduce the relic density even at zero $\lambda_{345}$.
Since the inverse processes, the $W$ and $Z$ boson decays to light inert scalars, would also take place,
this region is already closed by the LEP data.
For large mass splitting, the dips disappear since $M_{h_1}+M_{h_2}$ or $M_{h_1}+M_{h^+}$ 
are above $M_Z$ and $M_W$, respectively. 
Coannihilation is suppressed, and the relic density can be brought below the Planck value 
only at large $|\lambda_{345}| \gsim 0.3$. 
However, these large values are already ruled out by the non-observation of the invisible Higgs decay at the LHC:
the upper limit of Br$(h_{125}\to \mathrm{inv}) < 0.24$ currently established by 
the ATLAS \cite{Aad:2015pla} and CMS \cite{Khachatryan:2016whc} collaborations 
translates to $|\lambda_{345}| \lsim 0.02$ to $0.04$
depending on the exact value of $M_{h_1}$. As a result, the {\em entire} small-mass region $m_{h_1} < 45$ GeV is ruled out 
by the combination of the Planck and LHC data \cite{Krawczyk:2013jta,Ilnicka:2015jba,Diaz:2015pyv,Belyaev:2016lok}.

In the funnel region, $M_{h_1} \approx 65$ GeV, one observes the same characteristic sharp dip in the relic density, which 
corresponds to the DM annihilation through the Higgs boson $h_1 h_1 \to h_{125}$.
At higher masses, there is a wider and more shallow dip at about 80-90 GeV which comes from 
$h_1 h_1\to W^+W^-$ and $h_1 h_1\to ZZ$ channels merged together. 
The last dip around 125 GeV corresponds to the DM relic density reduction due to the opening 
of the $h_1 h_1\to h_{125}h_{125}$ annihilation channel and is sensitive to the values of $\lambda_{345}$.
The positive (solid lines) and negative (dashed lines) values of $\lambda_{345}$ differ due to interference between 
the $s$-channel Higgs boson process with the rest of annihilation diagrams.

For large DM masses, there is a qualitative difference in the asymptotic behavior of the DM relic density for small and large mass splitting.
In this region, the main contribution comes from the $h_1 h_1$ annihilation into (longitudinal) vector bosons.
The quartic coupling $h_1h_1Z_LZ_L$ depends on the parameters as $2(M_{h_2}^2-M_{h_1}^2)/v^2+\lambda_{345}$
and is very different for small and large mass splitting.
For small splitting, this coupling stays almost constant vs. $M_{h_1}$ and generates
a low $h_1 h_1$ annihilation cross section leading to the relic density rising with $M_{h_1}$. 
For large mass splitting, this vertex grows with $M_{h_1}$ boosting the annihilation cross section,
so that the DM density stays well below the experimental limit.
Similar qualitatively different regimes for small and large mass splittings can be derived for 
DM scattering cross section on matter. This means that if one wishes to exactly fit the Planck result
in the high-mass region, the mass splitting between inert Higgses must be small, a few GeV at most.
It cannot be too small, though: if $M_{h^+} - M_{h_1} < 200$ MeV, the charged inert Higgs
will be sufficiently metastable to produce (disappearing) charged tracks \cite{Belyaev:2016lok}.

Additional constraints on the parameter space and DM candidate properties
can be also deduced from the theoretical assumption of full stability of the IDM potential up to Planck scale
\cite{Chakrabarty:2015yia,Khan:2015ipa}.
IDM can also produce signals for direct \cite{Arina:2009um} and indirect DM search experiments 
via heavy inert scalar annihilation, which can be detectable via $\gamma$-rays 
\cite{Modak:2015uda,Queiroz:2015utg,Garcia-Cely:2015khw}
or via its neutrino \cite{Agrawal:2008xz,Andreas:2009hj} and cosmic-ray signals \cite{Nezri:2009jd}.
Collider signatures of IDM include mono-$Z$ or mono-jet + missing $E_T$ signatures
\cite{Arhrib:2013ela,Ilnicka:2015jba,Belyaev:2016lok}, dijets or dijets plus dileptons 
with missing $E_T$ \cite{Poulose:2016lvz,Hashemi:2016wup},
multilepton signals \cite{Miao:2010rg,Gustafsson:2012aj,Hashemi:2016wup,Datta:2016nfz},
modified rates of the SM-like Higgs decays to $\gamma\gamma$ and $\gamma Z$ 
\cite{Arhrib:2012ia,Swiezewska:2012eh,Krawczyk:2013jta,Krawczyk:2013pea},
long-lived charged scalars and other exotic signals.
Prospects of detecting and exploring IDM at ILC were discussed in \cite{Aoki:2013lhm,Hashemi:2015swh}.

Finally, models in which the IDM is accompanied by yet more scalars have also been considered,
such as IDM with a complex singlet \cite{Bonilla:2014xba} or (1+2)IDM with one usual and two inert
doublets \cite{Fortes:2014dca,Keus:2014jha,Ahriche:2015mea}. 
The 3HDM based on the exact order-4 GCP symmetry \cite{Ivanov:2015mwl,Aranda:2016qmp}
mentioned in section~\ref{section-NHDM-CPV} also bears the IDM features and 
contains two mass-degenerate non-coannihilating DM candidates.
All such models strongly differ in the dark sector, but whether the non-trivial scalar dark sector 
can be probed experimentally remains an open question.



\subsection{Phase transitions in early Universe}

Extended scalar sectors can have profound cosmological implications for the early Universe.
Extra scalars modify the Higgs potential not only at zero but also at finite temperatures,
which echoes in the strength of the electroweak phase transition (EWPT),
which is indispensable for generation of the observed baryon asymmetry of the Universe.
In general, 
thermally activated sphaleron processes 
tend to wash out preexisting or newly generated baryon asymmetry.
In order for the baryon asymmetry to persist, the phase transition to the Higgs phase must be
of the first order and sufficiently strong to make the sphalerons heavy and to strongly suppress the wash-out processes.
To efficiently cut-off the rate of these processes in the expanding Universe, 
the sphaleron mass needs to be $M_{sp}(T_c)/T_c \gsim 45$ \cite{Shaposhnikov:1986jp,Shaposhnikov:1987tw}.
Being proportional to the Higgs vev at the critical temperature,
it translates to the widely adopted criterion for EWPT to be strong: 
\be
{v(T_c) \over T_c} \gsim 1\,.\label{strong-PT}
\ee
The SM Higgs mechanism is unable to fulfill this requirement.
The large value of $v(T_c)$ require small self-coupling $\lambda$, which makes the Higgs boson rather light.
The criterion (\ref{strong-PT}) could be satisfied if $m_h \lsim 45$ GeV
\cite{Shaposhnikov:1987tw}, which since long ago was known to be inconsistent with the collider searches.
Non-perturbative studies on the lattice confirmed this conclusion \cite{Kajantie:1995kf,Rummukainen:1998as,Csikor:1998eu}
by finding that the line of the first-order EWPT on the SM phase diagram ends at about $m_h \sim 70$ GeV.

Extra scalars modify the shape of the Higgs potential and loosen the relation between the Higgs mass
and the jump of the vev at the transition point.
Even the simplest models with a single extra scalar singlet can quite easily satisfy (\ref{strong-PT}) 
in an appropriate parameter range, the cubic terms being especially useful for this purpose \cite{Choi:1993cv}.
The literature on this topic is vast, see e.g.
\cite{Espinosa:2007qk,Profumo:2007wc,Profumo:2014opa,Jiang:2015cwa,Huang:2015tdv,Huang:2016cjm,Hashino:2016rvx,Hashino:2016xoj,Darvishi:2016tni} 
for recent in-depth studies which combine the requirement of the strong EWPT with the collider results.
A detailed review of this field together with a compilation of various Higgs portal models 
can be found in the recent publication \cite{Assamagan:2016azc}.

Just to mention one specific example of the interplay between collider physics and cosmology, 
a recent work \cite{Huang:2015tdv} explored in detail the relation
between the triple Higgs self-coupling and the strength of the EWPT. Starting with an effective Higgs potential
with various higher-order non-renormalizable terms, the authors found that the two quantities indeed correlate
and that the pattern of their correlation depends on the specific higher-order terms added.
The results suggest that an insight into the strength of EWPT can be gained via collider measurements.
The authors also used the simplest singlet extension of the SM scalar sector to construct an ultraviolet completion,
but the conclusions they get do not rely on this particular example.
A similar conclusion was reached in \cite{Hashino:2016rvx}.

Starting from early 1990s, many authors investigated the issue of EWPT within generic 2HDMs
\cite{Bochkarev:1990fx,Turok:1991uc,Cottingham:1995cj,Andersen:1998br,Fromme:2006cm,Cline:2011mm,Dorsch:2013wja}
and within the Inert doublet model \cite{Ginzburg:2010wa,Chowdhury:2011ga,Borah:2012pu,Gil:2012ya,Cline:2013bln,Blinov:2015vma},
as well as in supersymmetric models \cite{Pietroni:1992in,Losada:1996ju,Funakubo:2005pu}.
The general conclusion was that a sufficiently strong phase transition is indeed possible,
allowing the lightest Higgs boson to lie in the experimentally preferred range and the masses of the heavier scalars 
to be of the order of few hundred GeV. For example, the recent detailed studies \cite{Dorsch:2013wja,Dorsch:2016nrg} 
performed a numerical scan over the parameter space of the 2HDM with softly broken $\Z_2$-symmetry, 
taking into account the electroweak precision data, 
the 125 GeV Higgs boson discovered at the LHC and other LHC Run 1 results, 
as well as the current flavor physics constraints. They confirmed that such a transition is well possible
and that it prefers $\tan\beta \sim 1$, an approximate alignment $\beta - \alpha \approx \pi/2$ 
(which reads $\alpha \approx \beta$ with the authors' redefinition of the angle $\alpha$),
and a rather heavy $A$, with $m_{A} \gsim 400$ GeV. 
An even more thorough analysis of the strength of the EWPT in $CP$-conserving 2HDM
was recently completed by Basler et al \cite{Basler:2016obg}. They studied both Type I and Type II models,
identified $h_{125}$ either with $h$ or $H$,
and used two procedures to minimize the loop-corrected effective potential.
The extensive numerical scan identified regions in the parameter space which generate sufficiently strong EWPT.
The LHC Higgs data were fully taken into account and played an important role in shaping these regions,
reiterating the synergy between collider data and astroparticle observations in constraining bSM models.
Haarr et al \cite{Haarr:2016qzq} looked into this issue in Type II softly broken $\Z_2$ model 
with a minimal amount of $CP$-violation and, with the help of Bayesian statistical fit, 
concluded that a strong phase transition is disfavored.
We also need to mention that the 2HDM potential can have several coexisting
minima with different vev alignments even at tree level. For a range of free parameters, they may lead 
to a chain of phase transitions when the Universe was cooling down.
Simplified analyses of this possibility in 2HDM were performed in \cite{Land:1992sm,Ivanov:2008er,Ginzburg:2009dp,Ginzburg:2010wa}.

Beyond two Higgs doublets, EWPT in the model with one usual and two inert Higgs doublets was investigated in \cite{Ahriche:2015mea}.
Some authors also looked into EWPT in models with triplets \cite{Patel:2012pi,Blinov:2015sna}, 
including the Georgi-Machacek model \cite{Chiang:2014hia}.
For example, in \cite{Patel:2012pi,Blinov:2015sna}, a rather minimalistic scalar potential with an additional real triplet $\Sigma$
was considered. To naturally satisfy restrictions coming form the $\rho$-parameter,
the model is constructed in such a way that $\lr{\Sigma^0} = 0$ at zero temperature.
In this case, $\Sigma^0$ serves as a scalar dark matter candidate protected by the $\Z_2$ symmetry.
However, there exists a temperature interval in which $\lr{\Sigma^0} \not = 0$ while the usual Higgs vev is zero.
As a result, while cooling down, the Universe evolved through a two-step phase transition:
first into the exotic phase with $\lr{\Sigma^0} \not = 0$, at which the baryon asymmetry was generated,
and only then into the SM phase. Both phase transitions must be of the first order and sufficiently strong:
the first one needs to produce baryon asymmetry, the second one cannot wash it out.

\begin{figure}[!htb]
   \centering
\includegraphics[height=8cm]{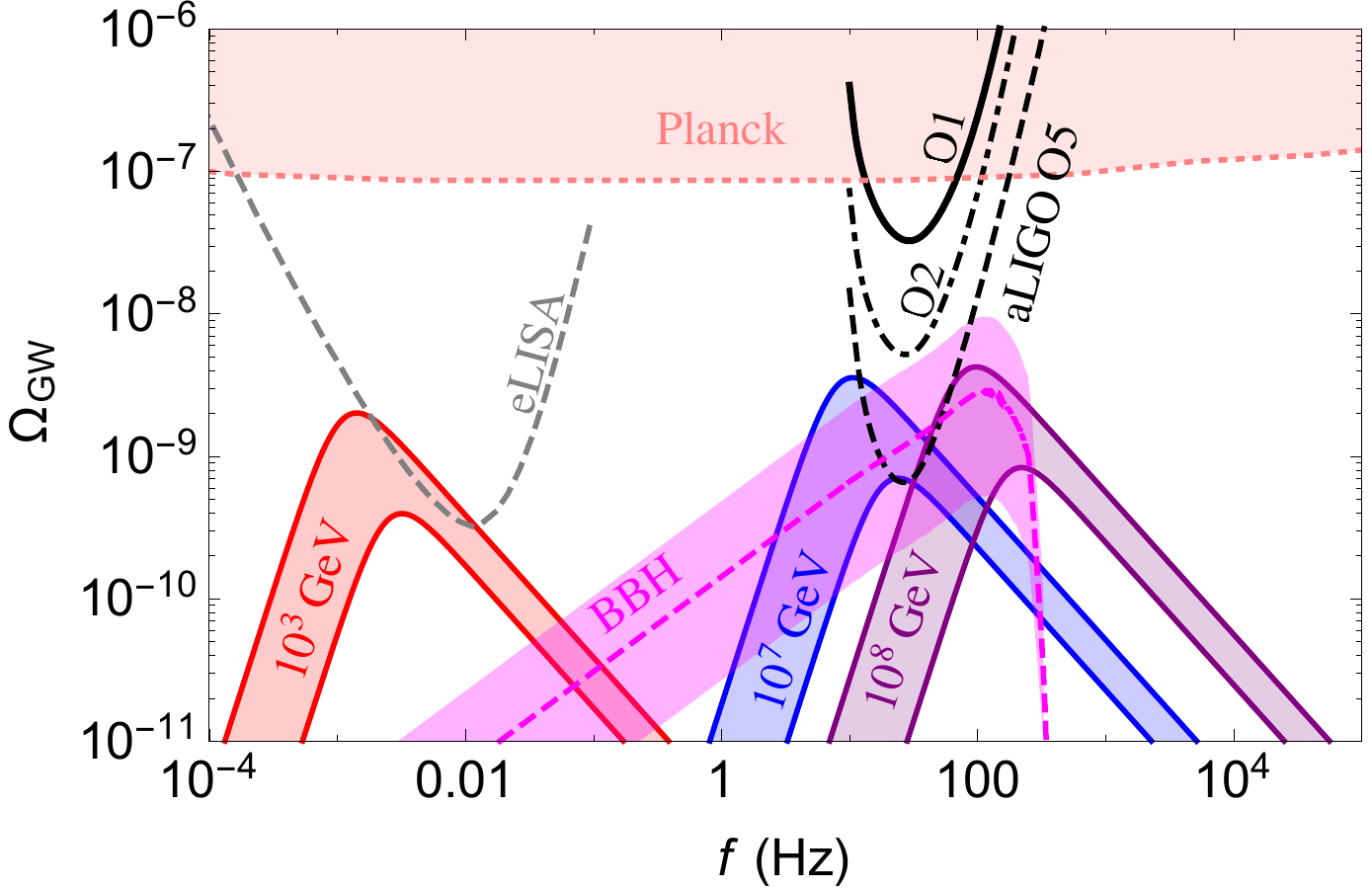}
\caption{Predicted stochastic gravitational-wave spectrum from a first-order cosmological phase transition 
occurring at various critical temperatures. The bands correspond to variations of 
parameters of the calculations. The expected stochastic background from binary black hole mergers 
is also shown for comparison. Reproduced with permission from \cite{Dev:2016feu}.}
   \label{fig-GW-bkgd}
\end{figure}

We wrap up this discussion by mentioning that there can exist a direct experimental window
into that epoch: the stochastic background of gravitational waves generated upon the violent collisions of
walls of the expanding bubbles of the new vacuum.
An initial analysis \cite{Kamionkowski:1993fg} indicated that they might be too weak for a direct detection, 
but several recent works 
\cite{Apreda:2001us,Grojean:2006bp,No:2011fi,Jinno:2015doa,Hashino:2016rvx,Hashino:2016xoj,Dorsch:2016nrg}
based on Higgs potentials with strong EWPTs were more optimistic.
With the typical frequencies peaking in the milli-Hz to Hz range, these gravitational waves become 
a target for future gravitational wave observatories.
Even more intriguing is the possibility that a rich scalar dark sector could also evolve through
a first order phase transition. Although not directly affecting the SM fields, it produces gravitational waves
or cosmic defects which could be observable by the gravitational wave experiments 
\cite{Jaeckel:2016jlh,Dev:2016feu,Chala:2016ykx,Balazs:2016tbi,Baldes:2017rcu,Chao:2017vrq}.
In particular, for the critical temperature of the first-order phase transition in the dark sector around $10^7$ to $10^8$ GeV,
the spectral peak is predicted to shift into the aLIGO operational range, see Fig.~\ref{fig-GW-bkgd}, 
so that the background noise of these primordial cosmic catastrophes
could be probed in aLIGO run O5 \cite{Dev:2016feu,Balazs:2016tbi}.


\section{Conclusions}

The HEP community is still navigating open seas in their quest for laboratory discovery 
of physics beyond the Standard Model.
Building models of New Physics with extended Higgs sectors evolved in the last decade 
into one of the few HEP directions of truly massive, vigorous theoretical and phenomenological activity.
The body of literature in this field is colossal. It keeps expanding at such a rate that one often feels
disconnected from developments in a nearby subfield. It also may represent a significant entry barrier
for those who start mastering the subject.
The goal of this review was to provide an aerial view of this subject,
with a selection of specific models for a more in-depth acquaintance.
With these examples, we illustrated the main objects, goals, methods, and challenges one needs to overcome
in successful models with non-minimal Higgs sectors.

Of course, what ultimately interests us is how Nature functions,
and not the mathematical exercises which the theorists have produced so far.
It is well possible that many of the models covered in this review will become merely of historic interest
in a decade or two. But in order to stay focused and safely interpret the results of many complementary
experiments and observations, we need to understand the spectrum of possibilities in advance.
The bSM community needs to stay sane, fit, and imaginative, and it must stimulate further experimental efforts ---
it is with this spirit that bSM model-building with extended scalar sectors moves on.

\bigskip 

{\bf Acknowledgements.} 
It is my pleasure to thank Gustavo Branco, Otto Eberhardt, Luis Lavoura, Axel Maas, Rui Santos, Avelino Vicente
for their useful comments and suggestions on various parts of the text.
This work was supported by the Portuguese
\textit{Fun\-da\-\c{c}\~{a}o para a Ci\^{e}ncia e a Tecnologia} (FCT)
through the Investigator Grant contract IF/00989/2014/CP1214/CT0004
under the IF2014 Program, and in part by contracts 
UID/FIS/00777/2013 and CERN/FIS-NUC/0010/2015,
which are partially funded through POCTI, COMPETE, QREN, 
and the European Union. 


\bibliography{extended-Higgs-sectors}
\bibliographystyle{elsarticle-num}
\end{document}